\documentclass[structabstract,longauth]{aa} 
\usepackage{graphicx}
\usepackage{amsmath, amssymb}
\usepackage{natbib}
\usepackage[farskip=0pt]{subfig}
\usepackage{bigstrut}
\usepackage{aas_macros}
\usepackage{multirow}
\usepackage[normalem]{ulem}

\def\degr{^{\circ}}
\def\arcmin{^\prime}
\def\arcsec{^{\prime\prime}}
\newcommand{\as}[2]{$#1''\,\hspace{-1.7mm}.\hspace{.1mm}#2$}
\newcommand{\un}[1]{\rm\ #1}

\makeatletter
\newcommand{\vast}{\bBigg@{3.5}}
\makeatother

\begin{document}

\title{M87 at metre wavelengths: the LOFAR picture}
\titlerunning{M87 at metre wavelengths}

\author{
       F.~de~Gasperin\inst{1,2}\and
       E.~Orr\'u \inst{3,5}\and
       M.~Murgia \inst{12}\and
       A.~Merloni \inst{2,4}\and
       H.~Falcke \inst{3,5,13}\and
       R.~Beck \inst{13}\and
       R.~Beswick \inst{10}\and
       L.~B\^irzan \inst{7}\and
       A.~Bonafede \inst{8,24}\and
       M.~Br\"uggen \inst{8,24}\and
       G.~Brunetti \inst{15}\and
       K.~Chyzy \inst{16}\and
       J.~Conway \inst{14}\and
       J.H.~Croston \inst{17}\and
       T.~En\ss lin\inst{1}\and
       C.~Ferrari \inst{9}\and       
       G.~Heald \inst{5}\and
       S.~Heidenreich\inst{17}\and
       N.~Jackson \inst{10}\and
       G.~Macario \inst{9}\and
       J.~McKean \inst{5}\and
       G.~Miley \inst{7}\and
       R.~Morganti \inst{5,6}\and
       A.~Offringa \inst{6}\and
       R.~Pizzo \inst{5}\and  
       D.~Rafferty \inst{7}\and
       H.~R\"ottgering \inst{7}\and
       A.~Shulevski \inst{6}\and
       M.~Steinmetz \inst{18}\and
       C.~Tasse \inst{11}\and
       S.~van~der~Tol \inst{7}\and
       W.~van~Driel \inst{11}\and
       R.~J.~van~Weeren \inst{5,7}\and
       J.~E.~van~Zwieten \inst{5}\and
       A.~Alexov \inst{19}\and
       J.~Anderson \inst{13}\and
       A.~Asgekar\inst{5}\and
       M.~Avruch \inst{20,6}\and
       M.~Bell\inst{17,21}\and
       M.~R.~Bell\inst{1}\and
       M.~Bentum \inst{5}\and
       G.~Bernardi\inst{22,6}\and
       P.~Best\inst{23}\and
       F.~Breitling\inst{18}\and
       J.~W.~Broderick\inst{17}\and
       A.~Butcher\inst{5,25}\and
       B.~Ciardi\inst{1}\and
       R.~J.~Dettmar\inst{26}\and
       J.~Eisloeffel\inst{27}\and
       W.~Frieswijk\inst{5}\and
       H.~Gankema\inst{6}\and
       M.~Garrett\inst{5}\and
       M.~Gerbers\inst{5}\and
       J.~M.~Griessmeier\inst{5,28}\and
       A.~W.~Gunst\inst{5}\and
       T.~E.~Hassall\inst{10,17}\and
       J.~Hessels\inst{5}\and
       M.~Hoeft\inst{27}\and
       A.~Horneffer\inst{13}\and
       A.~Karastergiou\inst{29}\and
       J.~K\"ohler\inst{13}\and
       Y.~Koopman\inst{5}\and
       M.~Kuniyoshi\inst{13}\and
       G.~Kuper\inst{5}\and
       P.~Maat\inst{5}\and
       G.~Mann\inst{18}\and
       M.~Mevius \inst{5}\and
       D.~D.~Mulcahy\inst{13}\and
       H.~Munk \inst{5}\and
       R.~Nijboer \inst{5}\and
       J.~Noordam \inst{5}\and
       H.~Paas\inst{6}\and
       M.~Pandey \inst{7,30}\and
       V.~N.~Pandey \inst{5,6}\and
       A.~Polatidis  \inst{5}\and
       W.~Reich \inst{13}\and
       A.~.P.~Schoenmakers \inst{5}\and
       J.~Sluman\inst{5}\and
       O.~Smirnov \inst{5,31}\and
       C.~Sobey\inst{13}\and
       B.~Stappers \inst{10}\and
       J.~Swinbank\inst{19}\and
       M.~Tagger \inst{28}\and
       Y.~Tang\inst{5}\and
       I.~van Bemmel\inst{5}\and
       W.~van Cappellen\inst{5}\and
       A.~P.~van Duin\inst{5}\and
       M.~van Haarlem \inst{5}\and
       J.~van Leeuwen\inst{5}\and
       R.~Vermeulen\inst{5}\and
       C.~Vocks\inst{18}\and
       S.~White \inst{1}\and
       M.~Wise \inst{5}\and
       O.~Wucknitz \inst{13}\and
       P.~Zarka\inst{11}
       }

\authorrunning{F. de Gasperin et al.}
\offprints{F. de Gasperin}

\institute{       
$^1$Max-Planck-Institut f\"ur Astrophysik, Karl Schwarzschild Str. 1, D-85741, Garching, Germany\\ \email{fdg@mpa-garching.mpg.de}\\
$^2$Exzellenzcluster Universe, Boltzmann Str. 2, D-85748, Garching, Germany\\
$^3$Department of Astrophysics, IMAPP, Radboud University Nijmegen, P.O. Box 9010, 6500 GL, Nijmegen, The Netherlands\\
$^4$Max-Planck-Institut f\"ur Extraterrestrische Physik, Giessenbach Str., D-85741, Garching, Germany\\
$^5$ASTRON, Postbus 2, 7990 AA, Dwingeloo, The Netherlands\\
$^6$Kapteyn Astronomical Institute, University of Groningen, P.O. Box 800, 9700 AV, Groningen, The Netherlands\\
$^7$Leiden Observatory, Leiden University, 2300 RA, Leiden, The Netherlands\\
$^8$Jacobs University Bremen, Campus Ring 1, D-28759, Bremen, Germany\\
$^{9}$Laboratoire Lagrange, UMR 7293, Universit\'e de Nice Sophia-Antipolis, CNRS, Observatoire de la C\^ote d'Azur, 06300 Nice, France\\
$^{10}$Jodrell Bank Centre for Astrophysics, School of Physics and Astronomy, University of Manchester, Oxford Road, M13 9PL, Manchester, United Kingdom\\
$^{11}$GEPI, Observatoire de Paris-CNRS, Universit\'e Paris Diderot, 5 place Jules Janssen, F-92190, Meudon, France\\
$^{12}$INAF - Osservatorio Astronomico di Cagliari, Strada 54, IT-09012, Capoterra (CA), Italy\\
$^{13}$Max-Planck-Institut f\"ur Radioastronomie, Auf dem H\"ugel 69, D-53121, Bonn, Germany\\
$^{14}$Onsala Space Observatory, Dept. of Earth and Space Sciences, Chalmers University of Technology, SE-43992, Onsala, Sweden\\
$^{15}$INAF - Istituto di Radioastronomia, Via P. Gobetti 101, IT-40129, Bologna, Italy\\
$^{16}$Jagiellonian University, ul. Orla 171, PL-30244, Krak\'ow, Poland\\
$^{17}$School of Physics and Astronomy, University of southampton, Highfield, SO17 1SJ, southampton, United Kingdom\\
$^{18}$Leibniz-Institut f\"ur Astrophysik Potsdam (AIP), An der Sternwarte 16, D-14482, Potsdam, Germany\\
$^{19}$Astronomical Institute Anton Pannekoek, University of Amsterdam, Science Park 904, NL-1098 XH, Amsterdam, The Netherlands\\
$^{20}$SRON Netherlands Institute for Space Research, P.O. Box 800, 9700 AV, Groningen, The Netherlands\\
$^{21}$Sydney Institute for Astronomy, School of Physics, The University of Sydney, Sydney, Australia\\
$^{22}$Harvard–Smithsonian Center for Astrophysics, Garden Street 60, MA 02138, Cambridge, USA\\
$^{23}$SUPA, Institute for Astronomy, Royal Observatory Edinburgh, Blackford Hill, EH9 3HJ, Edinburgh, United Kingdom\\
$^{24}$University of Hamburg, Gojenbergsweg 112, D-21029, Hamburg, Germany\\
$^{25}$Mt Stromlo Observatory, Research School of Astronomy and Astrophysics, Australian National University, A.C.T. 2611, Weston, Australia\\
$^{26}$Astronomisches Institut, Ruhr-Universit\"at Bochum, D-44780, Bochum, Germany\\
$^{27}$Th\"uringer Landessternwarte, Sternwarte 5, D-07778, Tautenburg, Germany\\
$^{28}$Laboratoire de Physique et Chimie de l'Environnement et de l'Espace, LPC2E UMR 7328 CNRS, F-45071, Orl\'eans Cedex 02, France\\
$^{29}$University of Oxford, Astrophysics, Denys Wilkinson Building, Keble Road, OX1 3RH, Oxford, United Kingdom\\
$^{30}$Centre de Recherche Astrophysique de Lyon, Observatoire de Lyon, 9 av Charles Andr\'e, F-69561, Saint Genis Laval Cedex, France\\
$^{31}$Centre for Radio Astronomy Techniques \& Technologies (RATT), Department of Physics and Elelctronics, Rhodes University, P.O. Box 94, 6140, Grahamstown, south Africa\\
}

\date{}
\abstract {\object{M87} is a giant elliptical galaxy located in the centre of the Virgo cluster, which harbours a supermassive black hole of mass $6.4 \times 10^9~M_{\sun}$, whose activity is responsible for the extended (80~kpc) radio lobes that surround the galaxy. The energy generated by matter falling onto the central black hole is ejected and transferred to the intra-cluster medium via a relativistic jet and morphologically complex systems of buoyant bubbles, which rise towards the edges of the extended halo.}
{In order to place constraints on past activity cycles of the active nucleus, images of M87 were produced at low radio frequencies never explored before at these high spatial resolution and dynamic range. To disentangle different synchrotron models and place constraints on source magnetic field, age and energetics, we also performed a detailed spectral analysis of M87 extended radio-halo.}
{Here we present the first observations made with the new Low-Frequency Array (LOFAR) of M87 at frequencies down to 20~MHz. Three observations were conducted, at $15-30$~MHz, $30-77$~MHz and $116-162$~MHz. We used these observations together with archival data to produce a low-frequency spectral index map and to perform a spectral analysis in the wide frequency range 30~MHz -- 10~GHz.}
{We do not find any sign of new extended emissions; on the contrary the source appears well confined by the high pressure of the intra-cluster medium. A continuous injection of relativistic electrons is the model that best fits our data, and provides a scenario in which the lobes are still supplied by fresh relativistic particles from the active galactic nuclei. We suggest that the discrepancy between the low-frequency radio-spectral slope in the core and in the halo implies a strong adiabatic expansion of the plasma as soon as it leaves the core area. The extended halo has an equipartition magnetic field strength of $\simeq 10\un \mu G$, which increases to $\simeq 13\un \mu G$ in the zones where the particle flows are more active. The continuous injection model for synchrotron ageing provides an age for the halo of $\simeq 40$~Myr, which in turn provides a jet kinetic power of $6-10 \times 10^{44}\un erg\ s^{-1}$.}{}

\maketitle

\section{Introduction}
\label{sec:introduction}

Accreting supermassive black holes in active galactic nuclei (AGN) can release enormous amounts of energy into their surroundings, which may profoundly influence the black hole's hosting environment up to the cluster scale. The way energy is transported to large distances, its amount and the typical time-scales of these processes are still not clear. New-generation radio-telescopes such as the Low-Frequency Array (LOFAR) allow us to study these mechanisms with unprecedented quality and resolution in a hitherto neglected wavelength range. One of the best studied examples of black hole -- host galaxy feedback in action is the AGN in the nearby giant elliptical galaxy M87 (NGC~4486), in the core of the Virgo cluster.

A particularly large amount of study, including hundreds of published papers, has been devoted to M87. This galaxy owes its popularity to several reasons, among others: it is one of the nearest\footnote{M87 is at a distance of $\sim 17\un Mpc$, where $1\arcsec$ corresponds to $\sim 85$~pc.} radio galaxies, it is at the centre of the nearest rich cluster of galaxies (the Virgo cluster), it is the fourth brightest radio source in the northern sky, and it hosts in its nucleus one of the most massive active black holes discovered so far \citep[$M_{\rm BH} \simeq 6.4 \pm 0.5 \times 10^9~M_{\sun}$,][]{Gebhardt09}.

The interaction between the AGN of M87 with its host galaxy and the intra-cluster medium (ICM) has been the subject of a large fraction of the aforementioned studies. The emission generated directly and indirectly by the AGN has been widely observed at radio \citep{BOLTON1949,MILLS1952,Baade1954,Owen2000}, infrared \citep{Shi2007}, optical \citep{Biretta1999a} and X-ray \citep{Fabricant1980,Feigelson1987,Bohringer1995a,Young2002,Forman07,Million2010} wavelengths. Theoretical and numerical models to interpret these observations have also been developed by e.g. \cite{Churazov2001} and \cite{Bruggen2002}.

The radio source associated with this galaxy is named Virgo~A (3C~274). Its inner region ($1.3\arcmin \times 0.5\arcmin$) contains a collimated relativistic jet, which points towards the north-west and is embedded in a halo with a diameter up to $15\arcmin$ ($\sim 80\un kpc$). The extended radio emission, discovered by \cite{MILLS1952} and \cite{Baade1954}, is responsible for much of the radio flux, especially at the lower frequencies. Due to the high surface brightness of the compact central region and the relatively faint surface brightness of the extended emission, high dynamic-range imaging of Virgo~A has always been a big challenge. In the past years \cite{Owen2000} presented a high-resolution ($7\arcsec$), high-dynamic range map of the halo of M87 observed at 327~MHz with the Very Large Array (VLA). At higher frequencies \cite{Rottmann96} mapped the extended Virgo halo at 10.55~GHz with the single-dish Effelsberg radio telescope at $69\arcsec$ resolution. At lower frequency (74~MHz) a $20\arcsec$ resolution map of Virgo~A was made by \cite{Kassim1993}. This paper will extend the high resolution imaging into the previously almost unexplored very low frequency range of $15-162$~MHz and present some of the highest-dynamic-range images ever made at these frequencies of extended source structures.

M87 lies at the centre of the Virgo cluster X-ray luminous atmosphere, first detected with the Einstein Observatory by \cite{Fabricant1980}. An asymmetry in the X-ray emission, in the form of two spectacular outflow-like structures extending from the nucleus towards the east and south-west, was discovered by \cite{Feigelson1987}, who also found a correlation between X-ray and radio emitting features. One of the first explanations for such a correlation was that the relativistic electrons that produce the synchrotron radio emission were also responsible for the inverse Compton scattering of cosmic microwave background (CMB) photons, thus producing X-ray radiation \citep{Feigelson1987}. However, \cite{Bohringer1995a} showed with a ROSAT PSPC observation, that the excess emission had a thermal spectrum and it is colder than the ambient gas which is at a temperature of $\gtrsim 2$~keV. This feature was explained by \cite{Churazov2001} as buoyant bubbles of cosmic rays, injected into the inner halo (or ``cocoon'') by the relativistic jet, which subsequently rise through the cooling gas at about half the sound speed. During their rise they uplift gas at the temperature of $\sim 1$~keV from the central regions. Sub-arcsecond \textit{Chandra} X-ray images \citep{Million2010} confirmed this picture and provided an unprecedented view of the physical and chemical properties of the ICM.

Although Virgo~A is a unique object because of its properties, close proximity and sheer quantity of available data, it remains a fundamental example to study the more general behaviour of AGNs located at the centre of galaxy clusters. Knowledge of its energetics and of the interaction between its jets and the ICM, may help solve open problems such as the suppression of the cooling flows \citep[for a review see][]{Peterson2006} and the AGN duty cycle. This, in turn, will provide important clues on the physical nature of AGN feedback in massive galaxies and on its relevance to a cosmological framework \citep{Croton06,Fabian2012}. Furthermore, it has been claimed that jet sources like Virgo~A and its southern sibling Centaurus~A are potential candidate sources for the production of ultra-high energy cosmic rays \citep[UHECR,][]{Auger07}.


In this paper we extend the study of Virgo~A to long, so far unobserved, wavelengths. We also retrieved available observations of Virgo~A at 1.4 and 1.6~GHz (VLA, from the data archive), at 325~MHz (VLA, provided by Frazer Owen) and at 10.55~GHz (Effelsberg radio telescope, provided by Helge Rottmann). This enabled us to assess the source energetics, the halo age and the main mechanisms which contributed to its spectral evolution. The paper is organized as follows: in the next section we outline the LOFAR features and characteristics. In Sect.~\ref{sec:observation} we present new LOFAR observations of Virgo~A and we describe the data reduction technique. In Sects.~\ref{sec:images} and~\ref{sec:spectral} we respectively present the outcome of these observations and perform a spectral analysis of them, discussing the physical interpretation of our results. In Sects. \ref{sec:discussion} and \ref{sec:conclusions}, we discuss the results and outline our conclusions.

\section{LOFAR}
\label{sec:LOFAR}

LOFAR (van Haarlem et al. in prep.) is a radio telescope optimized for the frequency range from 30 to 240~MHz, but also with the ability to observe down to 10~MHz. LOFAR does not have any moving parts, the telescope receivers are two different kind of dipoles: the low-band antennas (LBA), which cover the frequency range $10-90$~MHz, and the high-band antennas (HBA), which cover the frequency range $110-240$~MHz. The LBAs are inverted-V crossed-dipoles oriented NE-SW and SE-NW, while the HBA are organized into tiles made of a $4\times4$ array of bowtie-shaped crossed dipoles. Dipoles are organized into stations yielding, for each station, effective aperture sizes that range from 30~m to 80~m, depending on the frequency. Each set of dipoles within a station works as a phased aperture array -- i.e. a delay is applied to the relative phases of the signals feeding the dipoles in such a way that the radiation pattern of the array is reinforced in a target direction 
and suppressed in undesired ones. By applying different delays, LOFAR can therefore ``point'' (create a beam) in more than one direction simultaneously and the number of beams is limited only by the bandwidth necessary to transfer the signal to the correlator and its computational power. 

\begin{figure*}
\centering
\subfloat[HBA u-v tracks]{\includegraphics[width=0.33\textwidth]{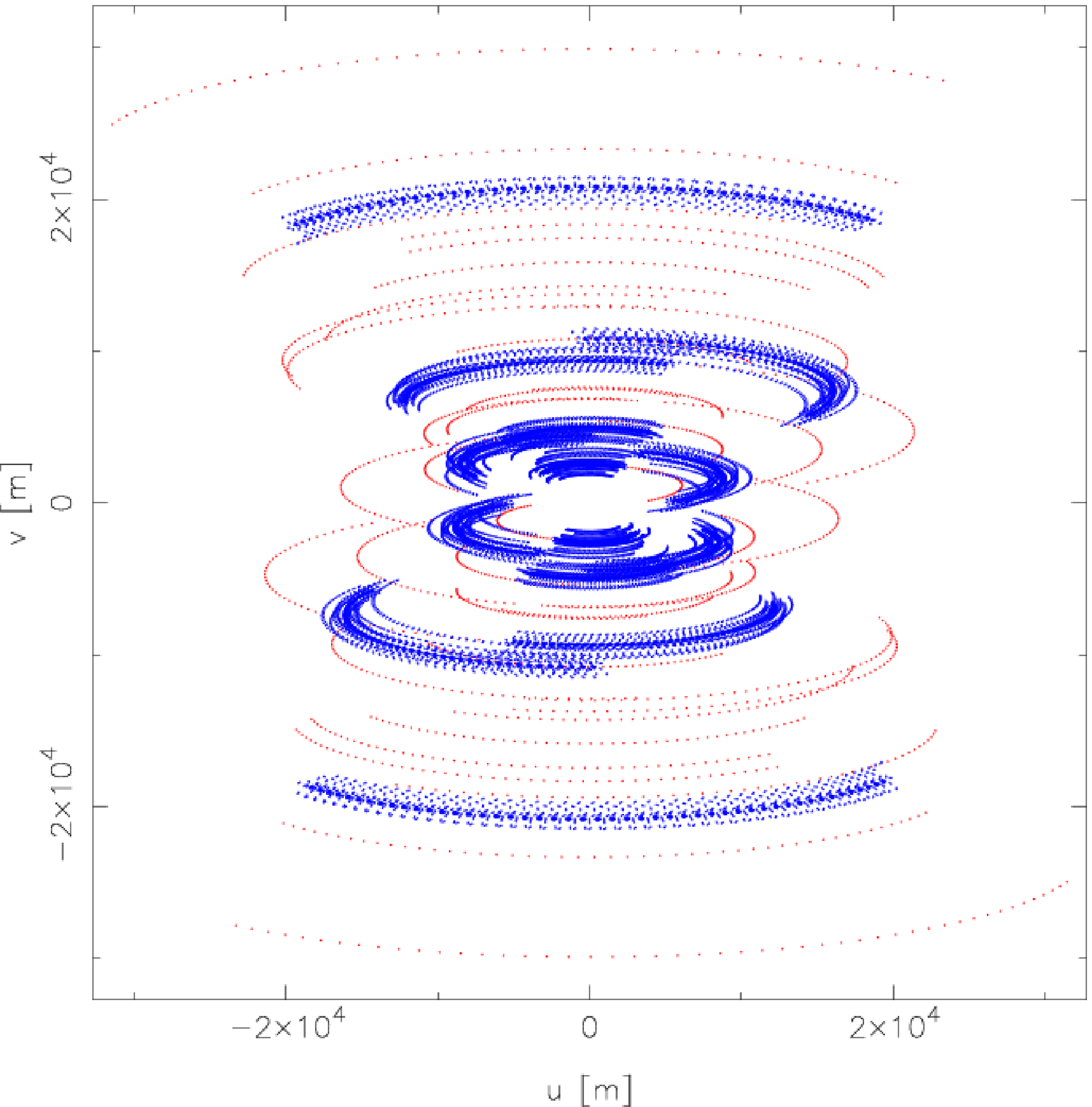}\label{fig:uvcovHBA1}}
\subfloat[HBA u-v tracks]{\includegraphics[width=0.33\textwidth]{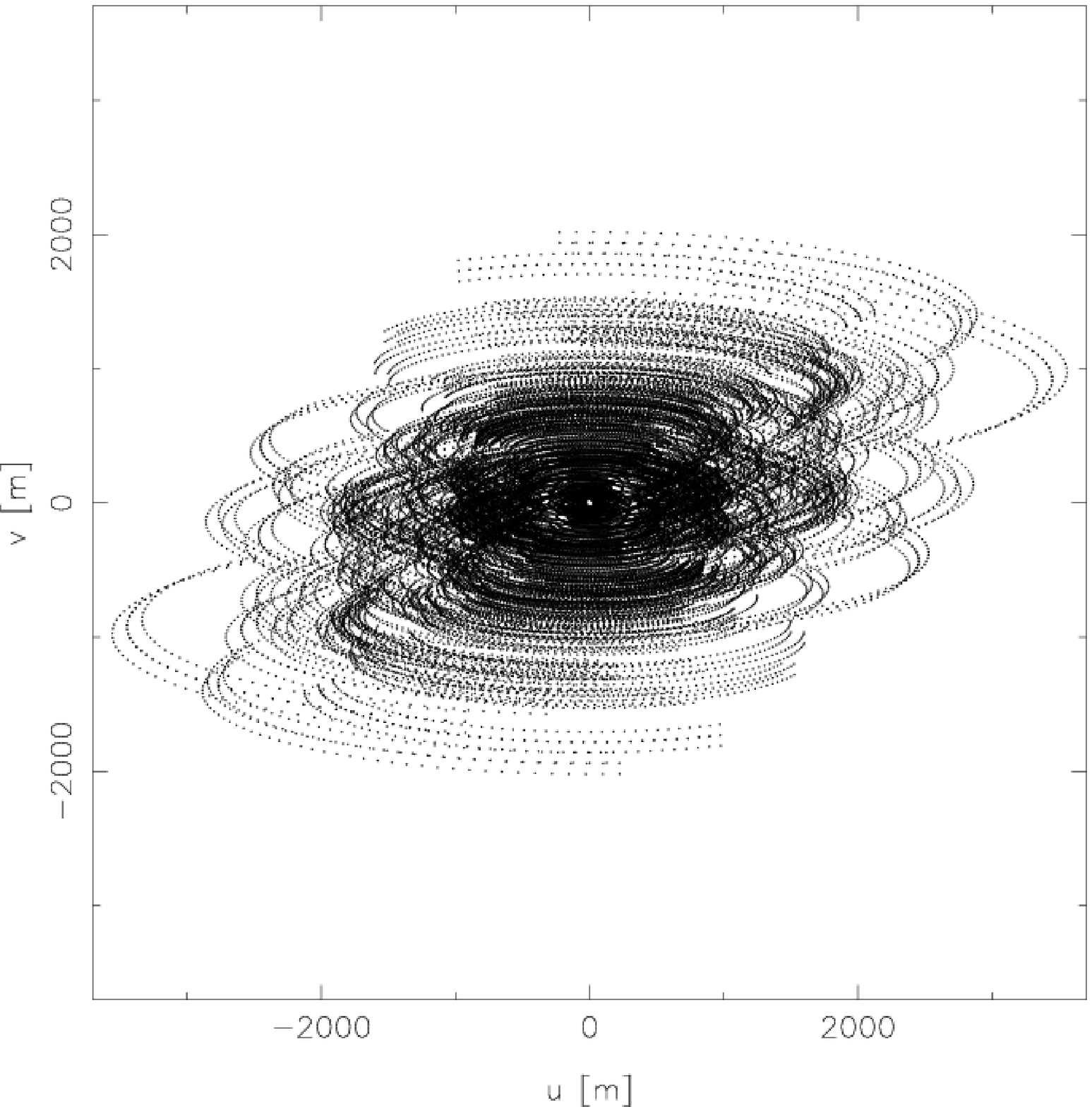}\label{fig:uvcovHBA2}}
\subfloat[HBA u-v tracks]{\includegraphics[width=0.33\textwidth]{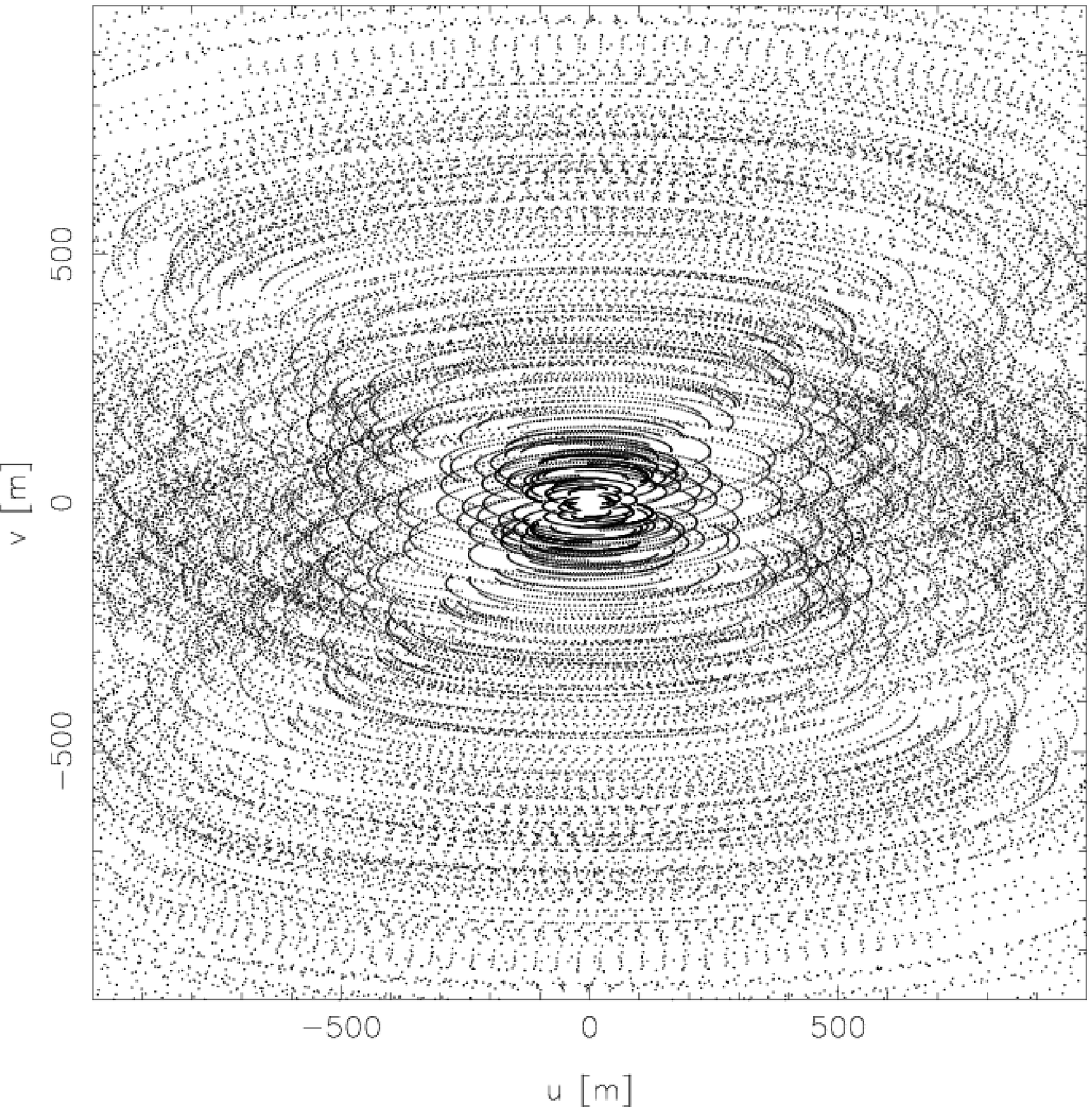}\label{fig:uvcovHBA3}}\\
\subfloat[LBA-high u-v tracks]{\includegraphics[width=0.33\textwidth]{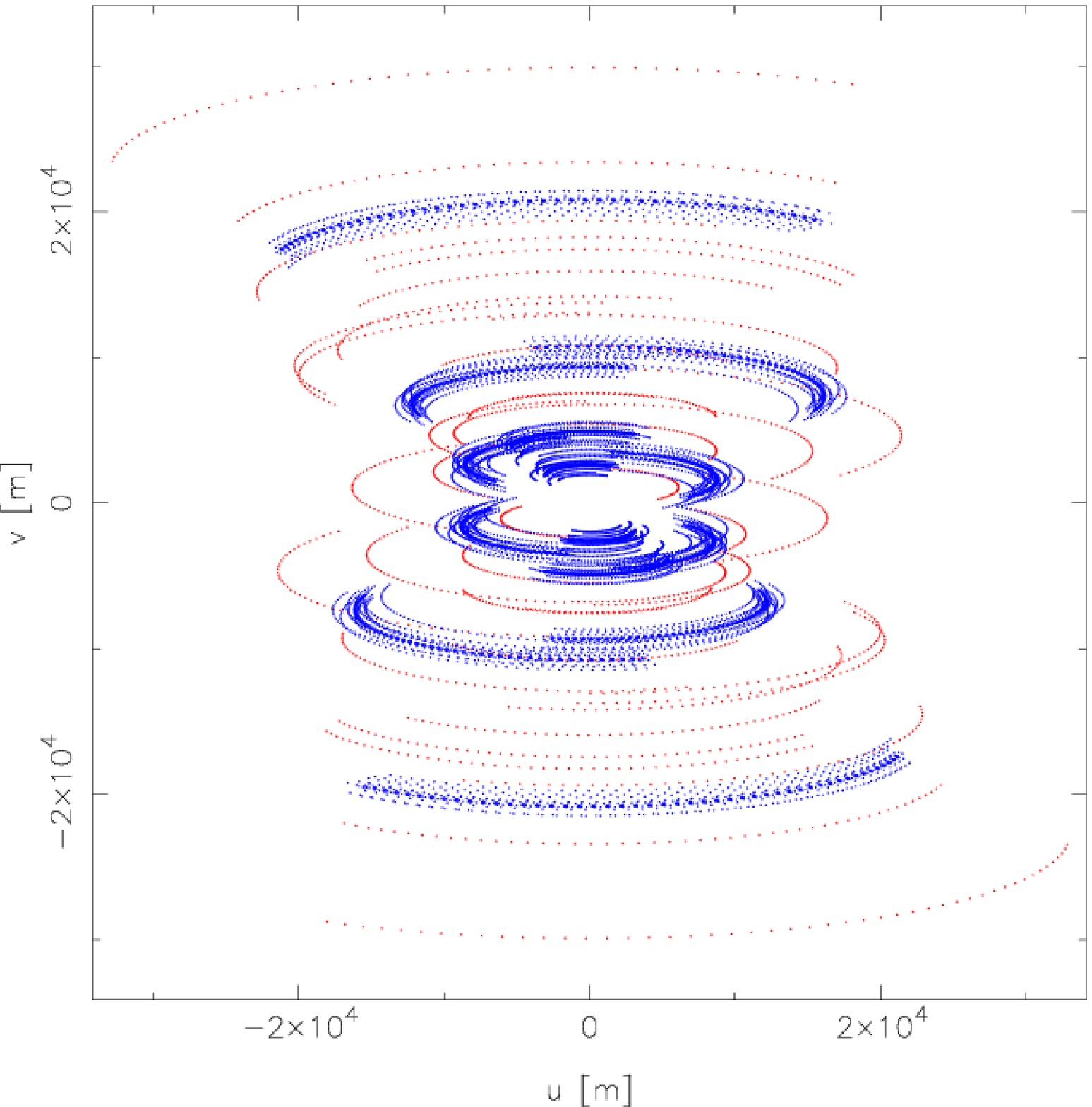}\label{fig:uvcovLBA1}}
\subfloat[LBA-high u-v tracks]{\includegraphics[width=0.33\textwidth]{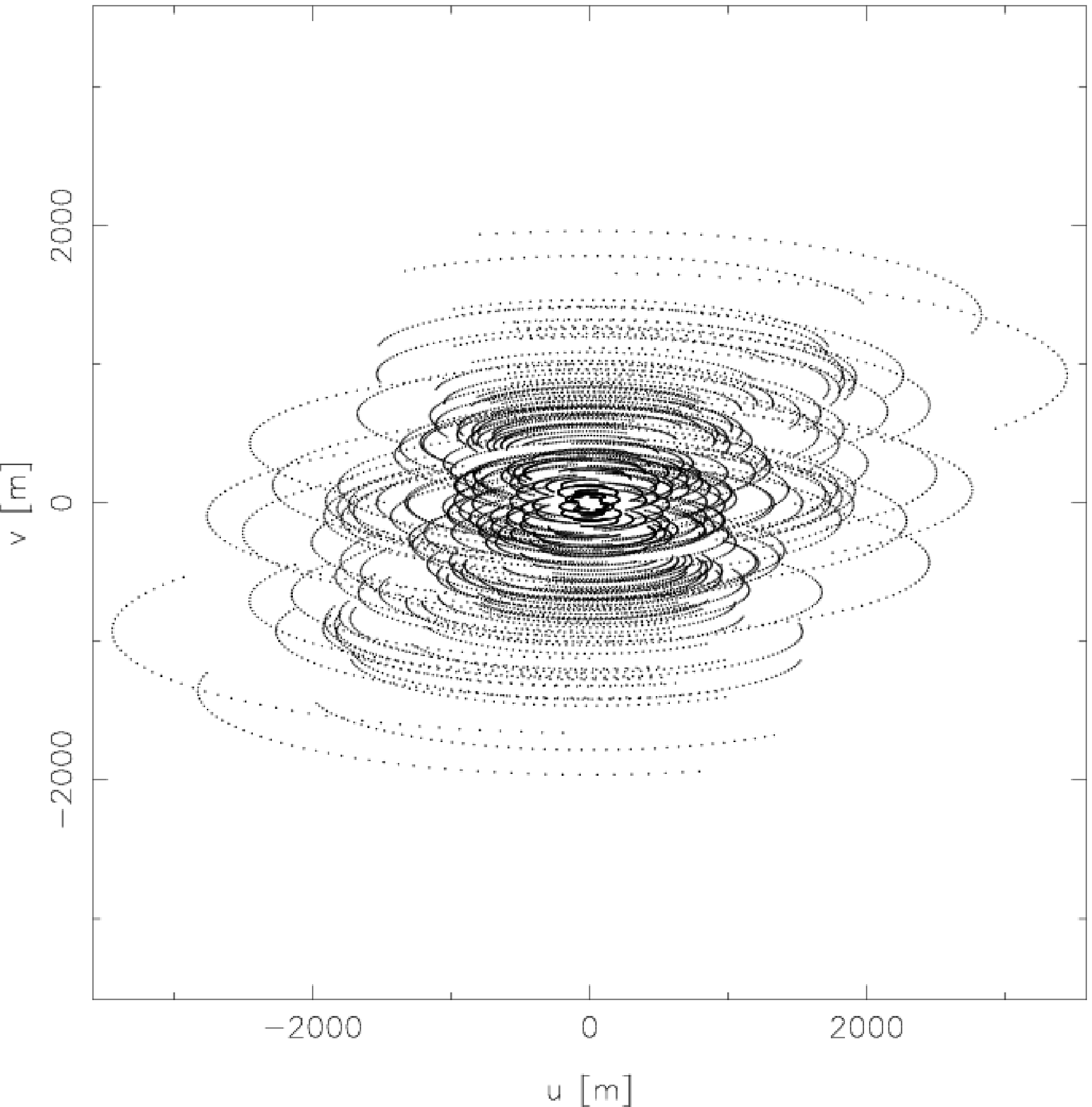}\label{fig:uvcovLBA2}}
\subfloat[LBA-high u-v tracks]{\includegraphics[width=0.33\textwidth]{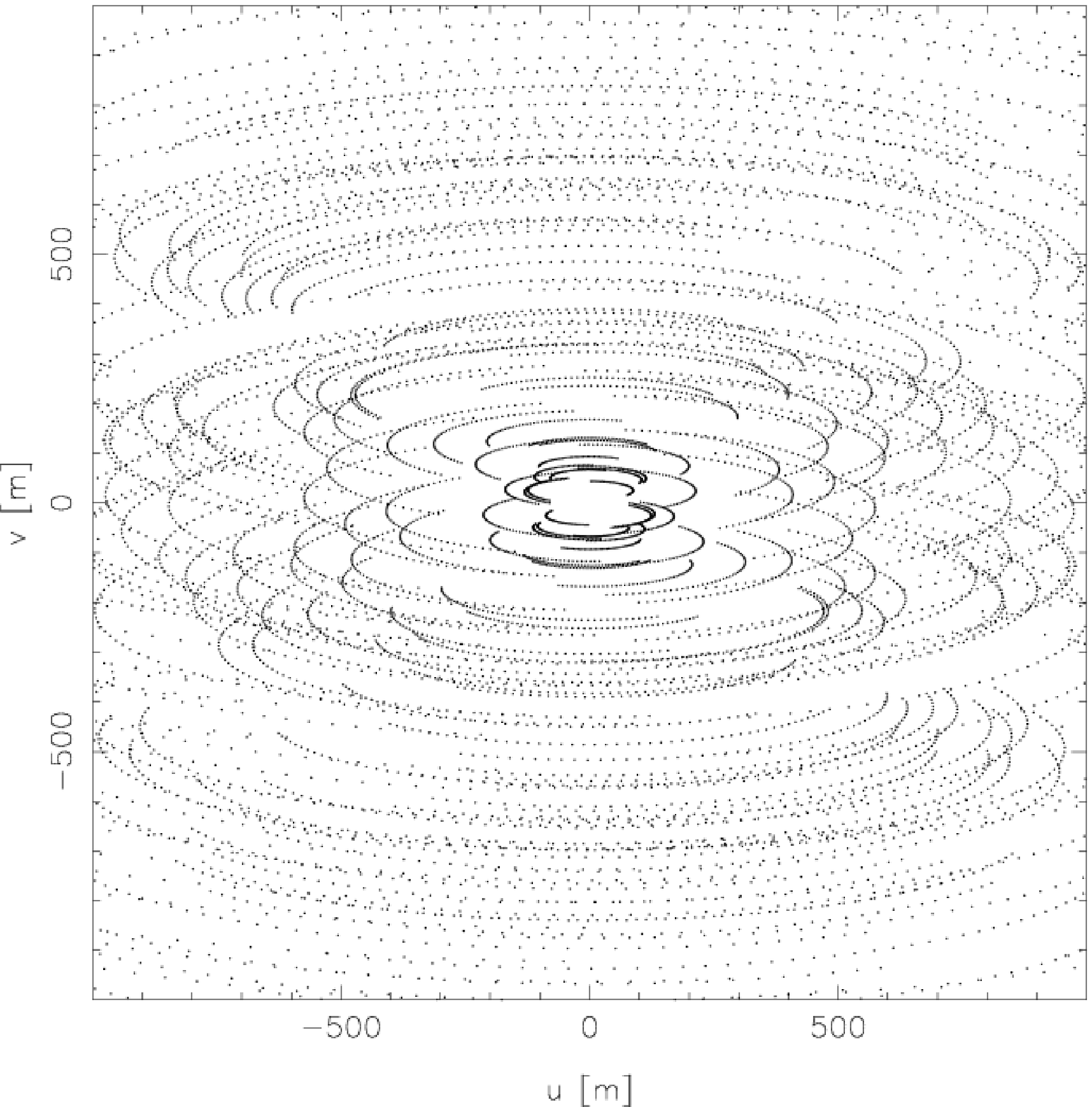}\label{fig:uvcovLBA3}}\\
\subfloat[LBA-low u-v tracks]{\includegraphics[width=0.33\textwidth]{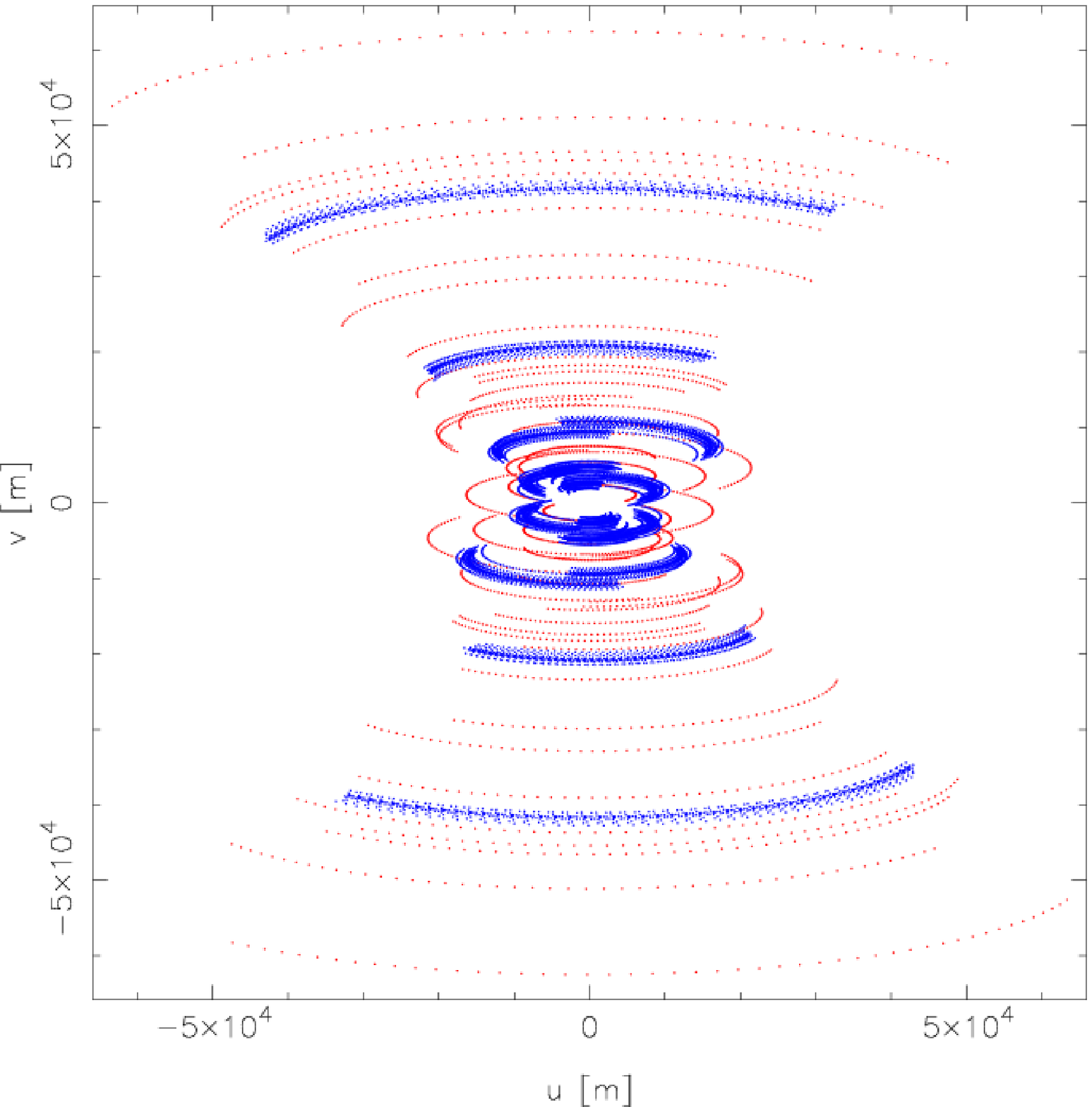}\label{fig:uvcovLLBA1}}
\subfloat[LBA-low u-v tracks]{\includegraphics[width=0.33\textwidth]{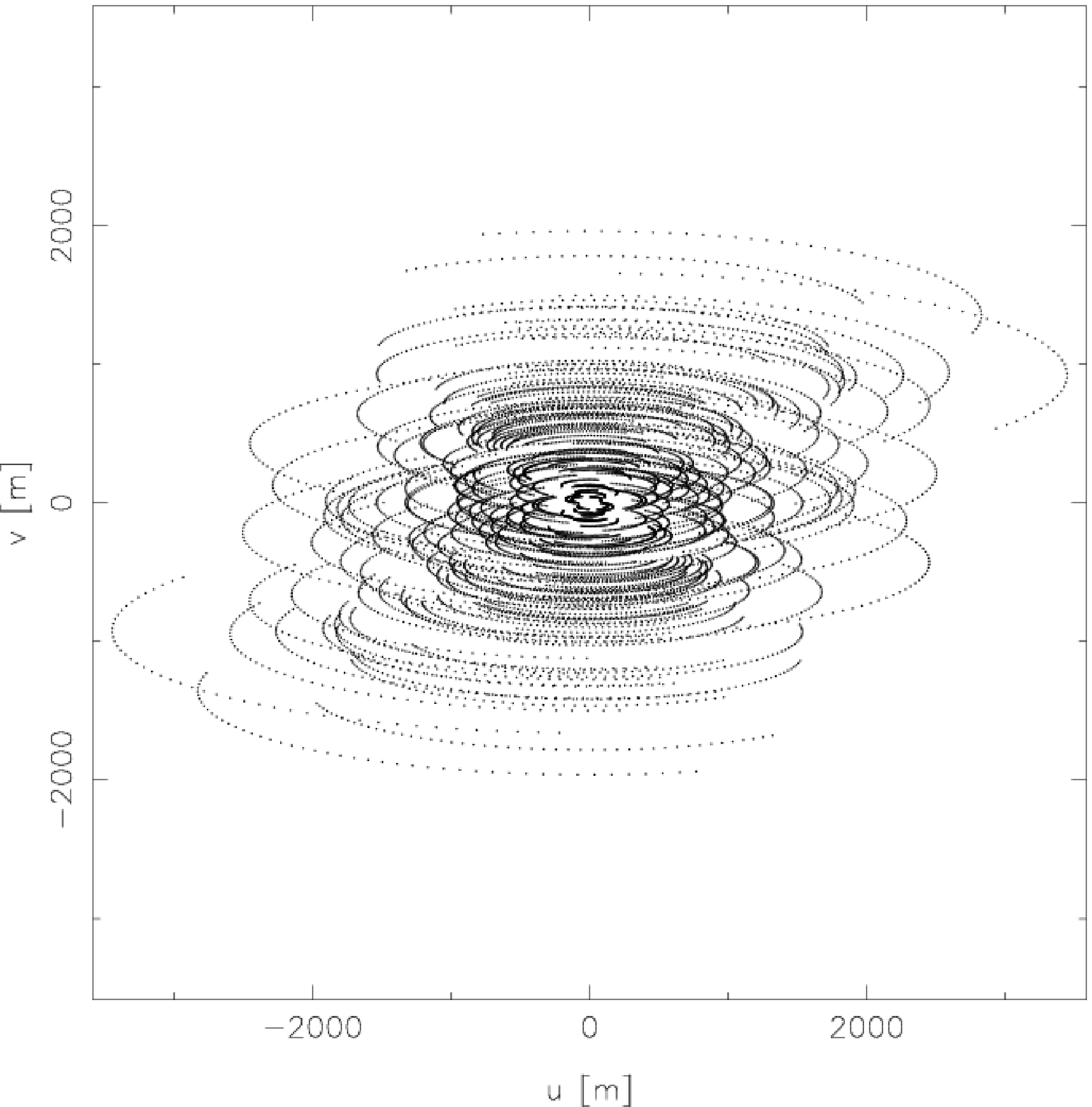}\label{fig:uvcovLLBA2}}
\subfloat[LBA-low u-v tracks]{\includegraphics[width=0.33\textwidth]{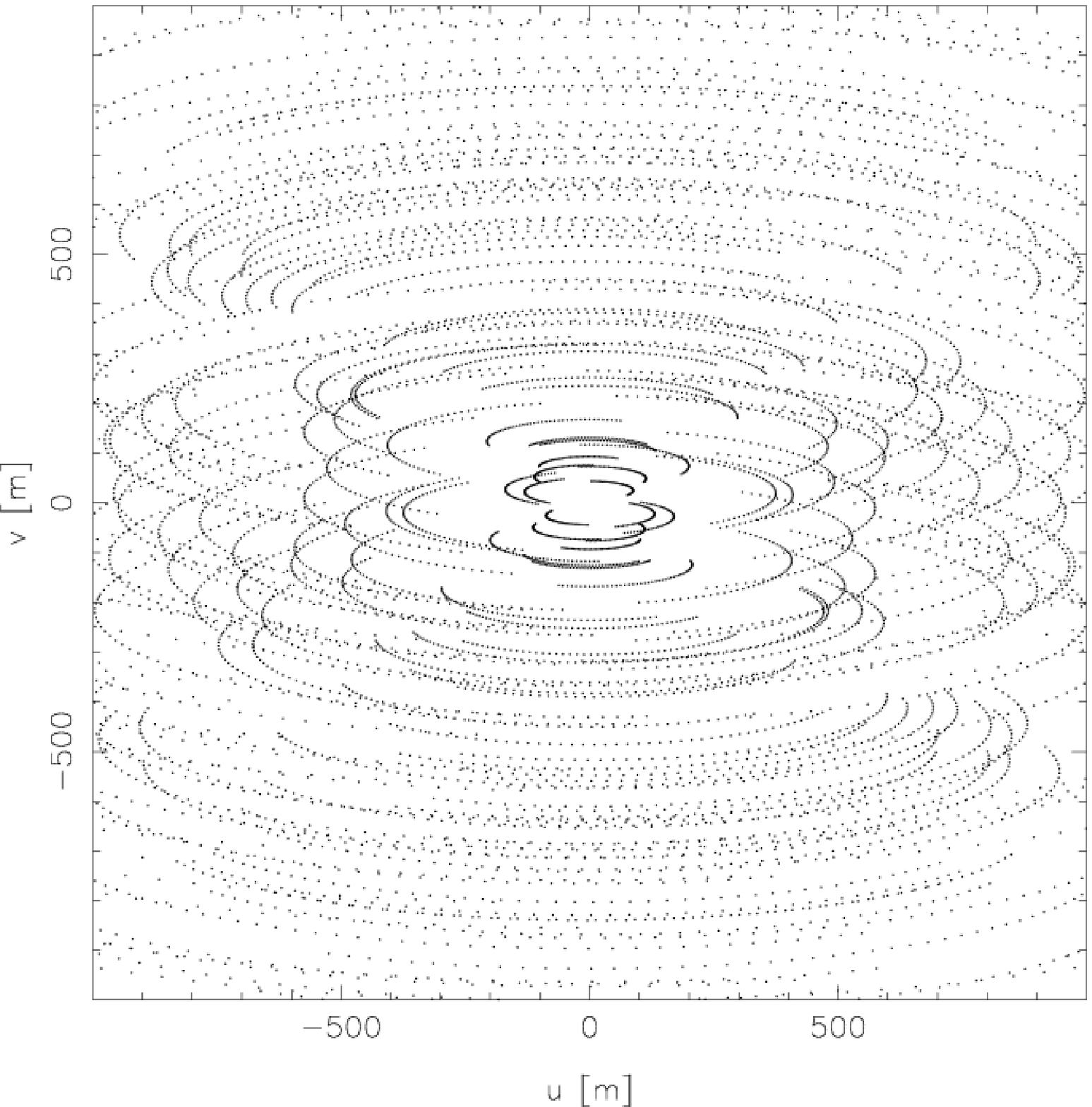}\label{fig:uvcovLLBA3}}\\
\caption{\textit{uv}-coverage for the three observations of Virgo~A: the first row is the HBA observation, the second row the LBA-high observation and the third row the LBA-low observation. In the first column are plotted only tracks involving remote stations (blue: core-remote baselines -- red: remote-remote baselines). In the second column only tracks of core-core baselines are plotted. The last column is a zoom-in on the centre of the \textit{uv}-plane.}
\label{fig:uvcov}
\end{figure*}

\begin{figure*}
\centering
\includegraphics[width=\textwidth]{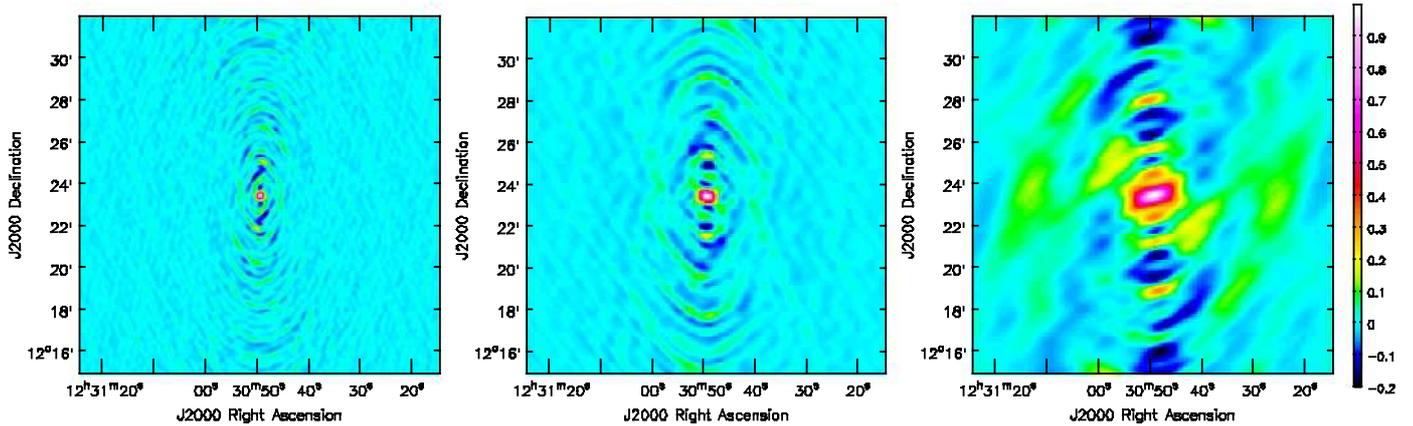}
\caption{\textit{Left}: dirty beam used in the deconvolution step for the HBA observation (at a frequency of $\sim 140$~MHz). \textit{Centre:} dirty beam used in the deconvolution step for the LBA-high observation (at a frequency of $\sim 54$~MHz). \textit{Right}: dirty beam used in the deconvolution step for the LBA-low observation (at a frequency of $\sim 22$~MHz).}
\label{fig:psf}
\end{figure*}

The complete configuration\footnote{An updated map of the station status can be found here: http://www.astron.nl/$\sim$heald/lofarStatusMap.html.} of LOFAR will consist of an array of stations distributed over 100 km within the Netherlands and out to 1000 km throughout Europe, which will provide sufficient resolution ($\approx 1\arcsec$ at 30~MHz) to allow optical identification of radio sources, even at low frequencies. At present, international stations provide LOFAR with an angular resolution of $\sim$~\as{0}{15} at 240~MHz and $\sim 1\arcsec$ at 30~MHz, while a dense core of 24 stations provides the necessary sensitivity to the extended emission.

For the stations in the Netherlands only 48 (out of 96) LBA dipoles can be currently used simultaneously. Among the different possibilities, mainly two possible configurations of the LBA dipoles are commonly used: LBA-INNER and LBA-OUTER. In these configurations the active dipoles are located respectively in the inner zone and in the outer zone of the station field. The more concentrated are the dipoles in the inner zone, the more the side-lobe levels are reduced and the field of view (FoV) is wider, but at the cost of a reduced sensitivity. Each core station has the HBA grouped into two different sub-stations located at the edge of the field. These sub-stations can be used together or as standalone stations (DUAL observing mode), to increase 
the number of baselines. 

Once the data are collected from the stations they are transported to the central processing location via a Wide-Area Network, using dedicated light paths. Data are then correlated by a Blue Gene/P computer that contains 12480 processor cores providing 42.4 TFLOPS peak processing power. For a detailed description of the correlator see \cite{Romein2010}.

\section{The observations}
\label{sec:observation}

\begin{table*}
\caption{Details of the observations}
\centering
\label{tab:observations}
\begin{tabular}{lcccccccc}
\hline\hline
Obs. ID & Antenna & Frequency & Date & Observation & Sampling    & FWHM\tablefootmark{1}  & Maximum    & Number \bigstrut[t] \\
        & type    & range     &      & length      & time        &       & resolution & of stations\\
        &         & [MHz]     &      & [s]         & [s]         & [deg] & [arcsec$^2$]   & \bigstrut[b] \\
\hline
L24923 & HBA-DUAL\tablefootmark{2} & $115 - 162$ & 2/3-Apr-2011 & 28810 ($\simeq 8$ h) & 1 & $\sim 5$ & $19 \times 14$ & 45 (7)\tablefootmark{2}\bigstrut[t]\\ 
L25455 & LBA-INNER & $30 - 77$ & 14/15-Apr-2011 & 28810 ($\simeq 8$ h) & 2 & $\sim 10$ & $37 \times 30$ & 24 (7)\tablefootmark{2} \\
L29694 & LBA-OUTER & $15 - 30$ & 16-Jul-2011 & 28805 ($\simeq 8$ h) & 2 & $\sim 10$ & $85 \times 44$ & 25 (8)\tablefootmark{2} \bigstrut[b]\\
\hline
\end{tabular}
\tablefoot{\tablefootmark{1}{FWHM of the primary beam when points at the zenith, its shape changes during the observation time and is not circular.} \tablefootmark{2}{``DUAL'' means that the two sub-stations of the core stations are treated separately (see text for details). This is why the number of stations in the HBA observation is higher with respect to the LBA observations.} \tablefootmark{3}{Enclosed in brackets the number of remote stations.}}
\end{table*}

In this paper we present a set of three observations performed during the LOFAR commissioning phase. The phase centre was set on the core of Virgo~A (RA: 12:30:49.420 -- DEC: +12:23:28.0 -- J2000) and the observational details are listed in Table~\ref{tab:observations}. Each observation was 8 hours in duration and all four polarization products (XX, YY, XY, and YX) were stored. Each observation had its frequency coverage divided into sub-bands (SB) of 0.1953~MHz of bandwidth and each SB was divided into 64 channels of $\simeq 3$~kHz of bandwidth. The following observations were performed:

\begin{description}
 \item[HBA ($115-162$~MHz):] we observed the target with the HBA on the 2nd and 3rd of April, 2011. The visibility sampling time was 2~s. Two stations (CS021HBA0 and CS021HBA1) were flagged by the correlator and their data were not used. All 244 SBs were correctly processed and stored by the correlator.
 \item[LBA-high ($30-77$~MHz):] a second observation was performed with the LBA system on 14th and 15th of April, 2011, using a 30~MHz high-pass filter. The visibility sampling time was 1 s. The LBA-INNER configuration was used. At the end of the data reduction procedure 36 SBs out of 244 (15\%, 7.2~MHz of bandwidth) were not usable due to computing-cluster or correlator failure.
 \item[LBA-low ($15-30$~MHz):] a third observation was performed with the LBA system on 16th of July, 2011, using a 10~MHz high-pass filter. The visibility sampling time was 1 s. Three SBs out of 77 (4\%) were corrupted during the data processing. We did a visual inspection of the 74 residual SBs and only 41 (55\%, 8.2~MHz of bandwidth) contained usable data, the others were unusable due to high RFI levels. One antenna (CS302) was flagged at correlation time. An LBA-OUTER configuration was used to keep the FoV comparable to that of the LBA-high observation.
\end{description}
International stations were not used in these observations, therefore the longest baseline available was about 80 km (for the observation at $15-30$~MHz) and 40~km (for the others), while the shortest was $\simeq 90$~m. A plot of the full \textit{uv}-coverage is shown in Fig.~\ref{fig:uvcov}, while the dirty beams used for deconvolution are shown in Fig.~\ref{fig:psf}.

\subsection{The pipeline}
\label{sec:pipeline}

Most of LOFAR data processing is done by pipeline software. The LOFAR processing system takes the data from the dipoles, forms the beams and correlates the output to ultimately generate images of the radio-sky. As a detailed description of the whole process is beyond the scope of this paper, here we will illustrate only the most important steps. Interested readers can refer to \cite{Heald2010}.

After correlation, data are recorded on storage nodes in the current LOFAR offline processing cluster. The first data processing step is to flag radio-frequency interference (RFI) and optionally compress the data in time and frequency. Automated flagging is required for the LOFAR data volume, which is performed by the AOFlagger \citep{Offringa2010a,offringa12}. This software estimates the astronomical signal by carrying out a surface fit in the time-frequency plane and flags outliers using combinatorial thresholding and morphological detection. Compression and further flagging is performed by the New Default Pre-Processing Pipeline, or NDPPP.

In cases where the contributions of other bright sources in the sky are not negligible, a subtraction of these sources directly from the visibilities is required. The technique we used is called \textit{demixing} and it is described in \cite{VanderTol2007}. This approach is computationally cheap and shows remarkably good results in some circumstances, where the source to \textit{demix} and subtract is not too close to the main target, but fails in more complex scenarios, for instance where a strong source is a few degrees away from a weak target. We note that the \textit{demixing} is just one of the possible approximate approaches that can be used to remove strong interfering sources. Another possibility, although computationally more expensive, is the \textit{peeling} procedure \citep{Noordam2004}.

The calibration step is performed with the BlackBoard Selfcal (BBS) software, developed explicitly for LOFAR. This calibration package is based on the Hamaker-Bregman-Sault Measurement Equation \citep{Hamaker1996,Smirnov2011}, which expresses the instrumental response to incoming electromagnetic radiation within the framework of a matrix formalism. BBS is thus able to handle complicated calibration tasks like direction-dependent effects and full polarization calibration as well as correcting for the time/position dependence of the element beam and the synthesized beam.

The imaging step is routinely performed using the AWimager software (Tasse et al. in prep.). The AWimager uses the A-projection algorithm \citep{Bhatnagar2008} to image wide fields of view, where data must be corrected for direction dependent effects varying in time and frequency (mainly beam and ionospheric effects). The software does not yet support multi-scale cleaning \citep{Cornwell08,Rau2011}, so for our purposes we decided to use the CASA\footnote{http://casa.nrao.edu.} imager which is adequate for the central portion of the image that we were interested in.

\subsection{Data reduction}
\label{sec:data reduction}

\begin{figure*}
\includegraphics[width=\textwidth]{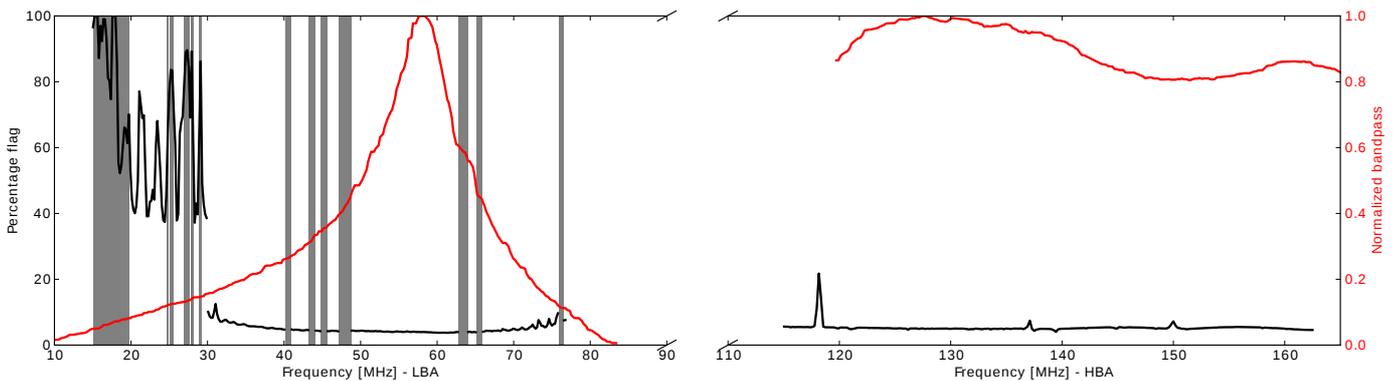}
\caption{Black line: percentage of flagged data. In the frequency range $15-30$~MHz, the last 37\% of the observation was manually flagged. Red line: normalized bandpass (for HBA it is only available for a slightly shifted frequency range). The SBs removed because of corrupted data or a computer failure are coloured in grey. Completely flagged stations are not taken into account to compute the percentage of flagged data.}
\label{fig:flag}
\end{figure*}
Although some steps were common, the data reduction procedure followed different schemes for the three observations. A preliminary common step is the use of the automated AOFlagger on the full resolution (in time and frequency) datasets. With a further visual inspection of the data, we did not recognise any visible RFI effects left in the raw data. After this step the procedures were different for each dataset, and therefore we explain them in detail:
\begin{description}
 \item[HBA ($115-162$~MHz):] first all baselines related to the two substations within the same station were flagged ($\sim 1$\% of data), this was necessary due to a possible cross-talk effect that was found in the intra-station baselines. Then, we applied the \textit{demixing} procedure to the dataset, subtracting in this way the two strongest sources in the sky, Cassiopeia~A ($\sim107\degr$ from Virgo~A) and Cygnus~A ($\sim98\degr$ from Virgo~A). This procedure was necessary only in the second half of the observation, where the two aforementioned sources were above the horizon. After that we compressed the dataset to one channel (excluding the first and last two channels) per SB and 20~s of sampling time. This reduced the data volume to the level of $\sim400$~MB per SB, where a cycle of self-calibration lasts $\sim 1$~h. The model for the self-calibration was extracted from VLA data at 325~MHz \citep{Owen2000}, which had a resolution high enough for our case ($\sim7\arcsec$). For each SB we rescaled the total flux of the model according to the source global spectral index value (see Sect.~\ref{sec:flux}). Several cycles of self-calibration (phase and amplitude) performed with BBS and using the imaging algorithm in CASA, were necessary to converge to the final image. The imaging step at these frequencies was performed using a standard CLEAN for the bright central region, followed by the use of a multi-scale cleaning.
 \item[LBA-high ($30-77$~MHz):] after \textit{demixing}, that was performed as described for the HBA dataset, the data were averaged to one channel (excluding the first and last two channels) per SB and to 10~s of sampling time. The model for self-calibration was extracted from a VLA observation at 74~MHz \citep{Kassim1993} with the total flux rescaled to the appropriate frequency (see Sect.~\ref{sec:flux}). We did several cycles of self-calibration (phase and amplitude) with BBS and using the imaging algorithm in CASA. The central region of Virgo~A was CLEANed using standard pixel-by-pixel cleaning while for the extended emission we used a multi-scale approach.
 \item[LBA-low ($15-30$~MHz):] the attempt to use the \textit{demixing} procedure failed since the data taken in the last 3 hours of observation were severely affected by ionospheric disturbances. Since this was the part of the observation also corrupted by Cassiopeia~A and Cygnus~A signals, we decided not to use it and simply average the rest of data to 5 s and one channel (excluding the first and last two channels) before the calibration procedure. Finally, several cycles of self-calibration (phase and amplitude), using BBS for the calibration and CASA for the imaging, were performed. The model for self-calibration was again extracted from a VLA observation at 74~MHz with the total flux rescaled to the appropriate frequency (see Sect.~\ref{sec:flux}). The imaging step was done in the same way as for the LBA-high dataset. At the end of the calibration procedure, a visual inspection of the images revealed that for 33 SBs (out of 74) we were unable to correctly calibrate the data due to the RFI level. The majority of these SBs are indeed concentrated in the frequency range $15-20$~MHz and where the RFI presence was critically high. We did not use those SBs for the following analyses. 
\end{description}

In Fig.~\ref{fig:flag} we plot the amount of flagged data for each SB, together with the shape of the bandpass functions. The amounts of flagged data reflect only partially the amount of RFI. Firstly because the RFI flagging is performed at full time-frequency resolution and during the subsequent data averaging flags are ignored if at least one datum is valid in the averaged block. Therefore, it was not possible to track those flags due to RFI which are narrow-frequency or shorter than the average time. Secondly because new flags are applied to remove outliers produced in the calibration phase. In the high frequency regime, the percentage of unusable data is more or less constant, at $\sim 5\%$, apart from a small increment at 118~MHz. Almost all of these are due to the manual-flagging of the first two hours of data from RS208 and RS307 and the last two hours from RS208. At lower frequencies the RFI is stronger and the peaks in Fig.~\ref{fig:flag} are related to it. Below 30~MHz all SBs had a flagged data percentage above 37\% because, as explained, we removed the last 3~h of observation. In the LBA-high frequency range ($30-77$~MHz) the amount of flagged data is rising towards the band edges, where it also presents some systematic oscillations. These behaviours are due to the lower sensitivity of the instrument at these frequencies, which produces some outliers during the calibration procedure and principally during the \textit{demixing} process. These outliers are due to a poor signal to noise in the calibration step and were flagged by an automated procedure through NDPPP after every selfcal cycle. This increases the amount of flagged data, but these flags are not RFI-related. The oscillating pattern is introduced by flagging outliers after the \textit{demixing} procedure, which may suggest that the \textit{demixing} is less effective at those frequencies where the strong (\textit{demixed}) sources are in particular configurations with respect to the beam side-lobe pattern.

\subsection{Absolute flux density}
\label{sec:flux}

\begin{figure}
\includegraphics[width=\columnwidth]{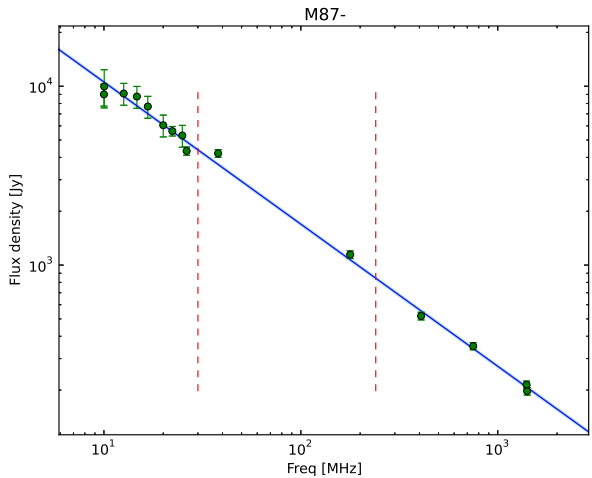}
\caption{Integrated flux of Virgo~A at different frequencies obtained from archival data. The line is a linear fit (slope: $-0.79$) obtained as described in the text. The two vertical dashed lines indicate the boundaries of the LOFAR observing band.}
\label{fig:scale}
\end{figure}

The flux density of Virgo~A integrated over all the extended emission was rescaled to its expected value to compensate for the absence of an absolute flux calibrator, while the relative fluxes of different components in the radio morphology is correctly recovered by self-calibration. To do that, we collected the total flux measurements available in the literature in the frequency range from 10 to 1400~MHz \citep{Braude1969,Bridle1968,Roger1969,Viner1975,kellermann69,Wright1990}. Each data-point was corrected to match the \cite{roger73} (RBC) flux scale with correction factors from \cite{Laing1980} and \cite{scaife12}. A model of the form
\begin{equation}
\begin{split}
  \log S = \log(A_0) + A_1 \log\ & \left(\frac{\nu}{150\ \rm MHz}\right)\\
  + A_2 \log^2 & \left(\frac{\nu}{150\ \rm MHz}\right) +\ ...
\end{split}
\end{equation}
where $\nu$ is the observing frequency and $S$ the observed flux, was used to fit this data set (see Fig.~\ref{fig:scale}). 

The model was applied in linear frequency space, i.e.
\begin{equation}
S[Jy] = A_0 \sum^{N}_{i=1} 10^{A_i \log^i \left[\nu/150\ \rm MHz\right]}
\end{equation}
in order to retain Gaussian noise characteristics. Parameters were fitted using a Maximum Likelihood (ML) approach through a Markov Chain Monte Carlo implementation \citep{scaife12}.

We tested polynomial fits up to the fourth order and found that a first order polynomial function ($A_0 = 1226 \pm 17$ and $A_1=-0.79 \pm 0.008$) is the best-fit model \cite[as already pointed out by a number of authors, e.g.][]{roger73}. We derived the expected total flux of Virgo~A at the frequency of each observed LOFAR SB and rescaled the model used for that SB to match it at the beginning of each cycle of self-calibration.

The primary beam attenuation at the edge of Virgo~A is $<3\%$ for HBA images and $<1\%$ for LBA images. We did not take this effect into account and included the systematic error in the error budget.

The map at 325~MHz was also rescaled to match the RBC flux scale, but at higher frequencies our first order polynomial model is probably no longer valid. However, at frequencies $\gtrsim 300\un MHz$, the RBC scale is in agreement with the KPW scale \citep{kellermann69}, for which we have conversion factors from the Baars scale \citep[][Table 7]{baars77}. Therefore, we used those factors to rescale the maps at 1.4, 1.6, and 10.55~GHz from the Baars scale to the RBC scale.

\section{Virgo A images}
\label{sec:images}

\begin{figure*}
\includegraphics[width=\textwidth]{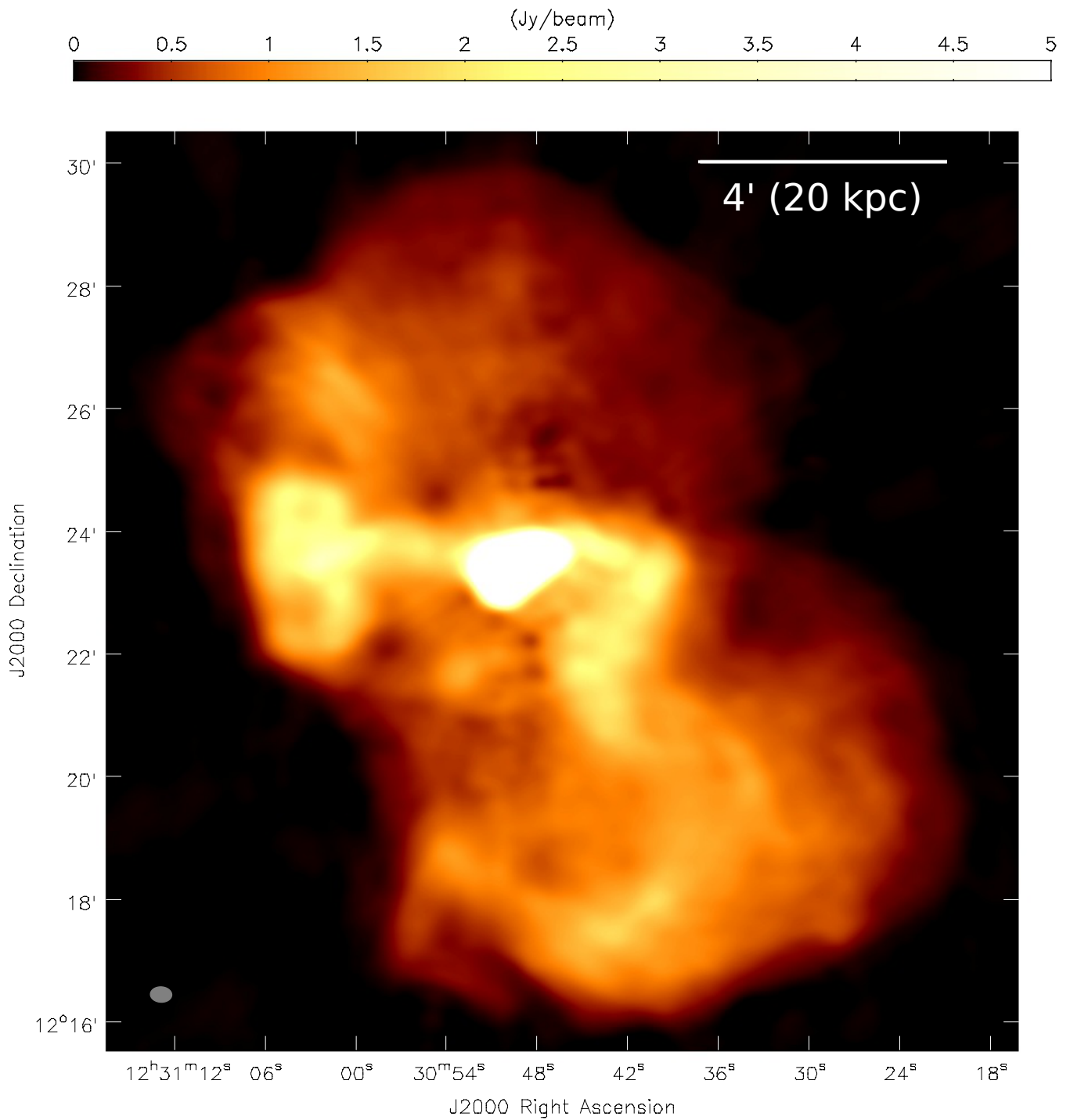}
\caption{LOFAR-HBA image of Virgo~A at 140~MHz. The rms noise level is $\sigma = 20\un mJy/beam$, the flux peak is $101\un Jy/beam$ and the beam size is $21 \arcsec \times 15 \arcsec$ (ellipse in the bottom-left corner).
Some deconvolution errors are visible as small holes slightly above and below the bright core.}
\label{fig:VirA-HBA}
\end{figure*}

\begin{figure*}
\centering
  \subfloat[36~MHz -- RMS: $0.2\un Jy/beam$ -- Beam: $73 \arcsec \times 58 \arcsec$]{
    \includegraphics[width=\columnwidth]{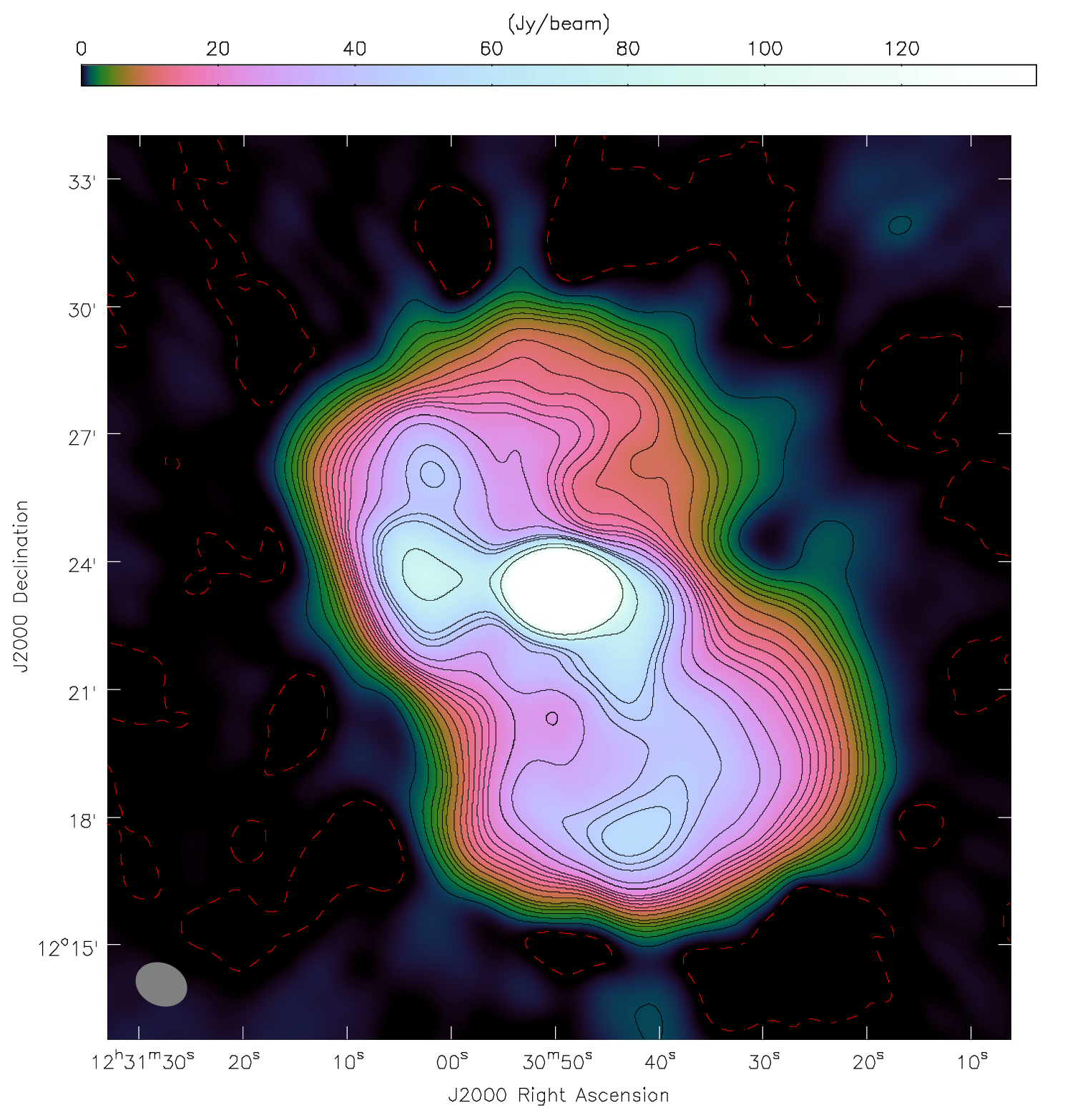}
    \label{fig:VirA-LBA1}}
  \subfloat[48~MHz -- RMS: $0.09\un Jy/beam$ -- Beam: $55 \arcsec \times 43 \arcsec$]{
    \includegraphics[width=\columnwidth]{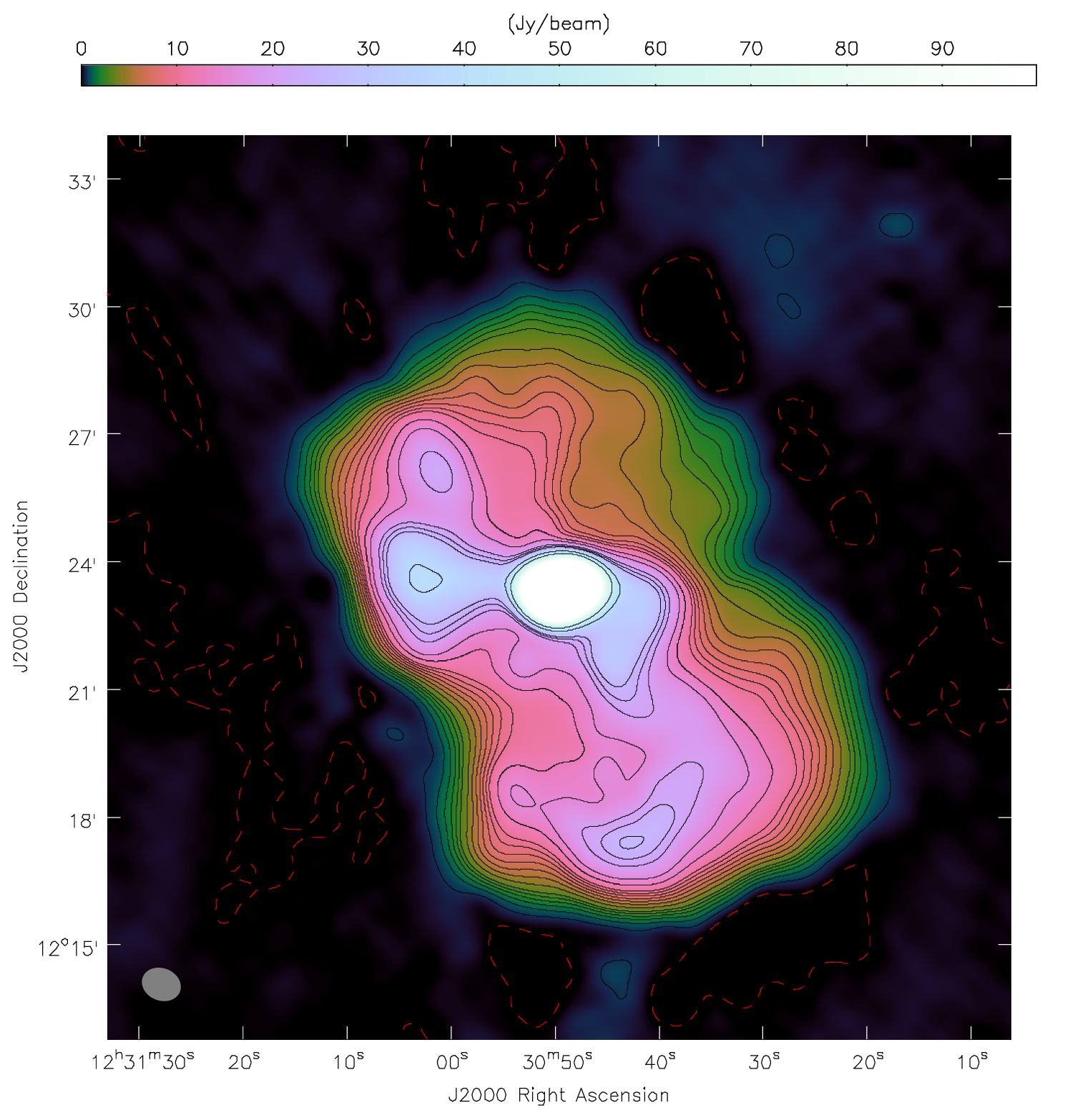}
    \label{fig:VirA-LBA2}}\\
  \subfloat[59~MHz -- RMS: $0.07\un Jy/beam$ -- Beam: $45 \arcsec \times 36 \arcsec$]{
    \includegraphics[width=\columnwidth]{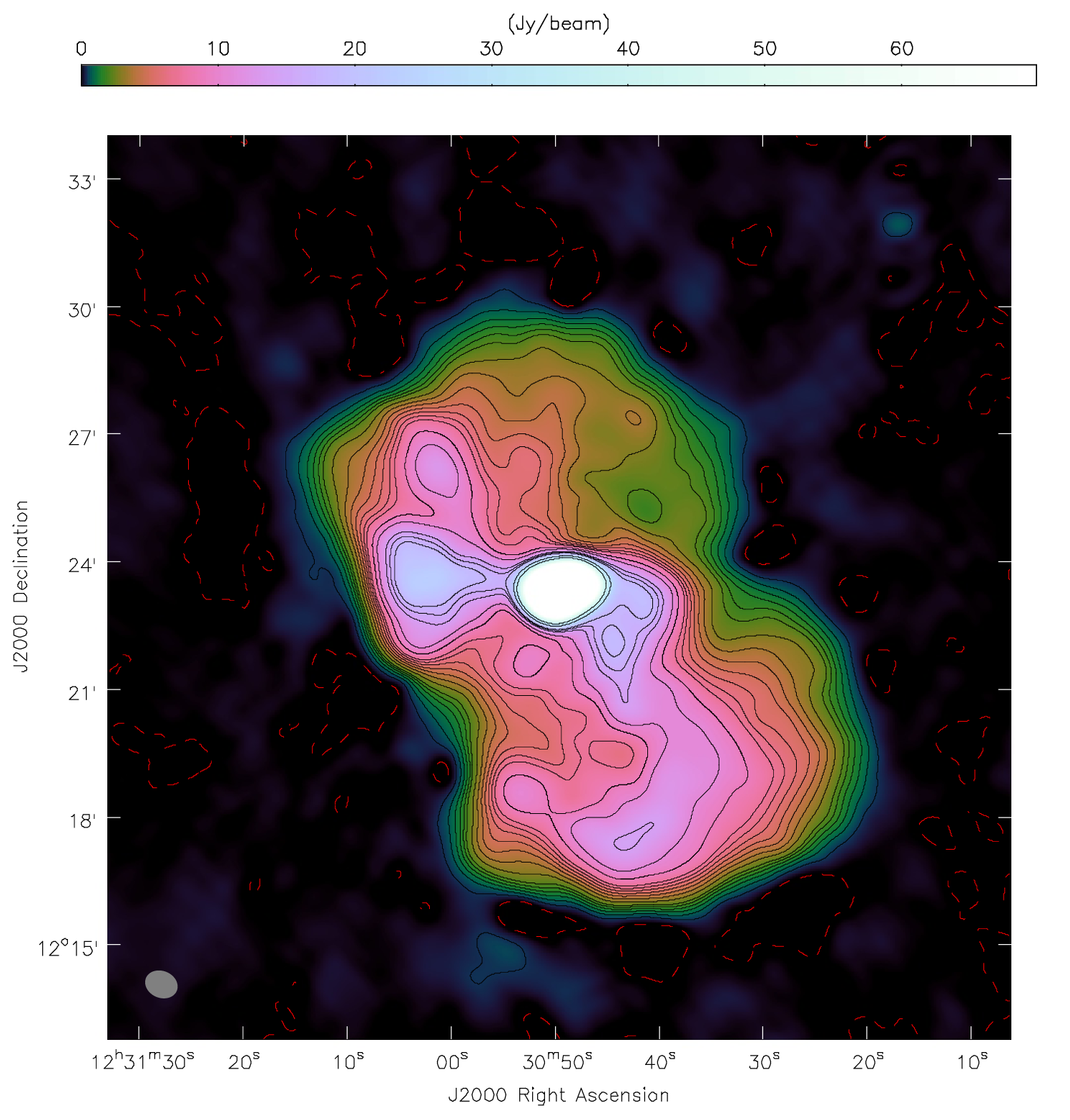}
    \label{fig:VirA-LBA3}}
  \subfloat[71~MHz -- RMS: $0.05\un Jy/beam$ -- Beam: $37 \arcsec \times 30 \arcsec$]{
    \includegraphics[width=\columnwidth]{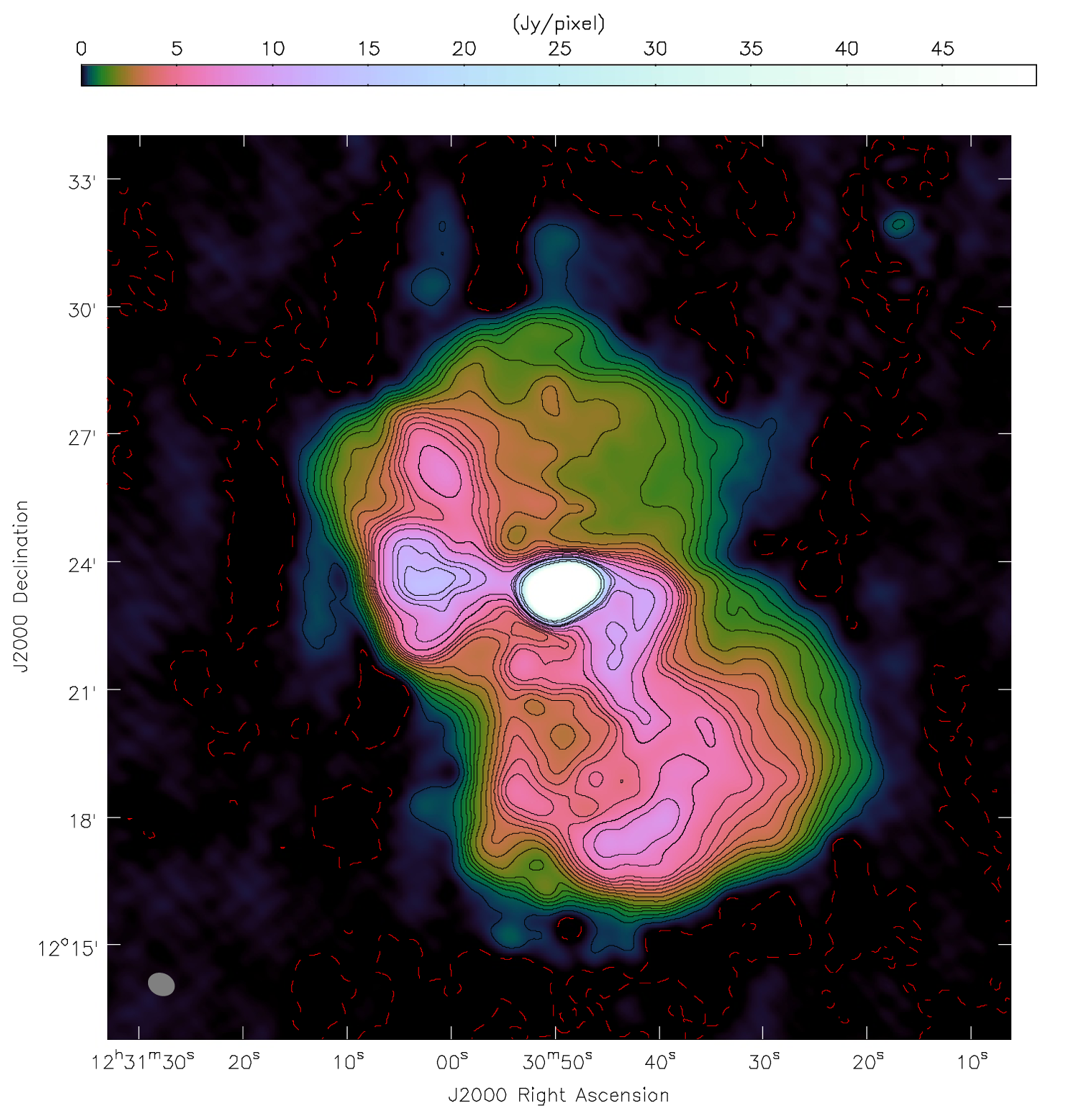}
    \label{fig:VirA-LBA4}}\\
\caption{LOFAR-LBA images of Virgo~A at frequencies ranging from 36 to 71~MHz. Each image is a result of a multi-scale multi-frequency cleaning on a subset of 60 SBs. The beam shape is visible in the bottom-left corner of each image. Positive contour levels are represented by black lines at (1, 2, 3, 4, 5, 6, 8, 10, 12, 14, 16, 18, 20, 25, 30, 25, 40, 45, 50, 75, 100) $\times 5\sigma$. Negative contour levels at $-1\sigma$ are represented by a red dashed line.}
\label{fig:VirA-LBA}
\end{figure*}

\begin{figure}
\includegraphics[width=\columnwidth]{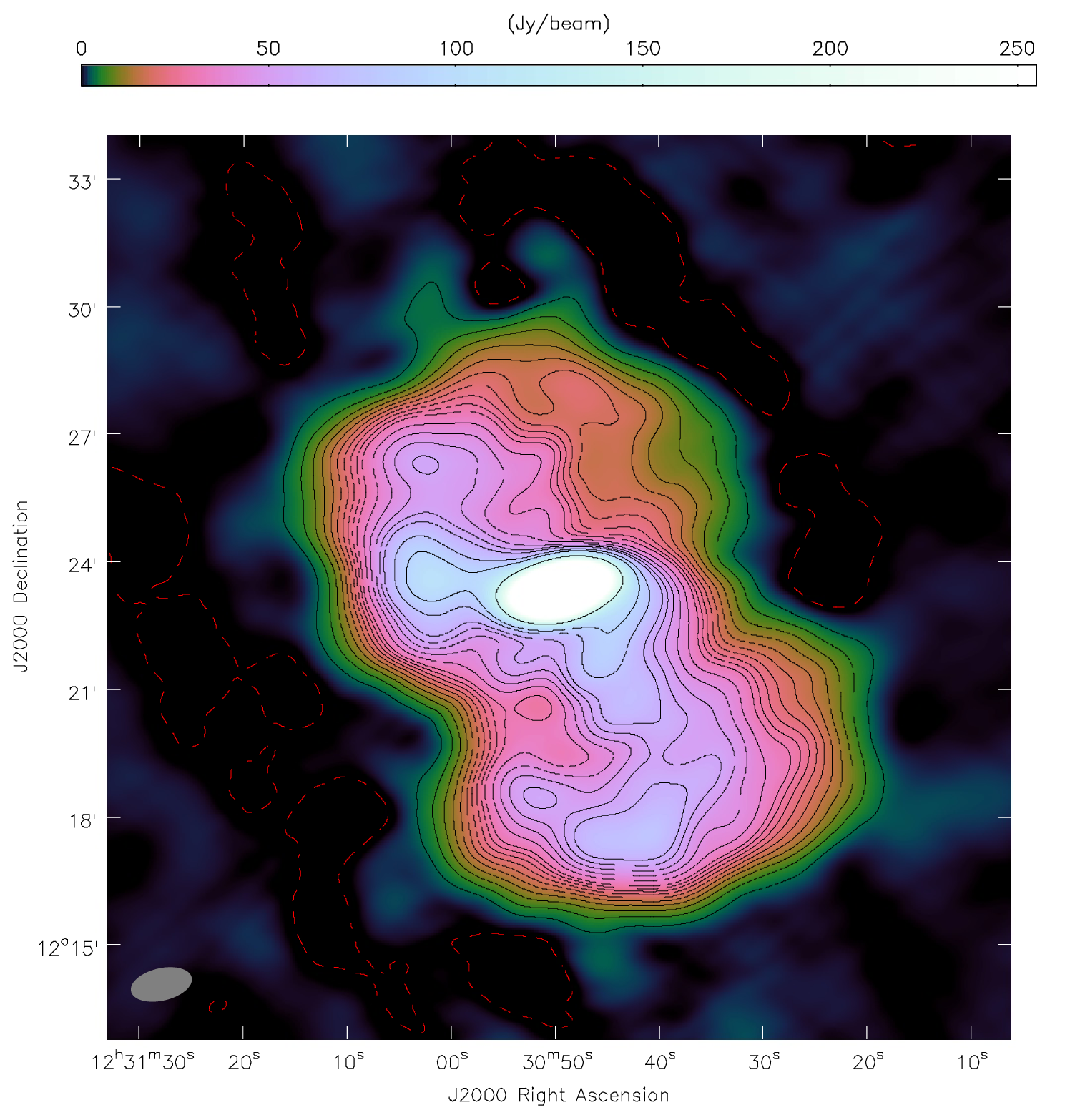}
\caption{Image of Virgo A at 25~MHz. The map noise level is $\sigma = 0.6\un Jy/beam$ and the beam size is $85 \arcsec \times 44 \arcsec$ (grey ellipse in the bottom-left corner). Positive contour levels are represented by black lines at (1, 2, 3, 4, 5, 6, 8, 10, 12, 14, 16, 18, 20, 25, 30, 25, 40, 45, 50, 75, 100) $\times 5\sigma$. Negative contour levels at $-1\sigma$ are represented by a red dashed line.}
\label{fig:VirA-LLBA}
\end{figure}

In Fig.~\ref{fig:VirA-HBA} we show the image of Virgo~A as seen with the LOFAR-HBA at an average frequency of 140~MHz. This image is an average over the entire 48~MHz of bandwidth and the imaging step has been performed with CASA using the multi-scale deconvolution algorithm with a Briggs weighting ($\rm robust=-0.5$). In Fig.~\ref{fig:VirA-LBA} we show four images of Virgo~A obtained with the LOFAR-LBA. Each image is realised with CASA using a multi-scale multi-frequency deconvolution algorithm \citep{Cornwell08,Rau2011} on a subset of 60 SBs (12~MHz of bandwidth) with uniform weighting. Finally in Fig.~\ref{fig:VirA-LLBA} we present a very low frequency (25~MHz) image of Virgo~A. This image was obtained in CASA using a multi-scale multi-frequency deconvolution algorithm with uniform weighting on all usable SBs of the LBA-low dataset ($20-30$~MHz). The rms error in the images is set by deconvolution errors which limit our dynamic range to $\sim 5000$ (for the HBA map). Our ability to recover flux not in the model was confirmed by the detection of several sources which were not included in it. We recovered $>50$ sources in LBA wide-field and $>300$ sources in the HBA wide-field (de Gasperin et al. in prep.). The morphological structure of Virgo~A instead appear similar from 20~MHz up to the GHz regime, a part from those differences caused by spatial changes in the spectral index.

\begin{figure}
\includegraphics[width=.5\textwidth]{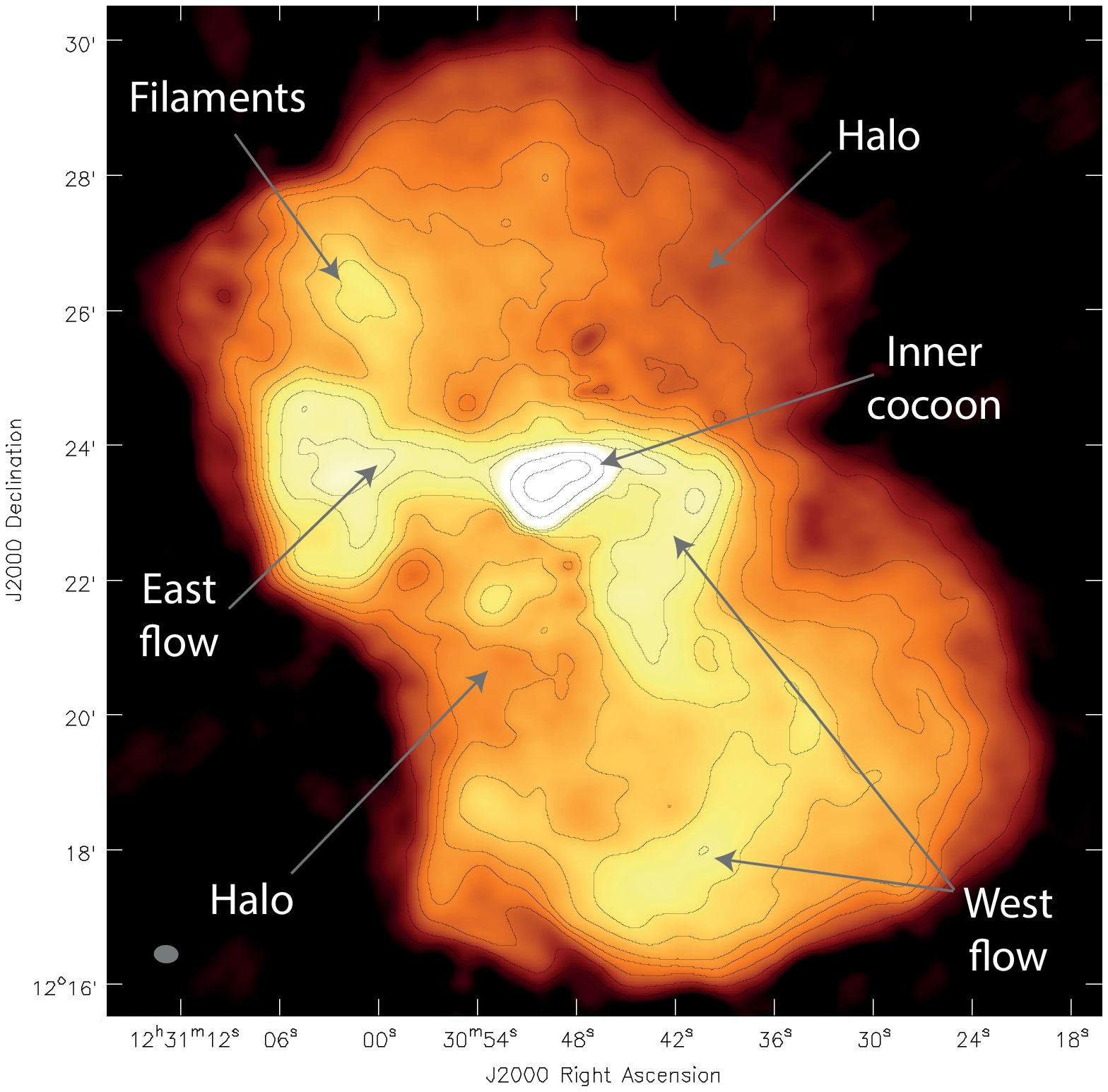}
\caption{Same as Fig.~\ref{fig:VirA-HBA} with most prominent features described in the text labelled. Contour lines are at (1, 2.5, 5, 7.5, 10, 15, 20, 25, 30, 35, 45, 250, 1000) $\times 3\sigma$.}
\label{fig:VirA-HBA-labels}
\end{figure}

Virgo~A has a $\sim 5$ kpc-wide inner cocoon, where a one-sided jet is visible. The jet, detected also in the optical and X-ray bands, points towards the north-west. The counter-jet, although not visible due to the effect of relativistic de-boosting, is probably responsible for the emission in the east part of the inner region. The inner region is surrounded by two much fainter, much larger ``bubbles'' ($\sim 40\un kpc$ wide, see Fig.~\ref{fig:VirA-HBA-labels}) that are overlapping in the central region because of projection effects. The inner and the outer haloes are connected by two large ``flows'' (see Fig.~\ref{fig:VirA-HBA-labels}). The first is oriented almost exactly east-west and the second slightly to the north of west, aligned with the inner jet. The eastward-flow proceeds straight, forming a well-defined cylinder, and ends in a pair of bright lobes, whose edges are brighter than their central part. The westward-flow, on the other hand, quickly changes its direction projected into the sky plane and twists as soon as it leaves the inner region. The flow then proceeds towards the south and is composed of a number of thinner structures that, following \cite{Owen2000}, we call ``filaments''. Both flows originate in the inner region and reach the border of the outer haloes. Once the halo edge is reached both flows disperse, although only the west-flow seems to fill the entire halo with its plasma-filaments. The presence of these flows which connect the inner halo to the outer edges indicate that the diffuse emission is not a simple relic of a previous outburst, but fresh energetic particles still flow from the central cocoon. Plasma ages derived from spectral fits along the flows, confirm this picture (see Sect.~\ref{sec:specfit}).

The 140~MHz image shown in Fig.~\ref{fig:VirA-HBA}, although less resolved than the $7 \arcsec$ resolution 325~MHz image presented in \cite{Owen2000}, confirms some characteristics of this source. First, the outer halo has a sharp edge and all the radio-emitting plasma seems to be confined within its boundaries. Second, although less visible than in the 325~MHz image, part of these edges are limb brightened, reinforcing the previous statement.

Interestingly, we can confirm that this picture is valid down to 25~MHz (see Fig.~\ref{fig:VirA-LBA} and~\ref{fig:VirA-LLBA}). If relic emission of past AGN activities extending beyond the sharp edges had been present, it would have been detectable thanks to its steep spectra. However, even at 25~MHz all visible emission is confined within the same boundaries that we see at higher frequencies. This fact also supports the picture that all of the emitting plasma is well confined by the strong pressure of the ICM, as we will discuss in more detail in Sect.~\ref{sec:specfit}. 

\section{Spectral analysis of the extended halo}
\label{sec:spectral}

\subsection{Spectral index map}
\label{sec:specmap}

\begin{figure*}
  \begin{minipage}[t]{.66\linewidth}\vspace{0pt}
    \includegraphics[width=\linewidth]{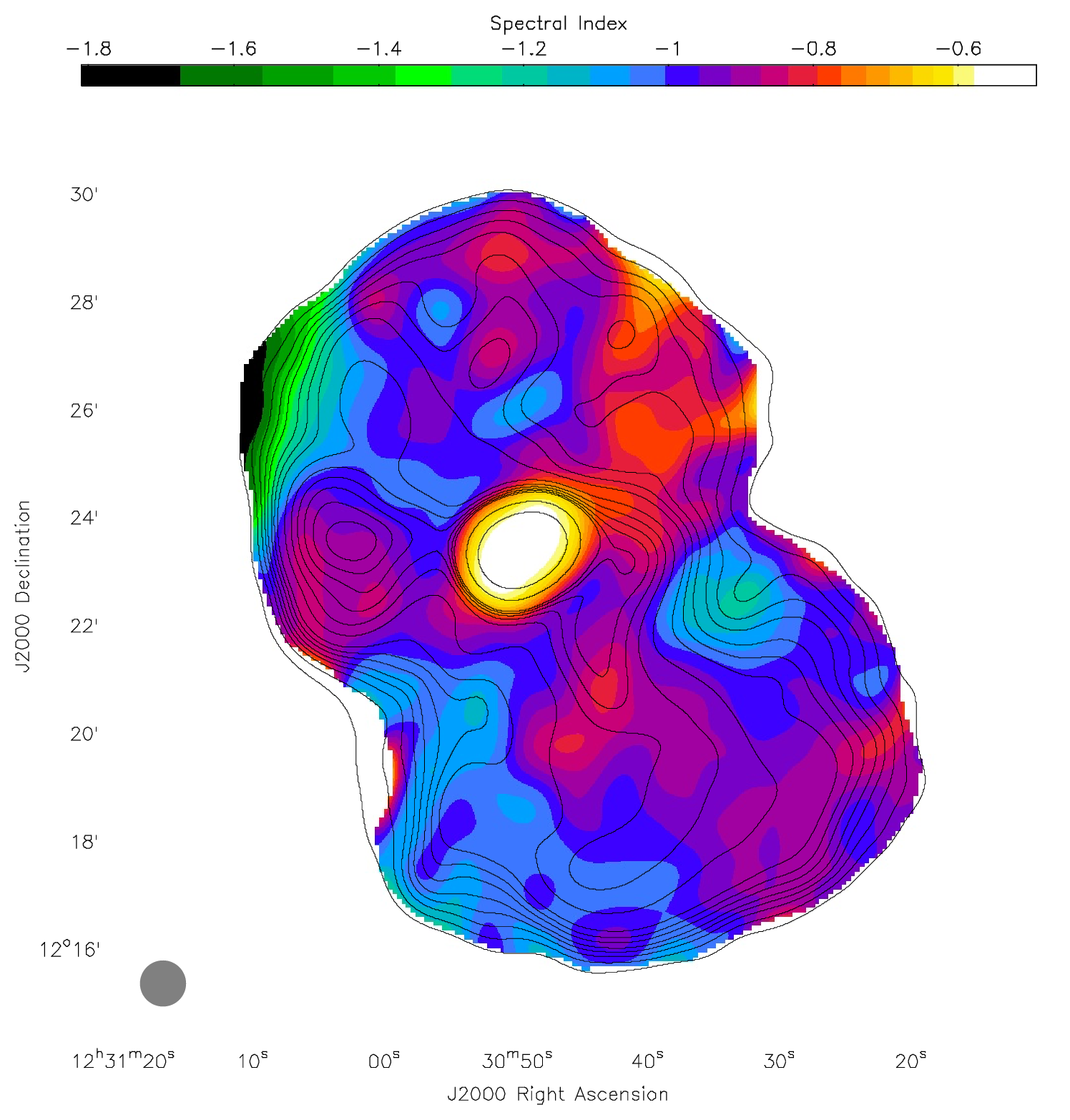}
    \label{fig:spidx-map}
  \end{minipage}
  \hfill
  \begin{minipage}[t]{.33\linewidth}\vspace{5pt}\raggedright
    \begin{minipage}[t]{\linewidth}\vspace{0pt}\raggedright
      \includegraphics[width=\linewidth]{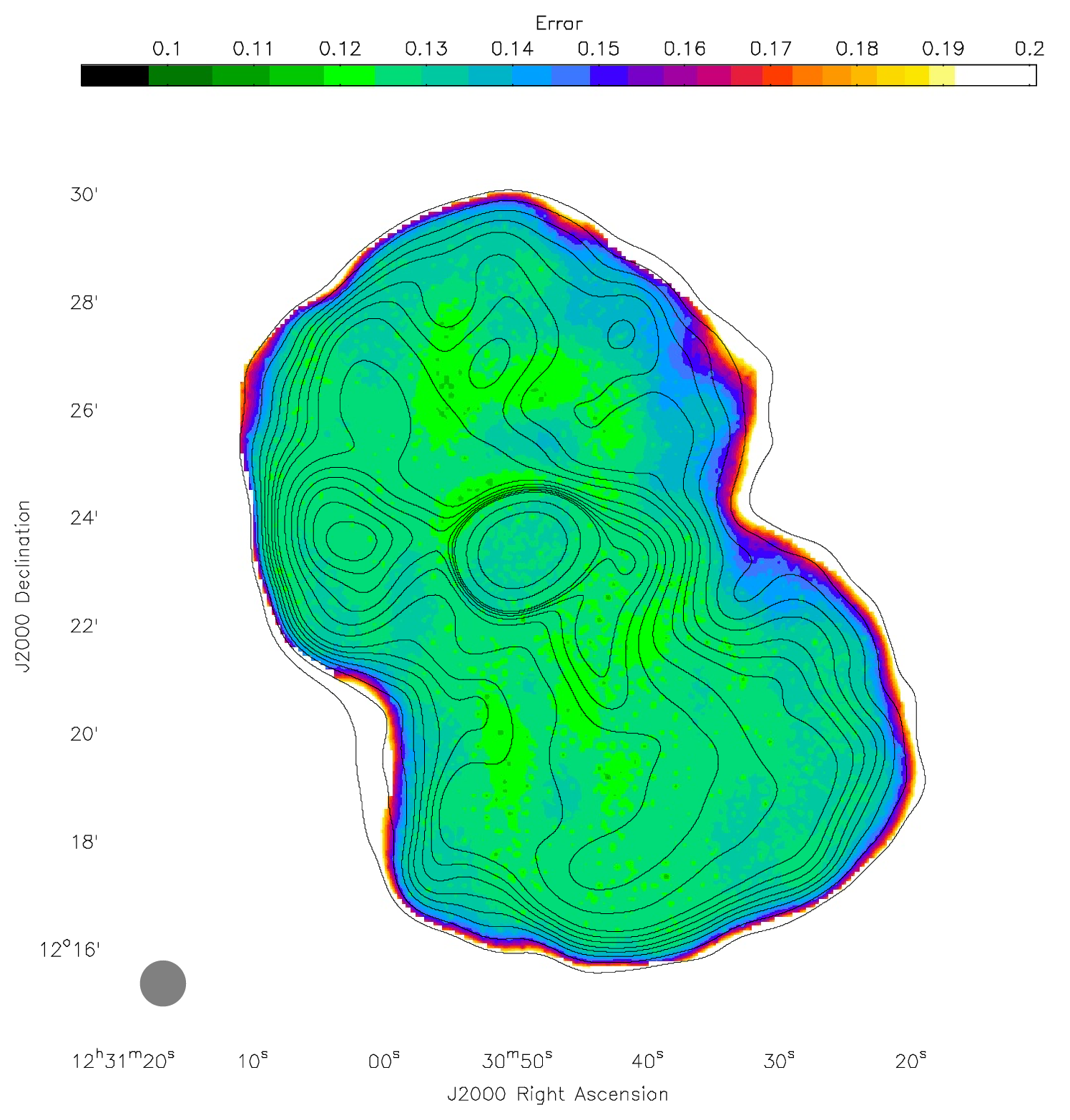}
      \label{fig:spidx-err}
    \end{minipage}

    \begin{minipage}[t]{\linewidth}\vspace{0pt}\raggedright
      \caption{\textit{Left figure}: Low-frequency spectral index map obtained from LOFAR-LBA ($45-71$~MHz only) and LOFAR-HBA ($115-162$~MHz) observations, together with VLA map at 325~MHz. All maps were convolved to a resolution of $50\arcsec$ (see circle in the lower left corner) and a pixel-by-pixel linear regression was extracted. Pixels where the error were above 3$\sigma$ are blanked. Contour lines are from the 325~MHz map. \textit{Top figure}: spectral index $1 \sigma$ error map.}
      \label{fig:spidx}
    \end{minipage}
  \end{minipage}

\end{figure*}

In this section we excluded the very low frequency part ($\leqslant 45\un MHz$) of the LBA observations to retain an angular resolution of $50\arcsec$. We produced an image from each SB and we convolved them to a resolution of $50\arcsec$. Then, we averaged all LBA and HBA images separately, obtaining one image in the HBA frequency range and one image in the LBA frequency range. Finally, the low-frequency spectral index map shown in Fig.~\ref{fig:spidx} was obtained by extracting a pixel-by-pixel linear regression using three images: one extracted from the LOFAR-HBA observation ($115-162$~MHz), one extracted from the LOFAR-LBA observation ($45-77$~MHz) and a third one from the VLA at 325~MHz.

\begin{figure}
\includegraphics[width=.95\columnwidth]{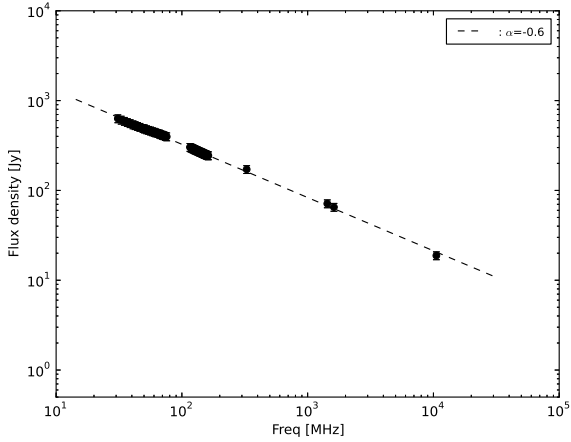}
\caption{Fit to the integrated flux spectrum of the central region. The dashed line is a linear fit with a slope of $\alpha = -0.6\pm0.02$.}
\label{fig:spectrum-core}
\end{figure}

The central cocoon of the source has a spectral index\footnote{Spectral index definition: $F_{\nu} \propto \nu^{\alpha}$.} ranging from $-0.55$ to $-0.6$, consistent with what has been observed by other authors at higher frequencies up to the optical band \citep{Biretta91}. The spectrum is a straight power-law down to 30~MHz (see Fig.~\ref{fig:spectrum-core}). No evidence of a turnover due to self-absorption is visible down to these frequencies. From the total integrated spectrum shown in Fig.~\ref{fig:scale} a possible sign of a turnover in the source integrated flux is visible at frequencies $\lesssim 20\un MHz$, so outside our frequency coverage. Features north of the bright core are likely affected by deconvolution errors and we do not consider them as real.

In the southern lobe the spectral index flattens by $10\%-20\%$ where the bright flow twists and bends with respect to the surrounding areas. The spectral index in the east lobe is comparable to what is observed in the southern lobe in the radio-brightest zones. The two prominent filaments to the north-east on the other hand, do not present any peculiar spectral index structure, although this can be related to the low resolution of our spectral index map. The faint extension towards the north-east is the steepest part of the halo, reaching in our map a spectral index of $-1.8$. Interestingly this feature is co-located with what \cite{Forman07} identify as an external cavity in the X-ray halo. The rest of the northern lobe has the lowest signal to noise ratio of the map, therefore it could be affected by spurious features.

The northern halo is related to the counter-jet, and therefore farther away from the observer. Differences between north and south lobes were also found by \cite{Rottmann96} at 10~GHz. They observed a higher degree of polarization in the southern lobe \citep[due to the Laing-Garrington effect,][]{Laing88,Garrington88} and a total flux from the southern lobe that is 20\% higher than from the northern one.

In general, there is no noteworthy relation between the spectral index and surface brightness maps, although a steepening of the spectral index is present (at the north-east and south-east edges of the map), where a reduction in the flow-related activity is present and a flattening is visible in some of the flow-active locations (the east lobe and the initial part of the west flow). Although in the lowest signal-to-noise zone of the map, we report a flattening of the spectrum by $\sim 20\%$ compared to the rest of the halo in the north-west part of it. This feature seems not to be related with any structure in the brightness maps.

\subsection{Spectral index fits}
\label{sec:specfit}

We will now make a detailed analysis of the source radio spectrum, in specific regions of interest. For this analysis we decided to retain all of the frequencies down to 30~MHz, which limits our angular resolution to $75\arcsec$. The LOFAR maps have been averaged in blocks of 10 (bandwidth of 2~MHz), resulting in 24 maps in the LBA and HBA frequency range each. We also used three archival VLA maps at 325, 1400 and 1600~MHz and a single-dish Effelsberg map at 10.55~GHz \citep{Rottmann96}. All of these maps were convolved to a resolution of $75\arcsec$ and spectral index fits were performed using Synage++ \citep{Murgia01}. At low-frequencies ($<325$~MHz) the spectra are almost straight and their slopes are the same reported in Fig.~\ref{fig:spidx}, although in the following analysis the resolution is lower compared to that used for the figure. To assess the reliability of a spectral index study with images produced by different interferometers, the same model of Virgo~A was simulated in the used LOFAR and VLA configurations and frequencies. We imaged those data using the same weighting scheme (uniform), cell-size and iterations and convolved the CLEANed maps at the same resolution to compare the outcomes of the different datasets. Virgo~A is never resolved out, but the different \textit{uv}-coverages and, to a lesser extent, the missing short baselines at higher frequencies, create artifacts. Although the ratio between maps produced with different instruments shows errors up to $10\%$ in a single pixel in the zones where the signal-to-noise ratio is low, we note that such errors are not in the form of an overall bias but of patches of higher/lower flux (the average error of the pixel fluxes across the whole source is $+0.3\%$). However, in our analysis we have always used the flux integrated over a certain solid angle, therefore we extracted the error for all the zones described in Sect.~\ref{sec:zonebyzone}, finding in every case an integrated flux discrepancy below 1\%.

In the following analysis of the spectral data, we tested three different models:
\begin{description}
 \item[The JP model \citep{Jaffe73}:] models spectral ageing as due to synchrotron and inverse Compton losses, with the pitch angles of the synchrotron emitting electrons continuously isotropized on a time-scale shorter than the radiative time-scale.
 \item[The KP model \citep{Kardashev62,Pacholczyk70}:] as in the JP model, but now the pitch angle of the electrons remains in its initial orientation with respect to the magnetic field. 
 \item[The CI model \citep{Pacholczyk70}:] in the ``continuous injection'' model, an uninterrupted supply of fresh particles is injected by the central source. These particles age following the JP model. This model is applicable only if the injected particles cannot escape from the selected region. Therefore, we used it only on the integrated flux from the whole halo.
\end{description}
Compared to the KP model, the JP model is more realistic from a physical point of view, as an anisotropic pitch angle distribution will become more isotropic due to scattering, magnetic field lines wandering in a turbulent medium, and changes in the magnetic field strength between different regions \citep[e.g.,][]{Carilli91}. Furthermore, since inverse Compton losses due to scattering by CMB photons can isotropise the electron population, a true KP model can be visible only for strong magnetic fields \citep[$\gtrsim 30\un \mu G$,][]{Slee01}, where inverse Compton losses are negligible. In all of these models the injected particles are assumed to have a power-law energy spectrum $N(\gamma) \propto \gamma^{\delta_{\rm inj}}$ (where $\gamma$ is the particles' Lorentz factor), which results in a power-law radiation spectrum with spectral index $\alpha_{\rm inj} = \left(\delta_{\rm inj} + 1\right)/2$. Finally, the magnetic field strength is assumed constant for the entire radiating period.

Since the most energetic particles radiate their energy more efficiently, they are the first to be depleted. Therefore, the source radio spectrum evolves and displays a break to a steeper slope at a break frequency $\nu_{\rm b}$, which relates to the time elapsed from the injection and to the magnetic field as $\nu_{\rm b} \propto B^{-3} t^{-2}$ \citep{Jaffe73}. Major differences between the models are visible at frequencies higher than $\nu_{\rm b}$, while at lower frequencies all models are expected to have a spectral index equal to $\alpha_{\rm inj}$. There are three free parameters in these models: the first is the spectral slope $\alpha_{\rm inj}$ of the synchrotron emission generated by an injected electron population, the second is the break frequency $\nu_{\rm b}$ and the third is the overall normalization.

A modification to the CI model that was also tested is the CIOFF model \citep{Komissarov1994}, which allows the source to switch off after a certain time, encoded in an extra free parameter $\eta_{\rm off} = t_{\rm relic}/t_{\rm source}$, where $t_{\rm source}$ is the time since the beginning of the outburst and $t_{\rm relic}$ is the time since the source switched off. In this case the spectrum would show a first break frequency that depends on $t_{\rm source}$, which separates two power-laws like in the standard CI model, and an exponential cut-off at higher frequencies which depends on $t_{\rm relic}$.

\subsubsection{Central cocoon and macro-regions}
\label{sec:coreandmacroregions}

First, a spectral fit was made to the central region (Fig.~\ref{fig:spectrum-core}) defined by C in Fig.\ref{fig:zones}. As the data appear to be described by a straight line down to 30~MHz, we fitted a simple power-law, obtaining a slope of $\alpha = -0.6\pm0.02$. In this and in the subsequent fits, the errors on the integrated fluxes are computed from the RMS on the individual maps (multiplied by the square root of the number of beams covering the area) plus a 10\% error due to systematics\footnote{The systematic errors considered are: a 1\% error due to the different \textit{uv}-coverages (Sec.~\ref{sec:specfit}), 3\% due to the uncorrected beam shape (Sec.~\ref{sec:flux}), and 4\% for the uncertainty in the absolute flux rescaling (Sec.~\ref{sec:flux}). A conservative total systematic errors budget of 10\% has been adopted. However, this value overestimates the real error in the brightest parts of the source, therefore providing particularly low $\chi^2_{\rm red}$ values.} combined in quadrature.

\begin{figure}
\centering
  \subfloat[Halo (without central cocoon)]{
    \includegraphics[width=.95\columnwidth]{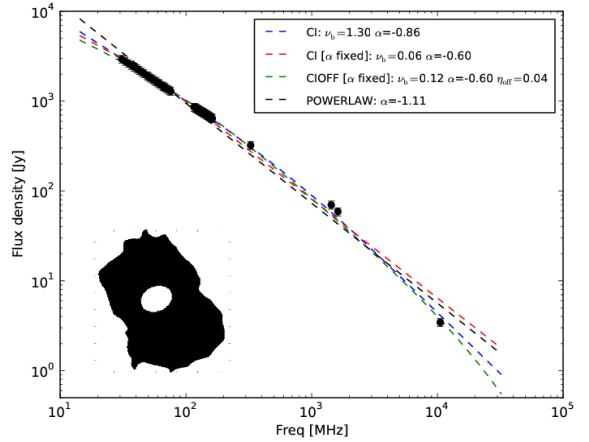}
    \label{fig:spectrum-halo-nocore}}\\
  \subfloat[Halo (without central cocoon and flows)]{
    \includegraphics[width=.95\columnwidth]{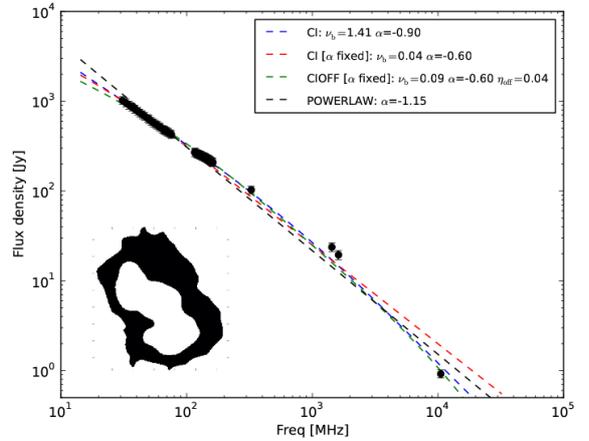}
    \label{fig:spectrum-halo-nojets-nocore}}\\
  \subfloat[Flows (without central cocoon)]{
    \includegraphics[width=.95\columnwidth]{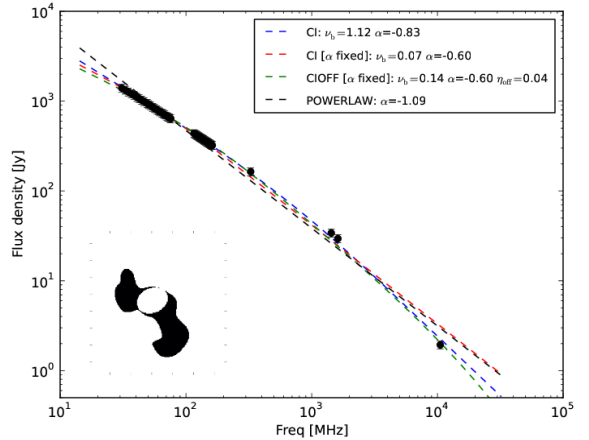}
    \label{fig:spectrum-jets-nocore}}\\
\caption{Spectral fits of the three macro-regions identified in Sect.~\ref{sec:coreandmacroregions}. In each panel the mask used to select the region is shown in black in the bottom-left corner, the white part was excluded from the analysis. A simple power-law fit (black line) is shown together with two fits of the CI and CIOFF models. Red line: fit obtained fixing $\alpha_{\rm inj}$ to $-0.6$. Blue line: with $\alpha_{\rm inj}$ allowed to vary. A CI model with $\alpha_{\rm inj}=-0.6$ is in general not able to fit the data. Finally, green lines show the fit of a CIOFF model with $\alpha_{\rm inj}$ fixed to $-0.6$. $\nu_{\rm b}$ is in GHz.}
\label{fig:spectra-large}
\end{figure}

\begin{table*}
\centering
\caption{Global spectral fits}
\label{tab:spfit-large}
\begin{tabular}{lcccccccccc}
\hline\hline
Region & \multicolumn{3}{c}{\sout{\hfill}\ CI Model\ \sout{\hfill}} \
       & \multicolumn{2}{c}{CI Model\ ($\alpha_{\rm inj} = -0.6$)} \
       & \multicolumn{3}{c}{CIOFF Model\ ($\alpha_{\rm inj} = -0.6$)} \
       & \multicolumn{2}{c}{\sout{\hfill}\ Power-law\ \sout{\hfill}}  \bigstrut[t]\\
       & $\chi^2_{\rm red}$ & $\nu_{\rm b}$ [GHz] & $-\alpha_{\rm inj}$ \
       & $\chi^2_{\rm red}$ & $\nu_{\rm b}$ [GHz] \
       & $\chi^2_{\rm red}$ & $\nu_{\rm b}$ [GHz] & $\eta_{\rm off}$ \
       & $\chi^2_{\rm red}$ & $-\alpha_{\rm inj}$ \bigstrut[b]\\

Central cocoon & -- & -- & -- & -- & -- & -- & -- & -- &  $0.060$ &  $0.60^{+0.02}_{-0.02}$ \bigstrut[t]\\
Halo (no core) &    $0.198$ &  $1.3^{+0.3}_{-0.7}$ &  $0.86^{+0.02}_{-0.06}$ &  $1.471$ &  $0.06^{+0.01}_{-0.01}$ & $0.282$ &  $0.12^{+0.03}_{-0.04}$ &  $0.04^{+0.01}_{-0.01}$ &  $1.997$ &  $1.11^{+0.02}_{-0.02}$ \\
Halo (no flows) &  $0.371$ &  $1.4^{+0.3}_{-0.8}$ &  $0.90^{+0.03}_{-0.06}$ &  $2.607$ &  $0.04^{+0.01}_{-0.01}$ & $0.439$ &  $0.09^{+0.02}_{-0.03}$ & $0.04^{+0.01}_{-0.01}$ &  $2.156$ &  $1.15^{+0.01}_{-0.02}$ \\
Flows (no core) &  $0.121$ &  $1.1^{+0.2}_{-0.7}$ &  $0.83^{+0.02}_{-0.07}$ &  $0.970$ &  $0.07^{+0.02}_{-0.01}$ & $0.182$ &  $0.14^{+0.03}_{-0.04}$ & $0.04^{+0.01}_{-0.02}$ &  $1.905$ &  $1.09^{+0.02}_{-0.02}$  \bigstrut[b]\\
\hline
\end{tabular}
\end{table*}

Then, we defined three macro-regions to obtain the average spectra of the halo and the flows.
\begin{itemize}
 \item The first region (Fig.~\ref{fig:spectrum-halo-nocore}) was obtained by cutting the 36~MHz map at the $5 \sigma$ level and removing the central cocoon. We chose this frequency to maximise the flux from the halo (which is higher at lower frequencies).
 \item The second region (Fig.~\ref{fig:spectrum-halo-nojets-nocore}) was obtained by removing from the previous map all the area with a surface brightness above $30\un Jy/beam$, i.e. the parts of the halo dominated by the flows.
 \item The final region (Fig.~\ref{fig:spectrum-jets-nocore}) was obtained by retaining only the flows-dominated part of the halo (surface brightness $> 40\un Jy/beam$), and removing the central region.
\end{itemize}
A spectral fit using the CI and the CIOFF models has been performed on each of these zones and the results are shown in Fig.~\ref{fig:spectra-large} and in Table~\ref{tab:spfit-large}. Firstly, we performed a standard linear regression, from which it can be seen that the spectra are curved at high frequencies and, to a lesser extent, also at low frequencies. Then, we fitted the data using the CI model and fixing $\alpha_{\rm inj} = -0.6$, equal to the core spectral index. In this case the model is not able to reproduce the data (see Fig.~\ref{fig:spectra-large}).

\begin{figure}
\includegraphics[width=\columnwidth]{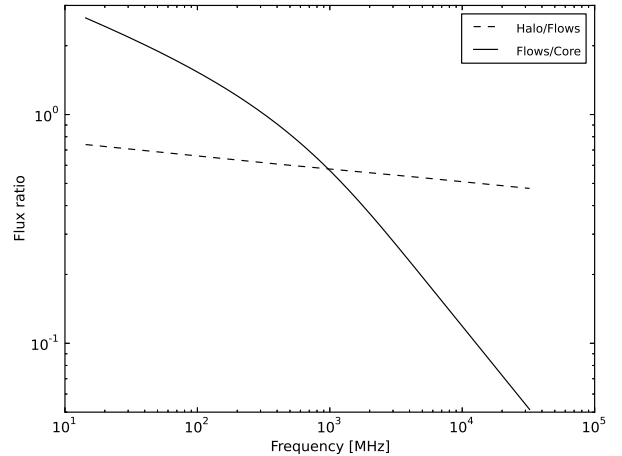}
\caption{The solid line shows the ratio between the power-law fit to the cocoon zone (Fig.~\ref{fig:spectrum-core}) and the CI-model fit to the flow zones (Fig.~\ref{fig:spectrum-jets-nocore}). The dashed line shows the ratio between the CI-model fits to the flow zones (Fig.~\ref{fig:spectrum-jets-nocore}) and to the halo (Fig.~\ref{fig:spectrum-halo-nojets-nocore}). Simple synchrotron ageing would have left untouched the low-frequency part of the spectrum, producing in this plot a horizontal line until the point where the break frequency occurs. We notice instead a steepening in the spectrum, going from the central region to the flows and, to a lesser extent, from the flows to the halo.}
\label{fig:specratio}
\end{figure}

We decided then to relax some constraints and we repeated the fit using a CI model with all three parameters ($\nu_{\rm b}$, $\alpha_{\rm inj}$ and the normalization) free to vary. In this case we found a $\nu_{\rm b}$ between 1.1 and 1.4 GHz and an $\alpha_{\rm inj}$ between $-0.83$ and $-0.9$. This result suggests that the extended halo is quite young: $\lesssim 50\un Myr$, assuming an average magnetic field strength of $10\ \rm \mu G$ (see Sect.~\ref{sec:syncrotron}) and using equation~\ref{eq:time}. We also observe a steepening of $\alpha_{\rm inj}$ moving from the central region ($\alpha_{\rm inj} = -0.6^{+0.02}_{-0.02}$), to the flows ($\alpha_{\rm inj} = -0.83^{+0.02}_{-0.07}$) and the halo ($\alpha_{\rm inj} = -0.90^{+0.03}_{-0.06}$), while taking the errors into account only a marginal steepening is detected moving from the flows to the halo. Although the core and the flows are surrounded by the halo, the projection effects should not alter these results, in fact the halo is on average $\simeq 20$ 
times and $\simeq 4$ times fainter in flux density than the core and the flow regions respectively. In Fig.~\ref{fig:specratio} we plot the ratio between the 
power-law fit to the central cocoon data and the CI-model fit to the flow zones, and the ratio between the latter and the CI-model fit to the halo without flows. If the emission in these zones were related to the same outburst of relativistic particles, simple synchrotron ageing would have left the low-frequency part of the spectrum untouched at $\alpha = -0.6$, producing a constant ratio until the break frequency. Therefore, some other mechanism must have steepened the spectra at the lowest frequencies.
We list here some possible explanations.

(1) Adiabatic expansion of the relativistic plasma will shift the spectra towards lower frequencies and lower intensities, therefore can also affect the low frequency end of the spectrum. A model was developed by \cite{Kardashev62} and revisited by \cite{Murgia1999}. They propose a continuous injection of particles which subsequently expand adiabatically (CIE model). Such a model produces a low-frequency slope of the spectrum that is dominated by the plasmas at different ages after adiabatic expansion which results in a steepening of the spectra compatible with that seen in the lobes of Virgo~A. A similar model was developed in \cite{Blundell99} for double radio sources. The authors proposed that adiabatic expansion of plasma may happen as soon as the plasma leak out from the hot-spot regions. Particles leaving the hot-spots have spent different amounts of time in these regions with high magnetic fields, experiencing different ageing. Consequently, the final particle spectral distribution will be again a sum of spectra with 
many break 
frequencies and adiabatically expanded. This idea was claimed by \cite{Blundell2000} to explain the ``injection index discrepancy'' discussed in \cite{Carilli91}. In this last paper, the authors observe that the low frequency spectral index measured in the lobe of Cygnus~A should reflect either the low-frequency spectral index of the hot-spot ($\alpha = -0.5$, marginal adiabatic losses) or the high-frequency spectral index of the hot-spot ($\alpha = -1$, strong adiabatic losses). They observe instead a low-frequency spectral index of $-0.7$. These numbers are not particularly different from what we observe in Virgo~A. Although these sources are remarkably different, the underlying physical effect which dominates at the low-frequency end of the spectrum might be similar.

(2) During their lifetime, relativistic electrons crossed a wide range of magnetic field strengths. As a result, their final spectrum is the sum of many spectra with many different break frequencies. Particles that have spent much of their life within strong magnetic fields will have a very low break frequency, which can modify the low frequency slope of the radio spectrum.

(3) A third scenario is that the radio spectrum is intrinsically curved even in the core \citep{Blundell2000}. In the source core the magnetic field strength is few mG while it is $\simeq 10\un \mu G$ in the halo (see Sect.~\ref{sec:syncrotron}), implying that the radio emission from the core is powered by electrons with energy $\sim 20$ times smaller than those emitting in the lobes, according to $\gamma \propto \left( \nu/B\right)^{1/2}$, $\gamma$ being the typical Lorentz factor of electrons emitting at the frequency $\nu$ in a magnetic field of strength $B$. Thus, it is possible that the low-frequency synchrotron spectrum of the extended halo reflects a curvature of the spectrum of the emitting particles in the core at higher energies; this is expected in the context of particle acceleration models \citep{Amato2006}. Fig.~\ref{fig:spectrum-core} does not provide compelling evidence in favour of this scenario, as the spectrum does not appear to steepen significantly at higher frequencies, yet future observations with higher resolution will test this possibility.

(4) Another scenario is that we are observing the relic emission of a source that had been active for an extended period and recently stopped injecting plasma in the halo. In this case the flatter spectral index ($\alpha_{\rm inj}\simeq-0.6$) tail of the spectrum lies at lower frequencies, where we cannot detect it, and the $\alpha_{\rm inj}\simeq-0.85$ slope is the steepened part of the spectrum. Following a simple CI prescription, we should expect the high-frequency part to have a slope of $-1.1$. However, to fit the data, a simple CI model is not enough and an exponential cut-off must occur at a frequency of $\sim 5\un GHz$, implying a recent switching off of the fresh particles flow. This model (CIOFF) retains an initial injection index of $\alpha_{\rm inj} = -0.6$, but fits the data much better than the standard CI-model with the same initial slope (see Fig.~\ref{fig:spectra-large} and Table~\ref{tab:spfit-large}). In the CIOFF case the halo is older ($t_{\rm halo} \simeq 150\un Myr$, assuming an average 
magnetic field of $10\un \mu G$ and using equation~\ref{eq:time}) than in the CI scenario and the source must have switched off only a few Myr ago. This ``switching off'' should be interpreted as the most recent detaching of a bubble from the source central region, while in the central cocoon a new bubble is forming. In this case, the steeper low-frequency slope is the consequence of the integration over a wide area of the source, which again implies the addition of many spectra with different break frequencies as a consequence of ages differences.

\subsubsection{Zone by zone spectral analysis}
\label{sec:zonebyzone}

\begin{figure}
\includegraphics[width=\columnwidth]{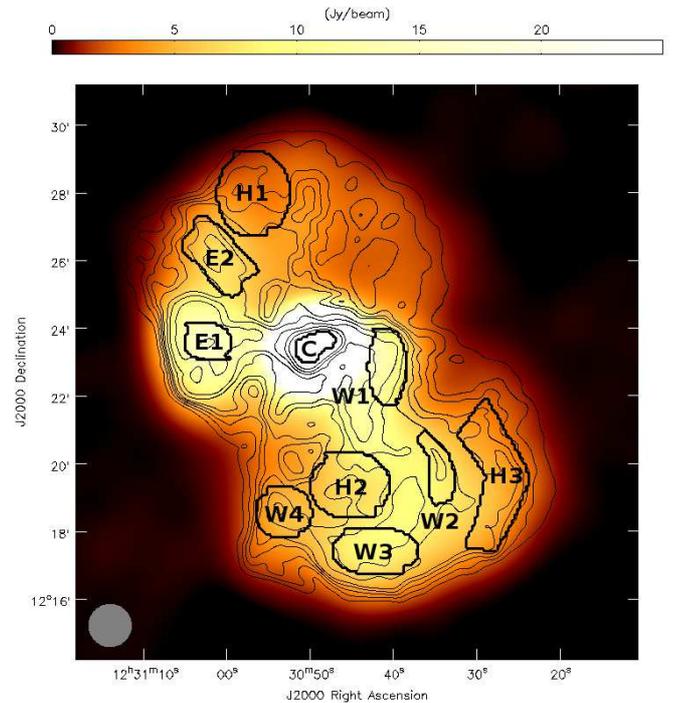}
\caption{Image of Virgo~A at 31~MHz convolved to a resolution of $75\arcsec$. The ten different zones we analysed have been outlined in the map.}
\label{fig:zones}
\end{figure}

\begin{figure*}
\centering
  \subfloat[Zone W1]{
    \includegraphics[width=.33\textwidth]{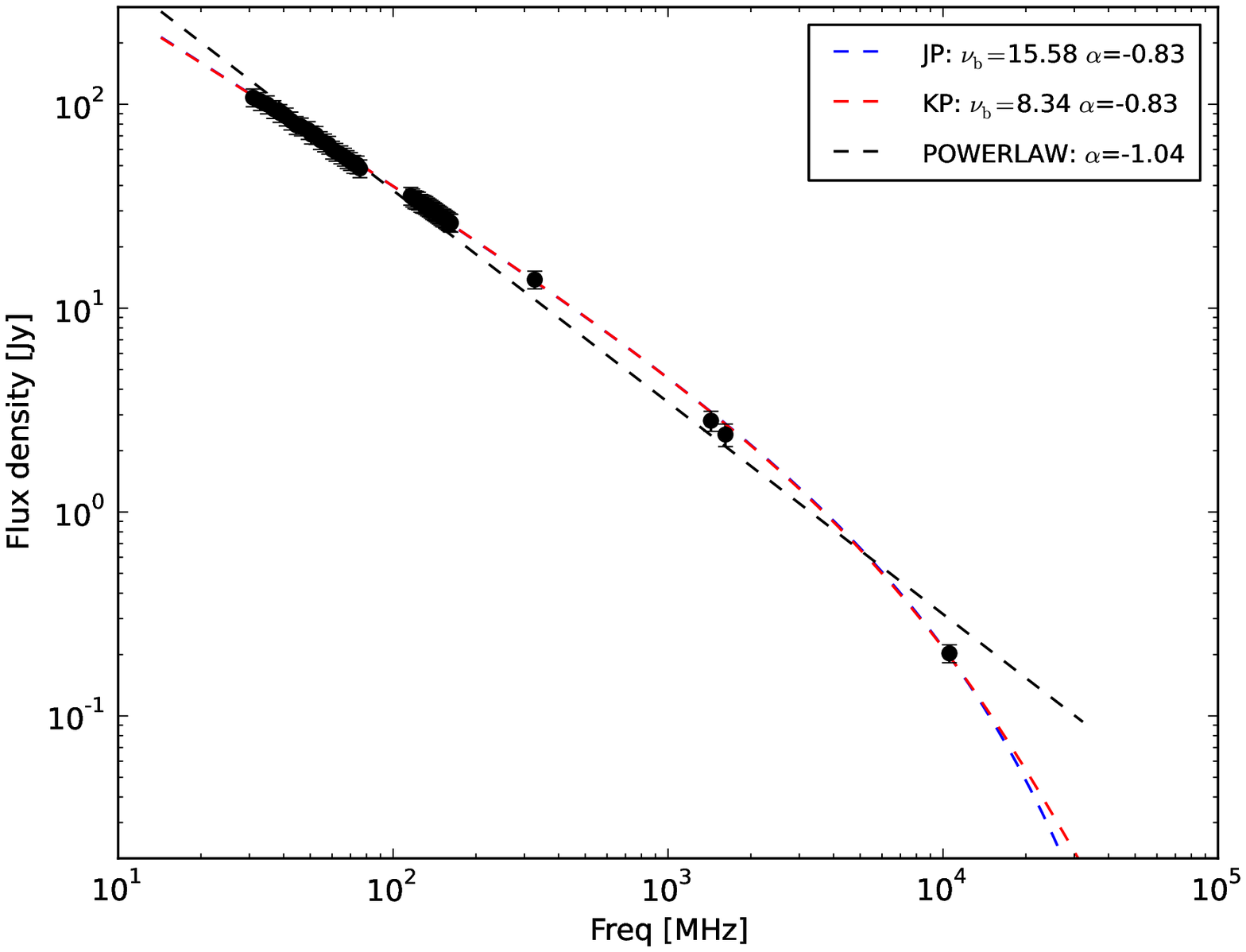}
    \label{fig:spectrum-W1}}
  \subfloat[Zone W2]{
    \includegraphics[width=.33\textwidth]{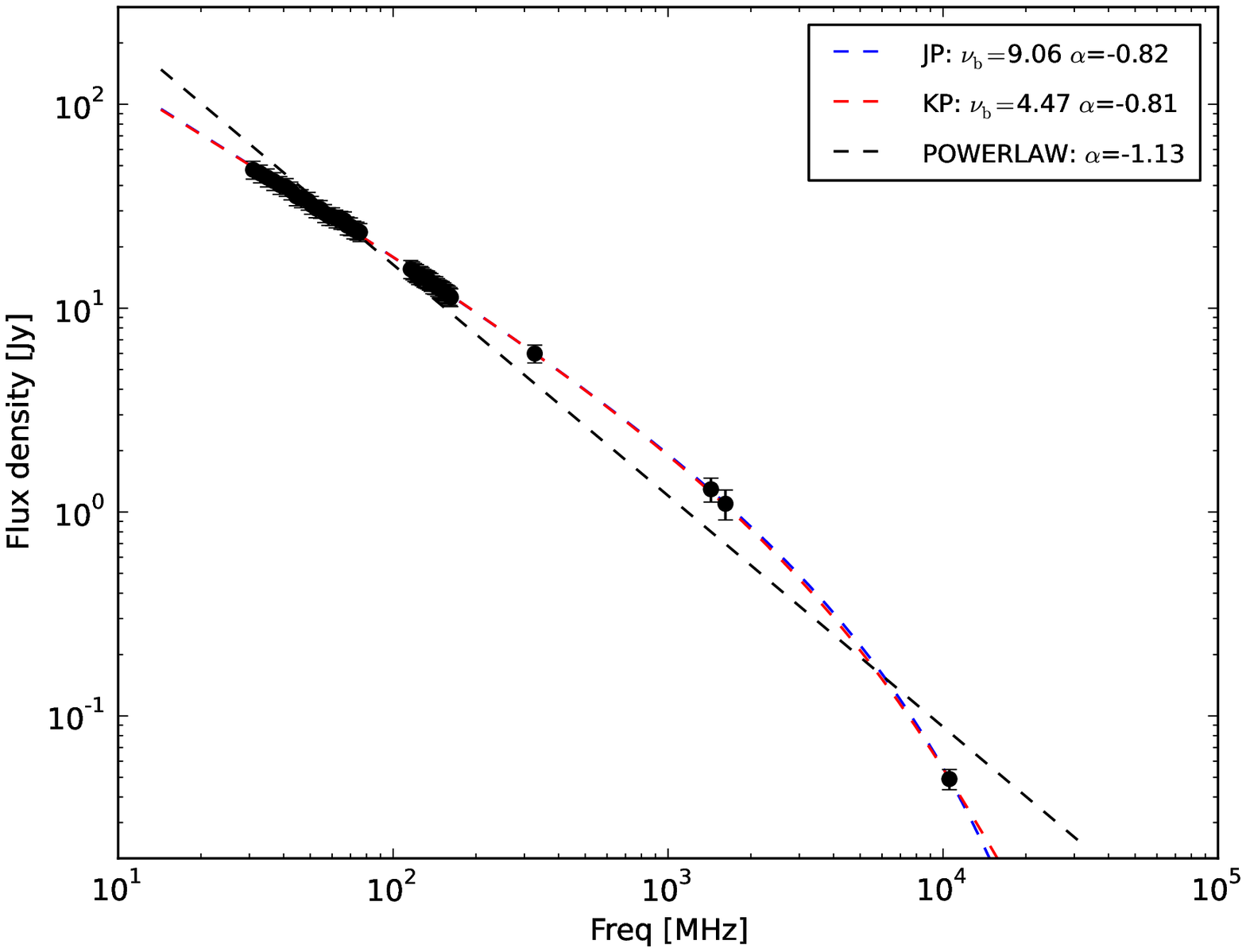}
    \label{fig:spectrum-W2}}
  \subfloat[Zone W3]{
    \includegraphics[width=.33\textwidth]{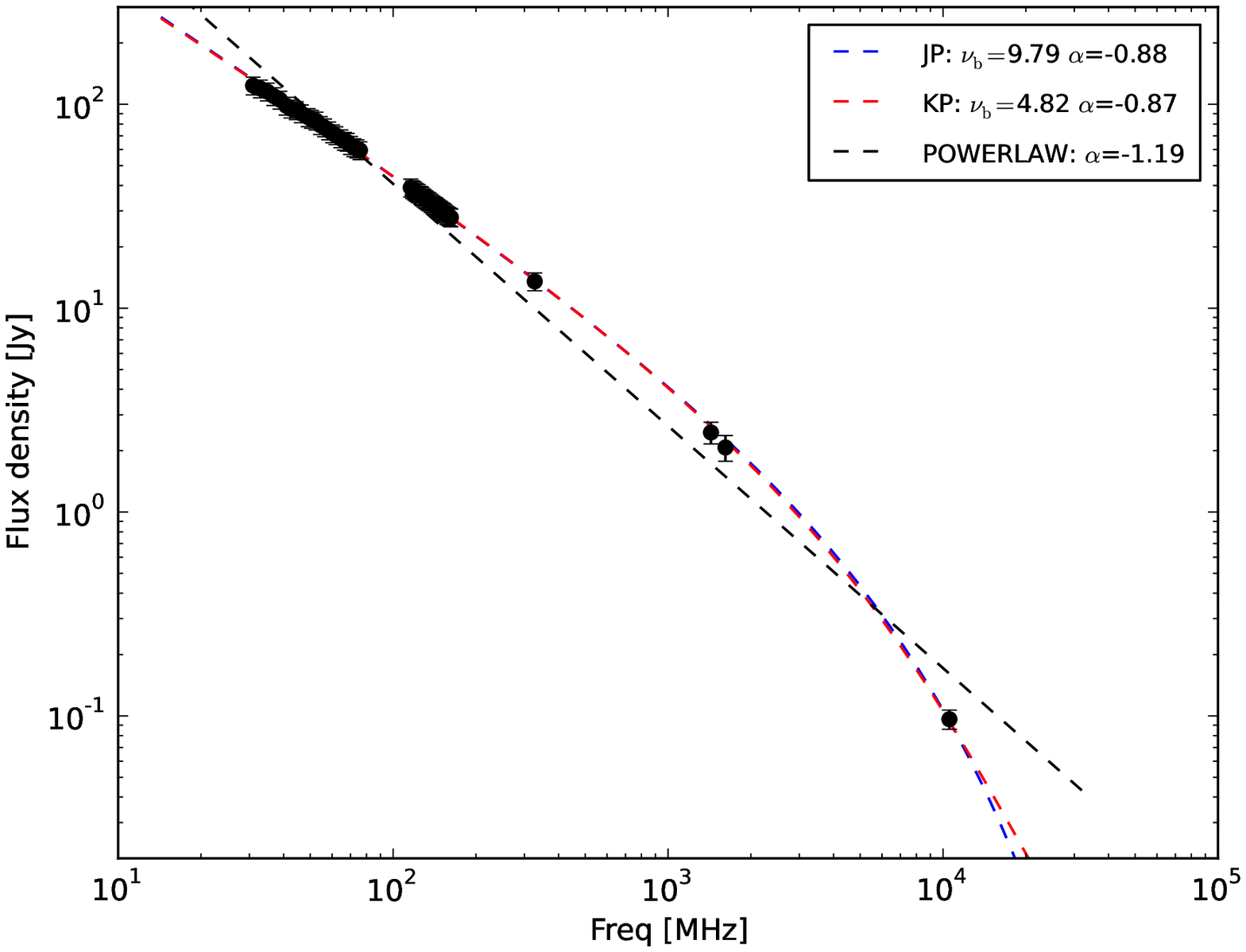}
    \label{fig:spectrum-W3}}\\  
  \subfloat[Zone W4]{
    \includegraphics[width=.33\textwidth]{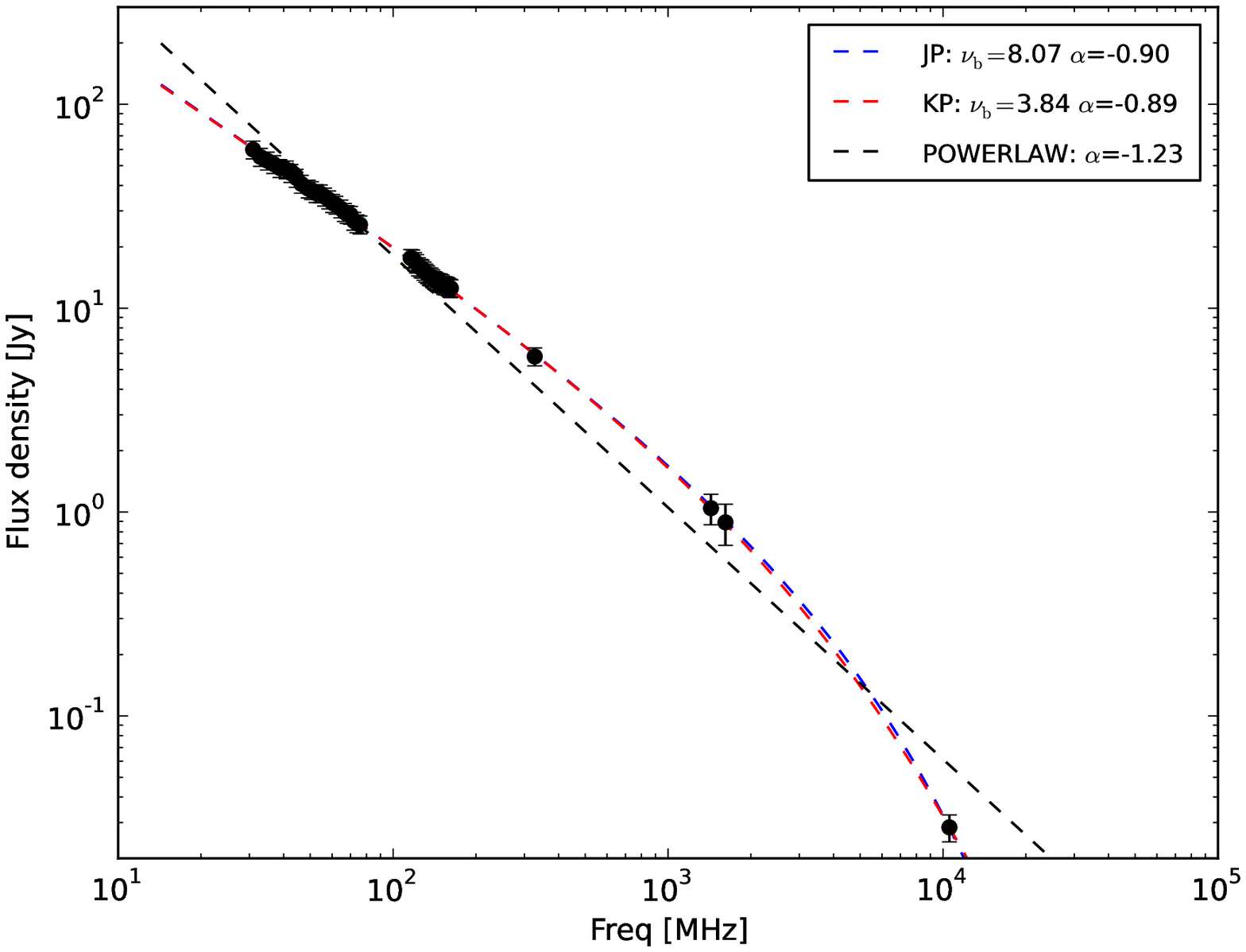}
    \label{fig:spectrum-W4}}
  \subfloat[Zone E1]{
    \includegraphics[width=.33\textwidth]{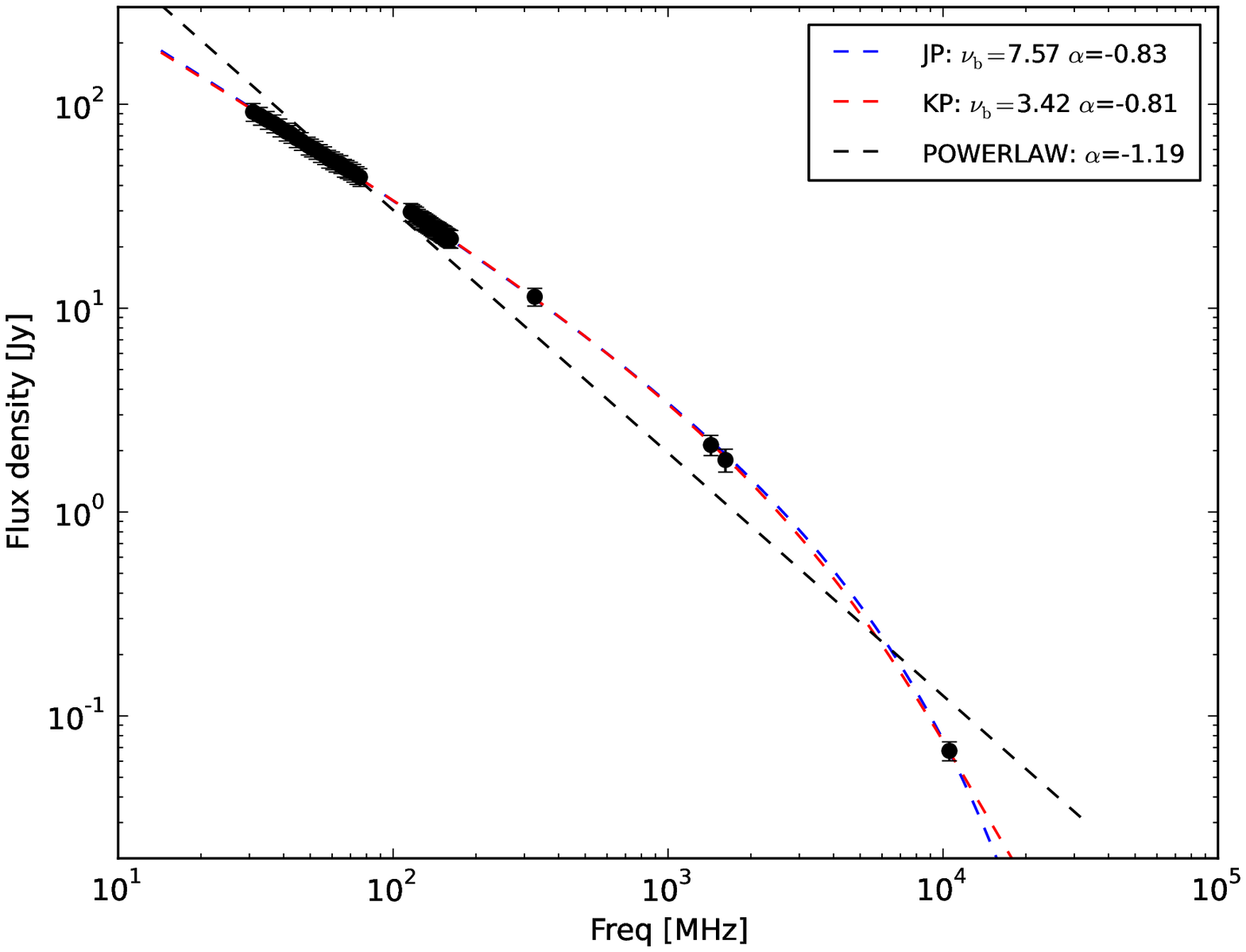}
    \label{fig:spectrum-E1}}
  \subfloat[Zone E2]{
    \includegraphics[width=.33\textwidth]{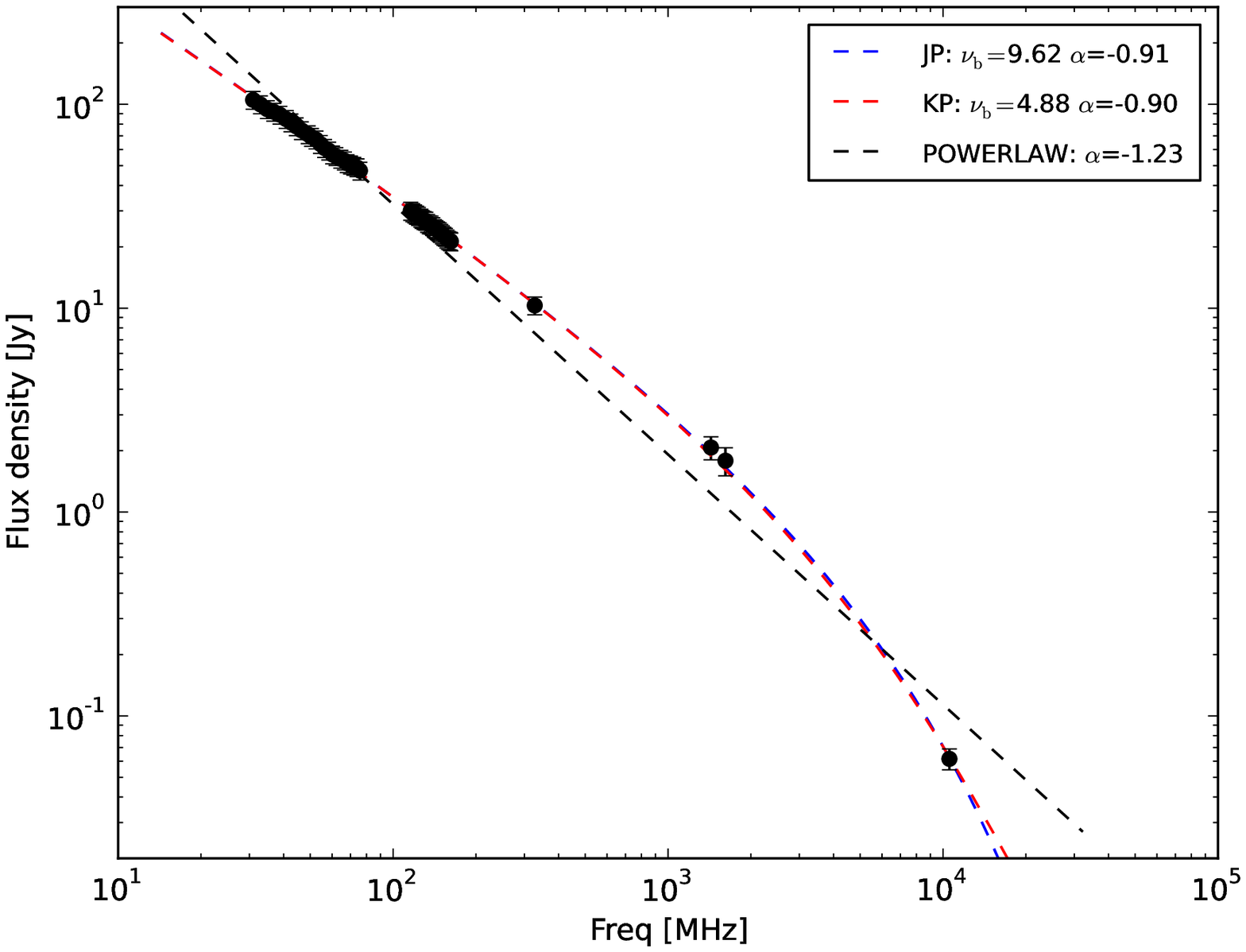}
    \label{fig:spectrum-E2}}\\
  \subfloat[Zone H1]{
    \includegraphics[width=.33\textwidth]{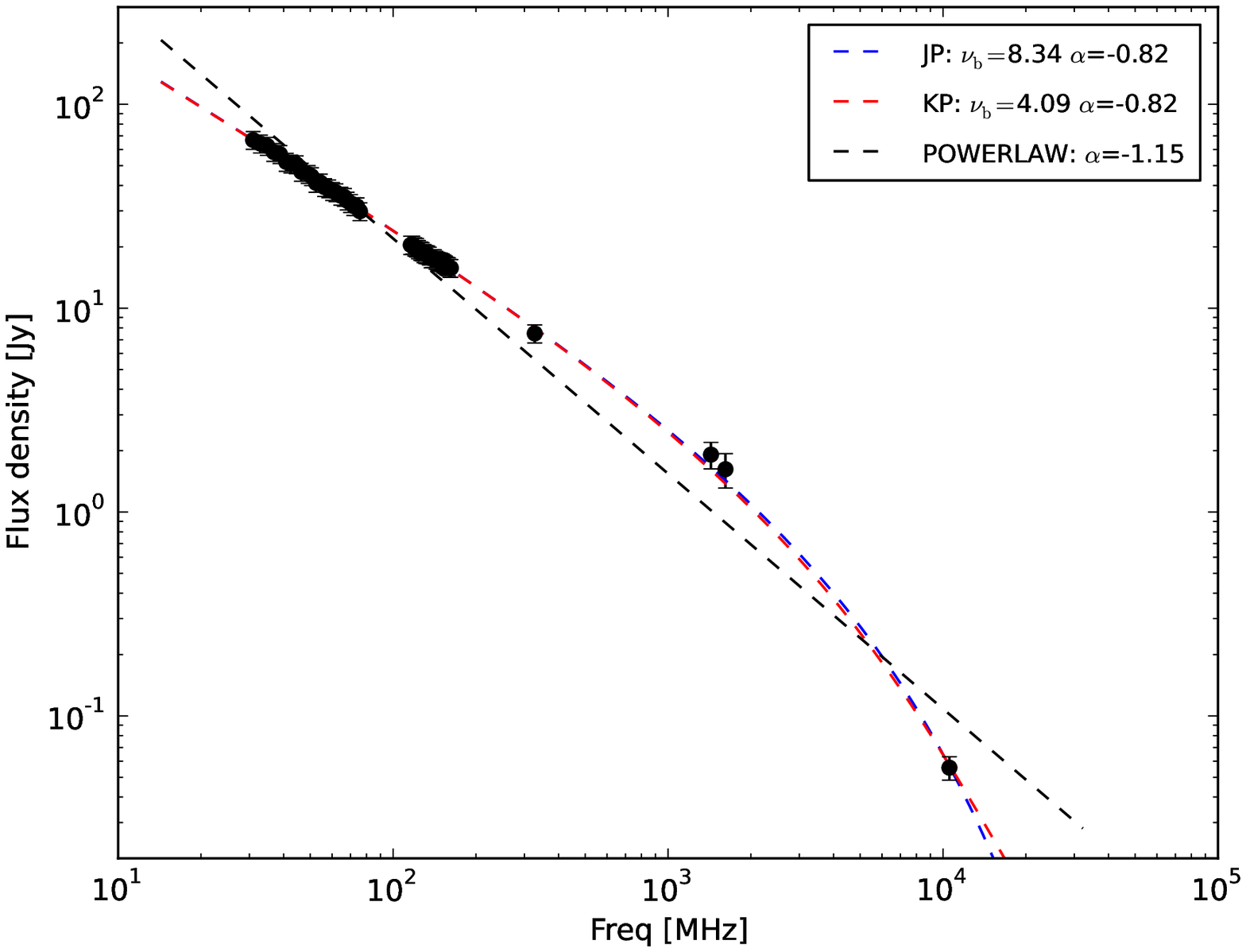}
    \label{fig:spectrum-H1}}
  \subfloat[Zone H2]{
    \includegraphics[width=.33\textwidth]{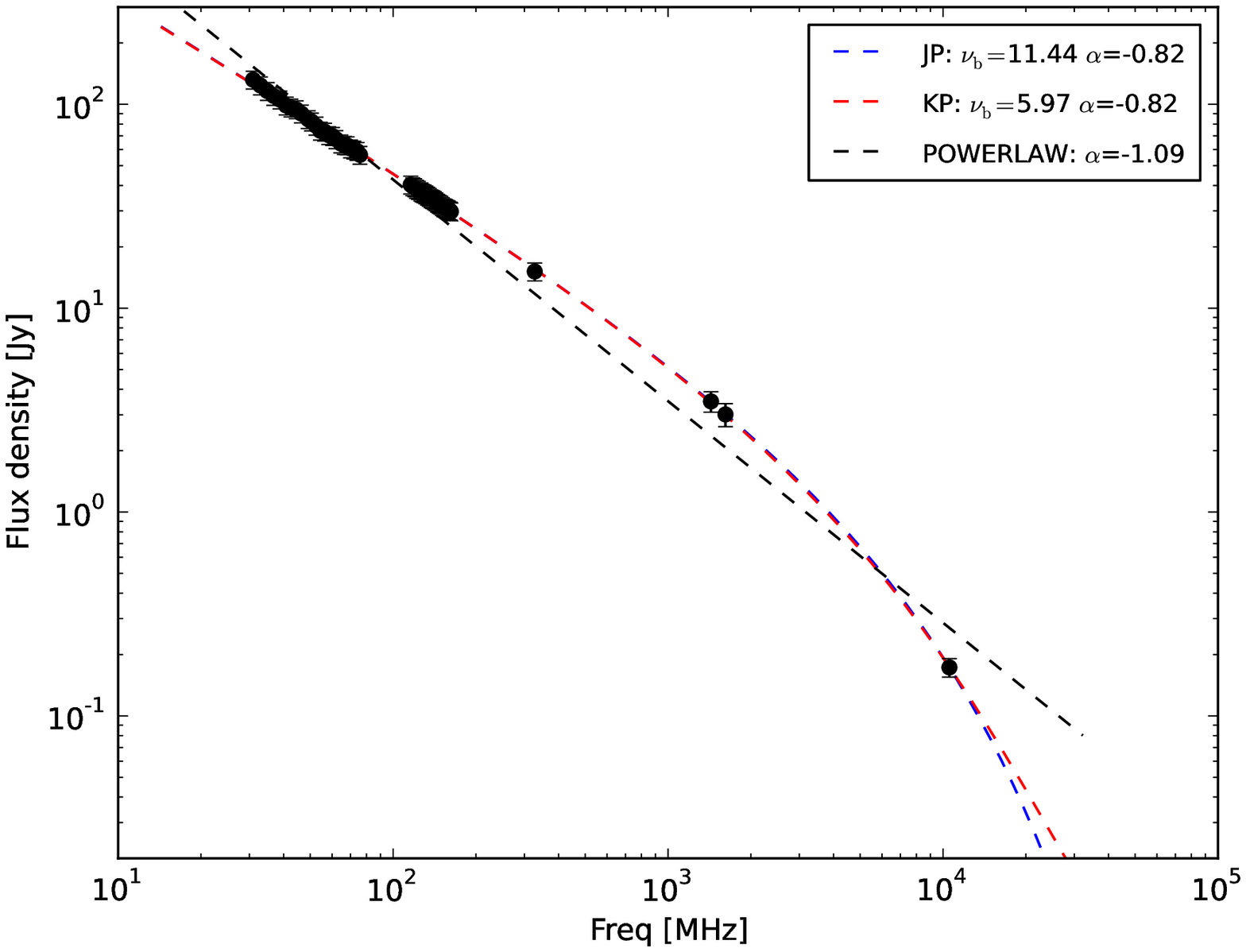}
    \label{fig:spectrum-H2}}
  \subfloat[Zone H3]{
    \includegraphics[width=.33\textwidth]{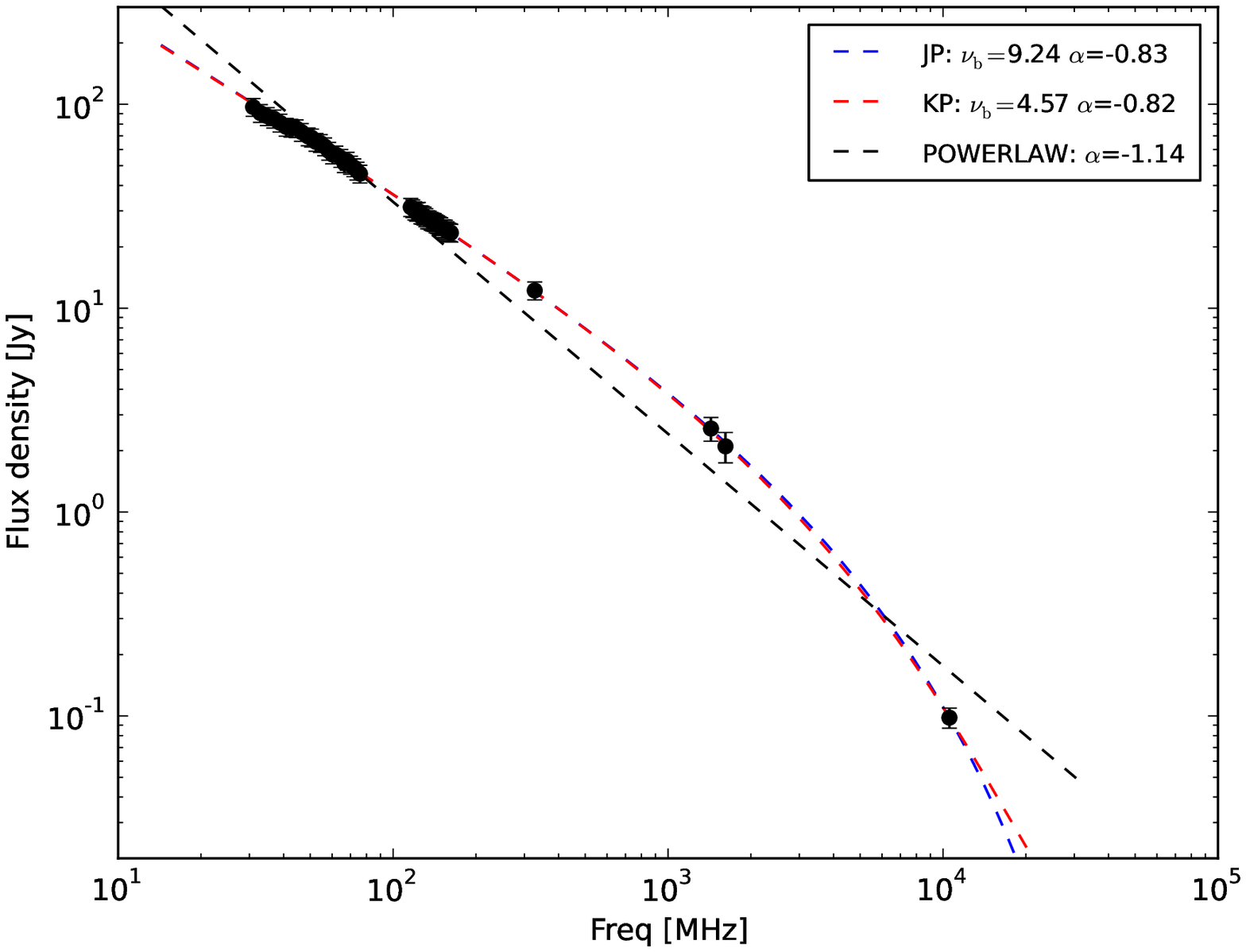}
    \label{fig:spectrum-H3}}\\
\caption{Fit of the JP (blue), KP (red) models to the zones related to the halo. The zones are defined in Fig.~\ref{fig:zones}. The black line is a simple linear regression fit to emphasize the curvature in the spectrum. $\nu_{\rm b}$ is in GHz.}
\label{fig:spectra}
\end{figure*}

\begin{figure*}
\centering
  \subfloat[Zone W1 (fixed $\alpha_{\rm inj}$)]{
    \includegraphics[width=.33\textwidth]{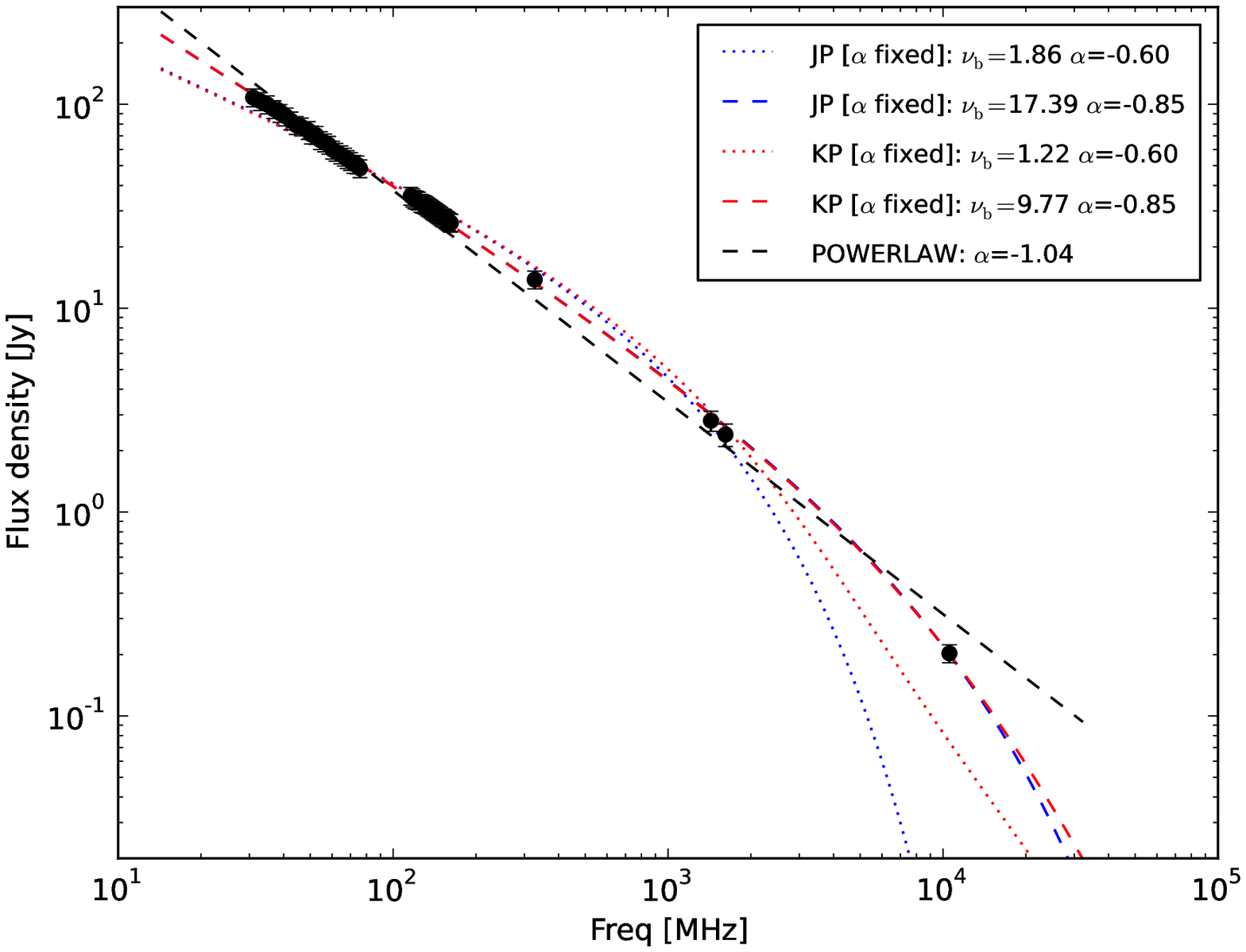}
    \label{fig:spectrum-alphafix-W1}}
  \subfloat[Zone W2 (fixed $\alpha_{\rm inj}$)]{
    \includegraphics[width=.33\textwidth]{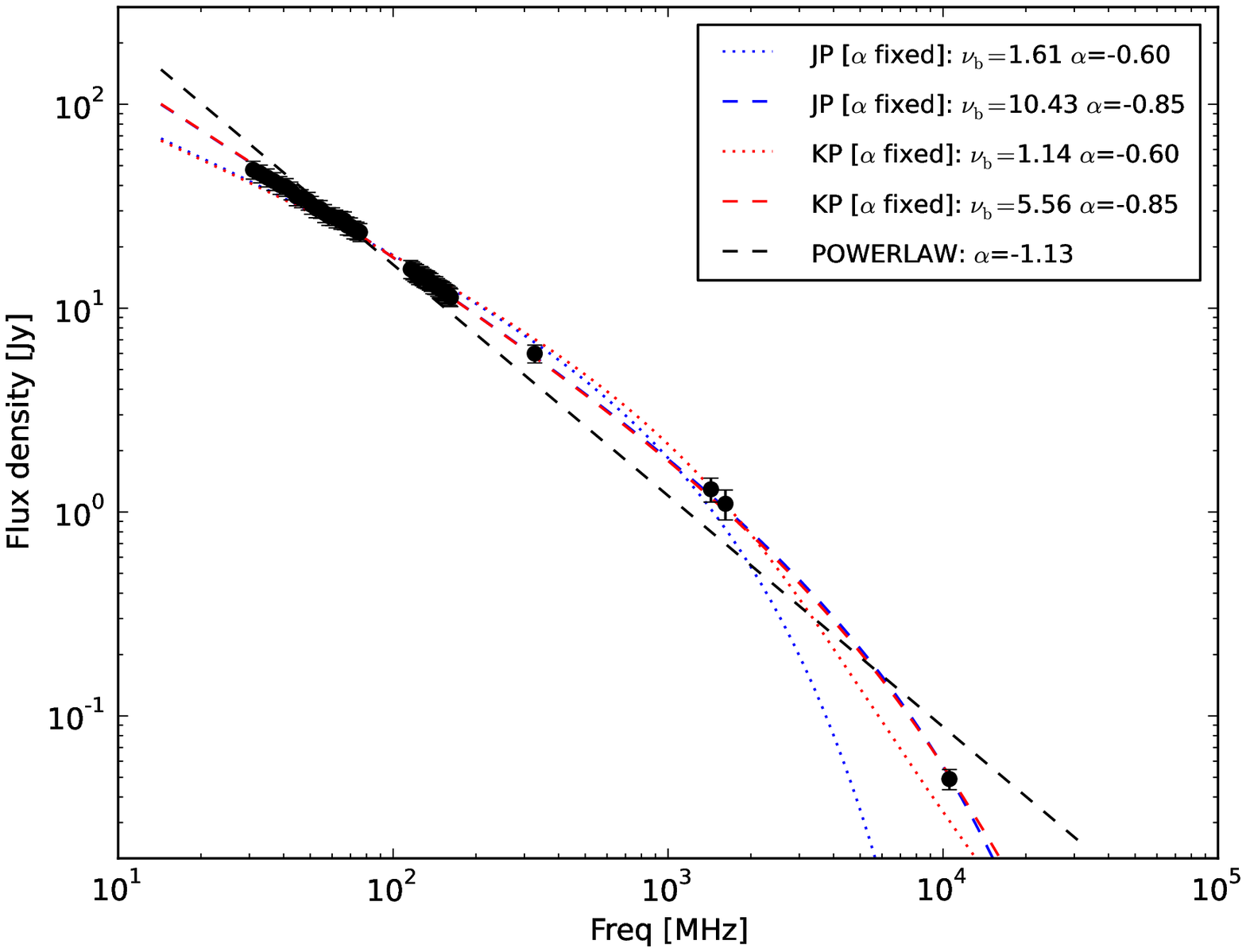}
    \label{fig:spectrum-alphafix-W2}}
  \subfloat[Zone W3 (fixed $\alpha_{\rm inj}$)]{
    \includegraphics[width=.33\textwidth]{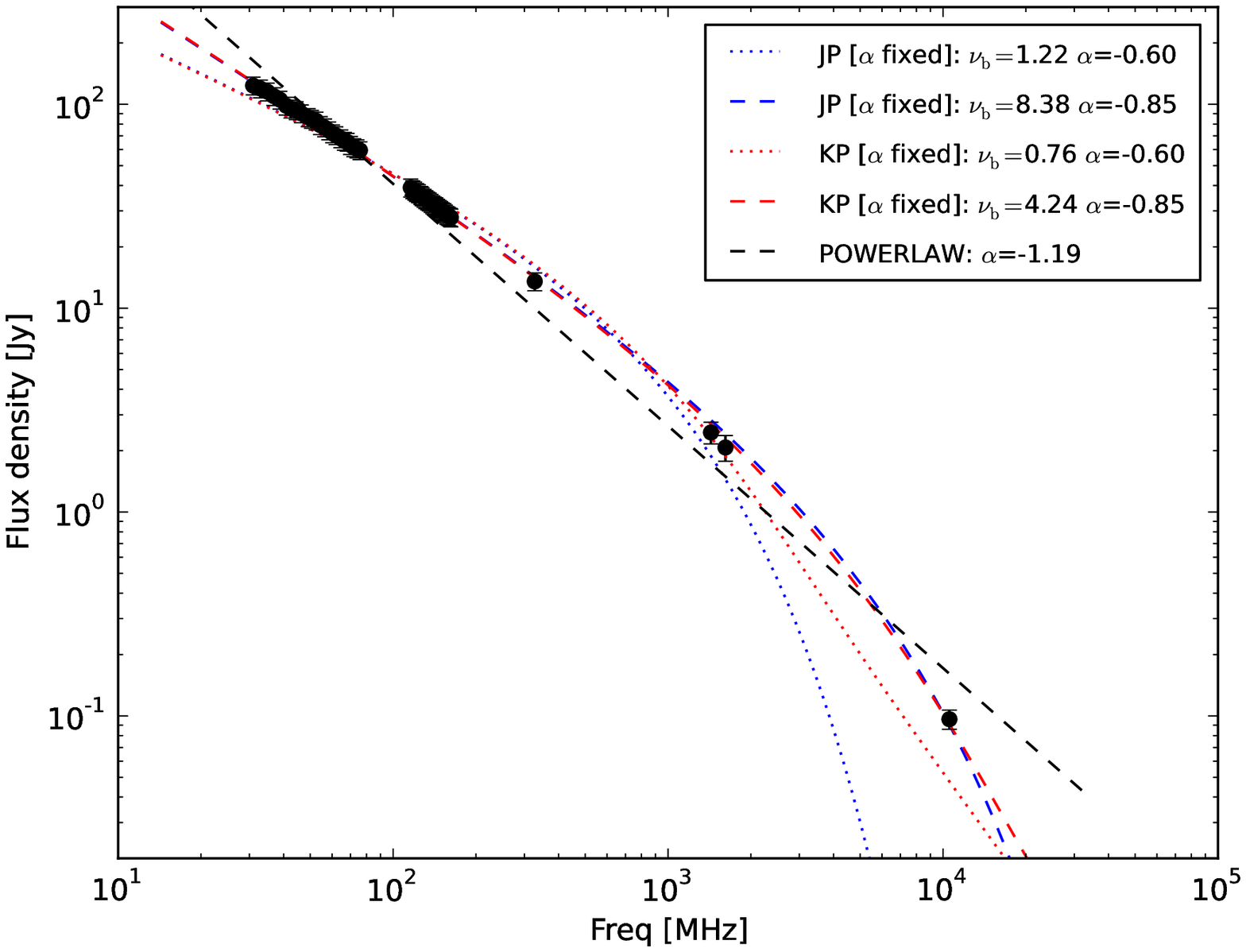}
    \label{fig:spectrum-alphafix-W3}}\\  
  \subfloat[Zone W4 (fixed $\alpha_{\rm inj}$)]{
    \includegraphics[width=.33\textwidth]{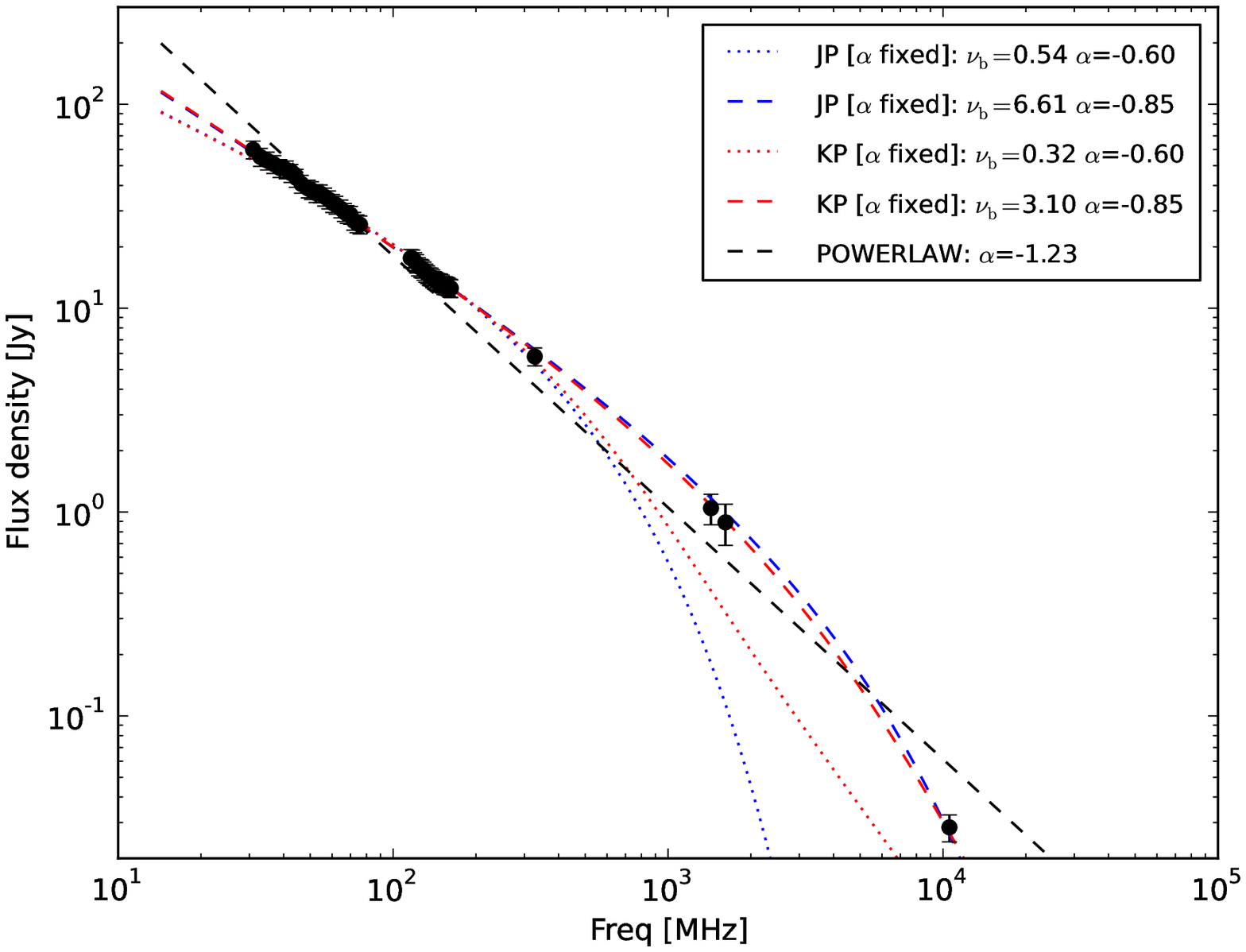}
    \label{fig:spectrum-alphafix-W4}}
  \subfloat[Zone E1 (fixed $\alpha_{\rm inj}$)]{
    \includegraphics[width=.33\textwidth]{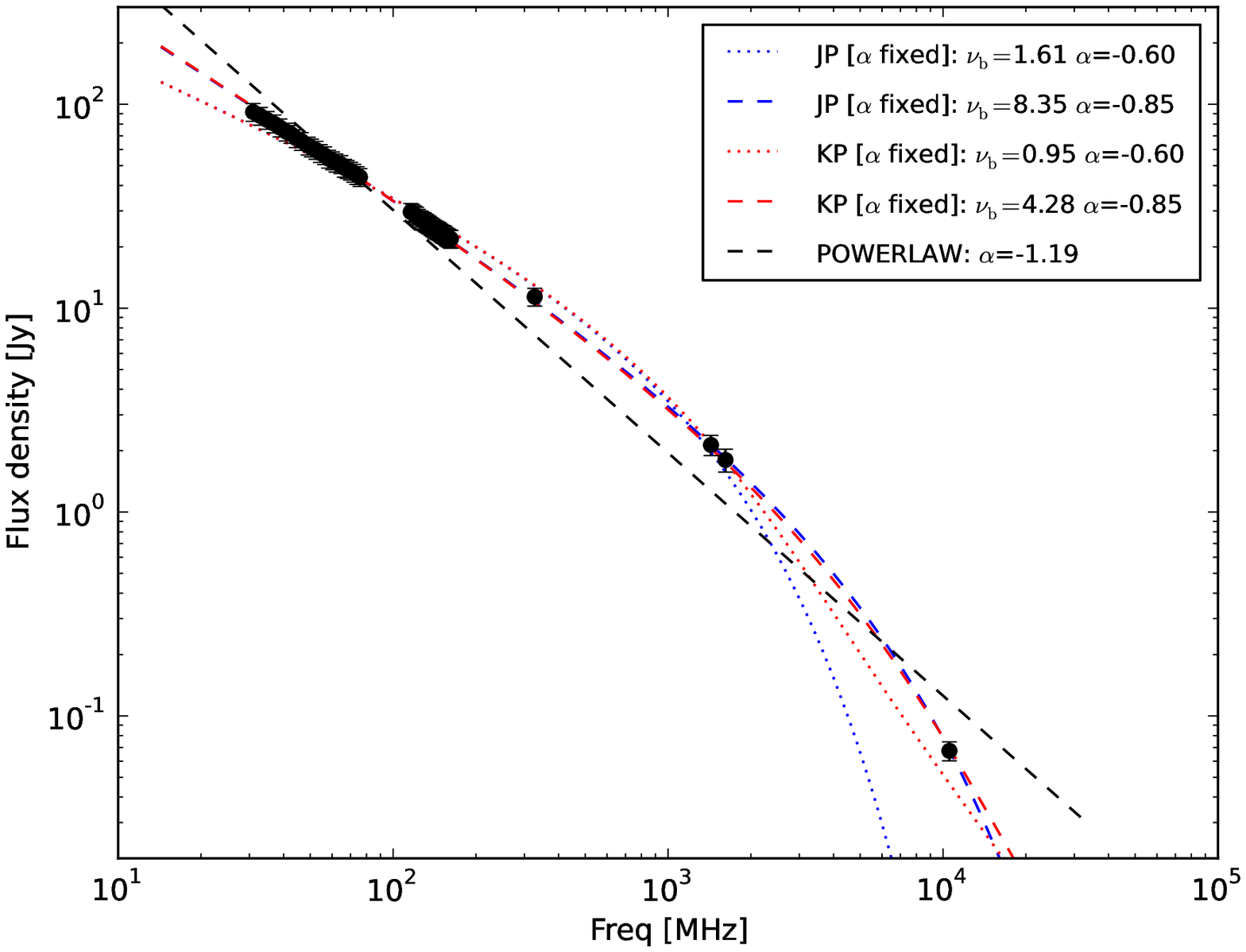}
    \label{fig:spectrum-alphafix-E1}}
  \subfloat[Zone E2 (fixed $\alpha_{\rm inj}$)]{
    \includegraphics[width=.33\textwidth]{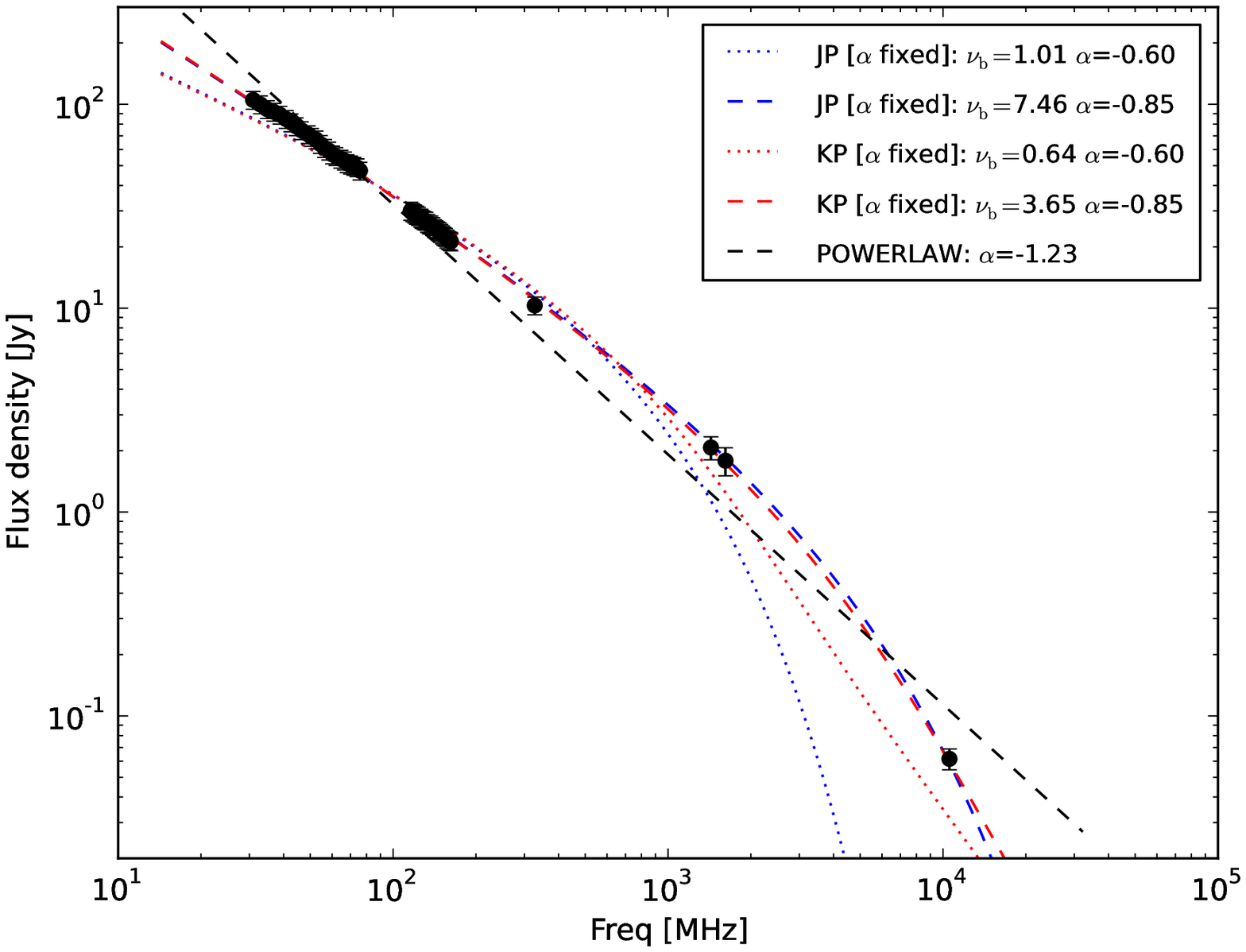}
    \label{fig:spectrum-alphafix-E2}}\\
  \subfloat[Zone H1 (fixed $\alpha_{\rm inj}$)]{
    \includegraphics[width=.33\textwidth]{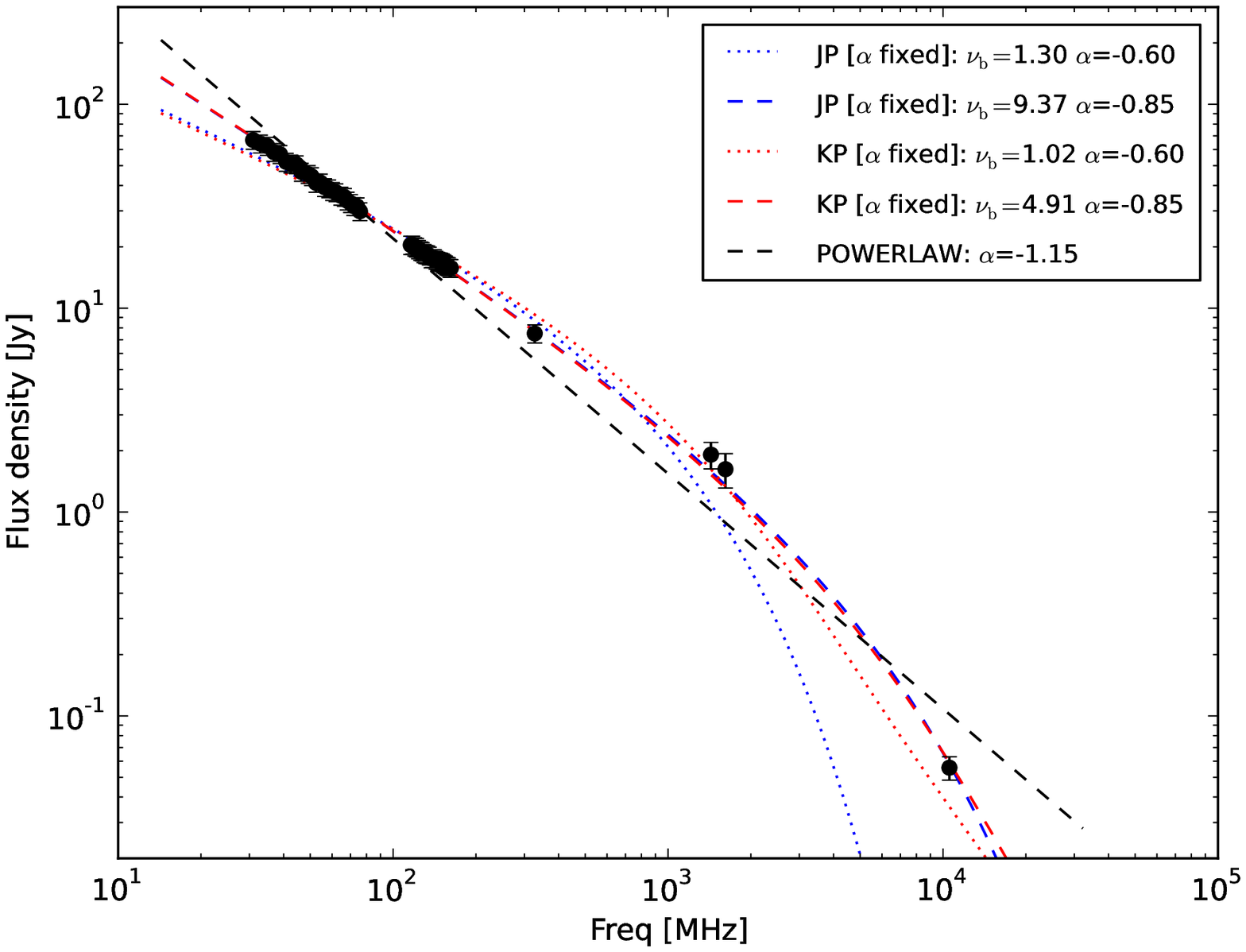}
    \label{fig:spectrum-alphafix-H1}}
  \subfloat[Zone H2 (fixed $\alpha_{\rm inj}$)]{
    \includegraphics[width=.33\textwidth]{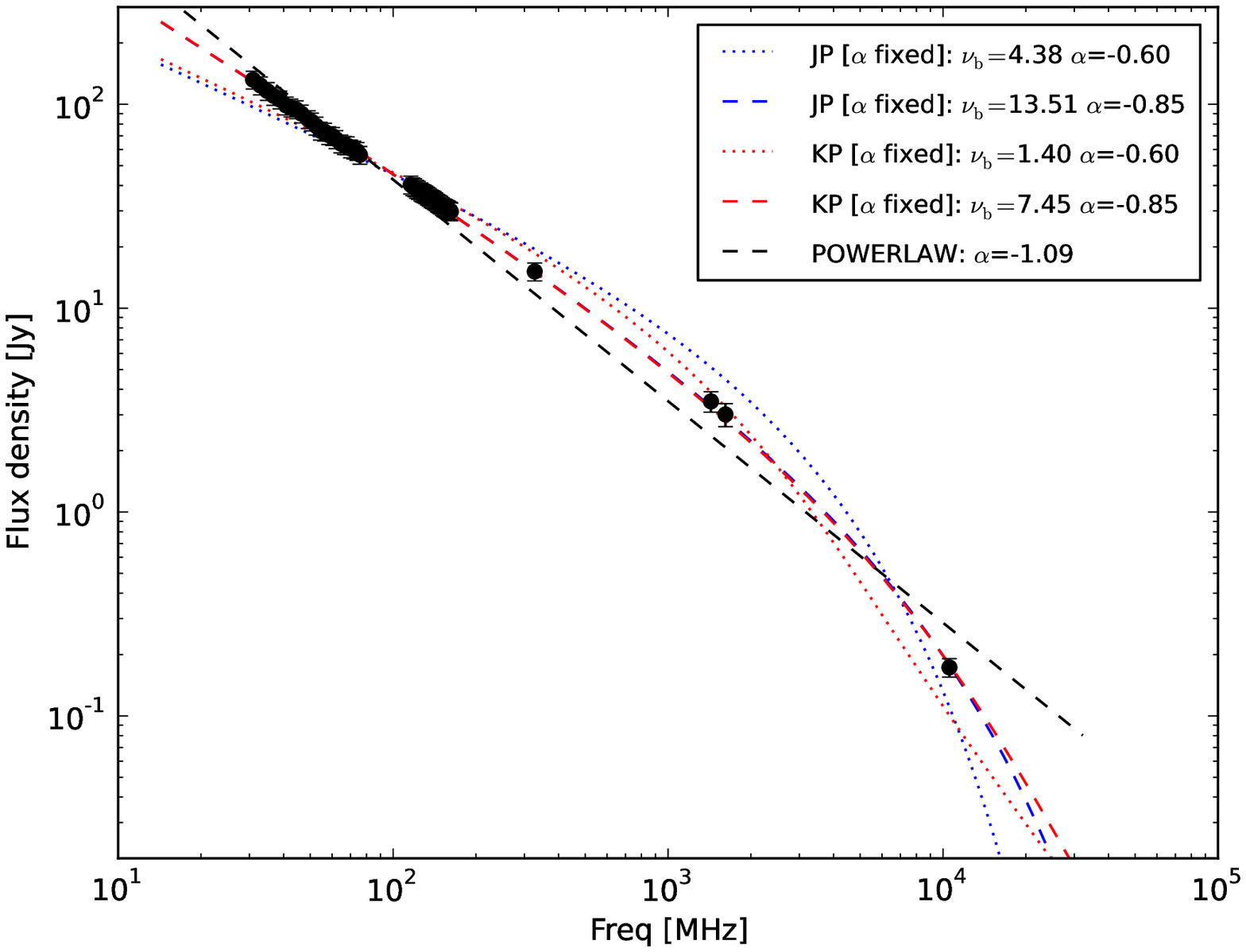}
    \label{fig:spectrum-alphafix-H2}}
  \subfloat[Zone H3 (fixed $\alpha_{\rm inj}$)]{
    \includegraphics[width=.33\textwidth]{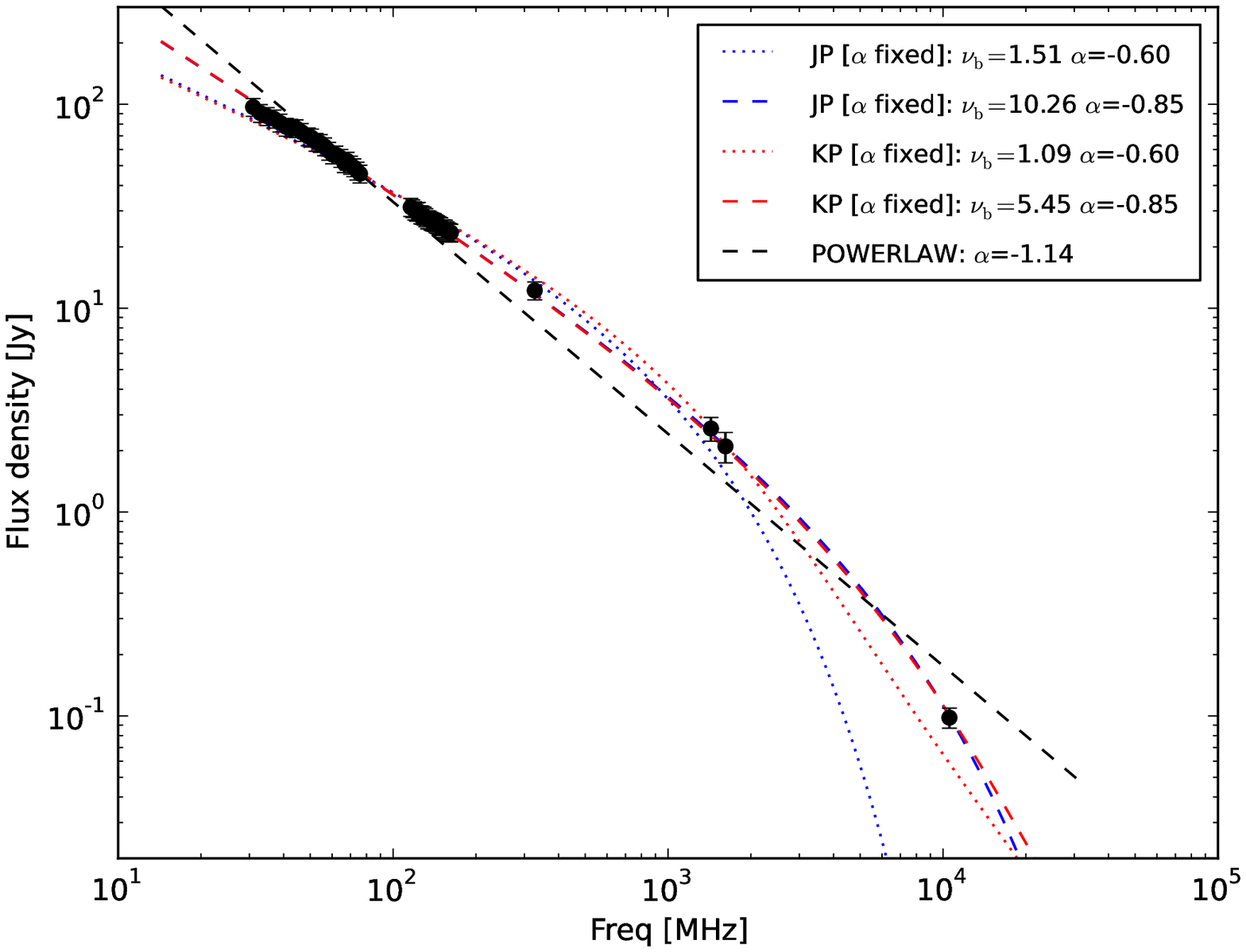}
    \label{fig:spectrum-alphafix-H3}}\\
\caption{Fit of the JP (dotted cyan and magenta lines) and KP (dashed blue and red lines) models for the zones related to the halo. The zones are defined in Fig.~\ref{fig:zones}. The slope of the injected electron population was fixed to $\alpha_{\rm inj} = -0.6$ (cyan and blue lines) or $\alpha_{\rm inj} = -0.85$ (magenta and red lines). The black line is a simple linear regression fit to emphasize the curvature in the spectrum. $\nu_{\rm b}$ is in GHz.}
\label{fig:spectra-alphafix}
\end{figure*}

To check the validity of these hypotheses through an in-depth spectral analysis, we selected ten relevant zones (defined in Fig.~\ref{fig:zones}) considered to be representative of the different parts of the source: central cocoon, flows/filaments (zones W1, W2, W3 and W4 for the west flow; zones E1 and E2 for the east flow), and halo (zones H1, H2 and H3). We deliberately avoided those zones where the signal to noise ratio is low and data could be dominated by calibration or deconvolution errors. For each zone we extracted the averaged flux at each frequency and performed a spectral analysis (see results in Table~\ref{tab:spfit}). Since our data are well approximated by a straight line in the LOFAR bands, with the curvature determined only by the three higher frequency values, both JP and KP models are able to fit the data comparably well (see Fig.~\ref{fig:spectra}) when leaving free to vary all the parameters. To reduce the degrees of freedom, we fit the data by fixing $\alpha_{\rm inj} = -0.6$ and $\alpha_{\rm inj} = -0.85$ (see Fig.~\ref{fig:spectra-alphafix}).

A fixed $\alpha_{\rm inj} = -0.6$ fails to fit our data, and when all of the parameters are left free to vary, we find $\nu_{\rm b} \gtrsim 5\un GHz$. These results are in contradiction with the CIOFF scenario (model 4 in Sect.~\ref{sec:coreandmacroregions}), where we would have expected a wide range of $\nu_{\rm b}$ and $\alpha_{\rm inj} = -0.6$. On the other hand, we can fit both these and the global spectra if we assume an $\alpha_{\rm inj}\simeq-0.85$. This steeper slope can be due to the intrinsically different and rather steep injection spectrum of a previous outburst. However, in that case, the presence of uninterrupted flows from the cocoon to the outer halo would be harder to explain.

Alternatively, we argue that a global steepening of the spectrum may occur at a very early stage, when the bubble detaches from the central cocoon. This can be related to an adiabatic expansion of the bubble, which happens as soon as it leaves the high-pressure central region \citep{Churazov2001, Carilli91}. In this case, a mix of plasmas at different ages would quickly expand and the high frequency parts of their spectra, which is indeed curved and steeper than $-0.6$ \citep{cotton09}, would be shifted towards lower frequencies. The sum of these spectra produces a resulting spectrum that has a low-frequency end steeper than the initial ones \citep{Murgia1999}. Furthermore, an abrupt lowering in the spectra normalization is also expected, this can account for the strong brightness contrast between the cocoon and the halo. After that no further strong expansions occurred, otherwise a gradient in the surface brightness across the halo would be visible.

Regardless of the mechanism responsible for the observed steepening, in what follows we assume $\alpha_{\rm inj} = -0.85$ at the point where the plasma bubbles leave the cocoon. This provides a $t_{\rm halo}\simeq 40\un Myr$ (from equation~\ref{eq:time}, based on the break frequency in the CI model fit to the entire halo and assuming an average magnetic field strength of $10\un \mu G$).

\begin{table*}
\centering
\caption{Spectral fits to representative regions}
\label{tab:spfit}
\begin{tabular}{lcccccccccc}
\hline\hline
\multicolumn{3}{c}{Region} & \multicolumn{3}{c}{\sout{\hfill}\ JP Model\ \sout{\hfill}} & \multicolumn{3}{c}{\sout{\hfill}\ KP Model\ \sout{\hfill}} &  \multicolumn{2}{c}{\sout{\hfill}\ Power-law\ \sout{\hfill}} \bigstrut[t]\\
& &    & $\chi^2_{\rm red}$ & $\nu_{\rm b}$ [GHz] & $-\alpha_{\rm inj}$ \
       & $\chi^2_{\rm red}$ & $\nu_{\rm b}$ [GHz] & $-\alpha_{\rm inj}$ \
       & $\chi^2_{\rm red}$ & $-\alpha_{\rm inj}$ \bigstrut[b]\\
& \multirow{4}{*}{$\vast\{$} & W1 & $0.090$ &  $15.6^{+0.7}_{-4.3}$ &  $0.83^{+0.02}_{-0.04}$ &  $0.083$ &  $8.3^{+0.3}_{-2.6}$ &  $0.83^{+0.02}_{-0.04}$ &  $1.551$ &  $1.04^{+0.02}_{-0.02}$  \bigstrut[t]\\
West & & W2 & $0.028$ &  $9.1^{+1.2}_{-1.9}$ &  $0.82^{+0.02}_{-0.04}$ &  $0.029$ &  $4.5^{+0.8}_{-1.2}$ &  $0.81^{+0.02}_{-0.05}$ &  $3.030$ &  $1.13^{+0.02}_{-0.02}$ \\
Flow & & W3 & $0.060$ &  $9.8^{+1.2}_{-2.1}$ &  $0.88^{+0.01}_{-0.04}$ &  $0.047$ &  $4.8^{+0.5}_{-1.3}$ &  $0.87^{+0.02}_{-0.05}$ &  $3.120$ &  $1.19^{+0.01}_{-0.02}$ \\ 
& & W4 & $0.034$ &  $8.1^{+1.3}_{-1.8}$ &  $0.90^{+0.02}_{-0.05}$ &  $0.033$ &  $3.8^{+1.1}_{-1.2}$ &  $0.89^{+0.03}_{-0.05}$ &  $3.259$ &  $1.23^{+0.01}_{-0.02}$ \\
East & \multirow{2}{*}{$\bigg\{$} & E1 & $0.022$ &  $7.6^{+0.2}_{-1.3}$ &  $0.83^{+0.02}_{-0.04}$ &  $0.010$ &  $3.4^{+0.2}_{-0.9}$ &  $0.81^{+0.02}_{-0.05}$ &  $4.370$ &  $1.19^{+0.01}_{-0.01}$ \\
Flow & & E2 & $0.034$ &  $9.6^{+0.4}_{-2.1}$ &  $0.91^{+0.03}_{-0.05}$ &  $0.042$ &  $4.9^{+0.9}_{-1.3}$ &  $0.90^{+0.01}_{-0.05}$ &  $2.943$ &  $1.23^{+0.02}_{-0.02}$ \\
& \multirow{3}{*}{$\Bigg\{$} & H1 &  $0.056$ &  $8.3^{+1.2}_{-1.7}$ &  $0.82^{+0.04}_{-0.05}$ &  $0.068$ &  $4.1^{+0.2}_{-1.2}$ &  $0.82^{+0.02}_{-0.05}$ &  $3.088$ &  $1.15^{+0.01}_{-0.02}$ \\
Halo & & H2 & $0.034$ &  $11.4^{+0.7}_{-2.6}$ &  $0.82^{+0.01}_{-0.04}$ &  $0.035$ &  $6.0^{+0.4}_{-1.7}$ & $0.82^{+0.03}_{-0.05}$ & $2.229$ &  $1.09^{+0.01}_{-0.02}$ \\
& & H3 & $0.045$ &  $9.2^{+0.9}_{-1.9}$ &  $0.83^{+0.01}_{-0.04}$ &  $0.044$ &  $4.6^{+0.2}_{-1.3}$ &  $0.82^{+0.02}_{-0.05}$ &  $2.997$ & $1.14^{+0.02}_{-0.02}$ \bigstrut[b]\\

\hline
\end{tabular}
\tablefoot{The assumed 10\% error in the flux densities overestimates the real random, normally distributed error providing artificially low $\chi^2_{\rm red}$ values.}
\end{table*} 

\subsection{Magnetic fields and synchrotron ageing}
\label{sec:syncrotron}

Equipartition magnetic field strengths were computed in the same regions shown in Fig.~\ref{fig:zones}. We made the standard assumption that each zone of the radio source contains relativistic particles and uniformly distributed magnetic fields under energy equipartition conditions \citep[e.g.][]{Miley1980}. The radio luminosities for the computation were extracted from the lowest frequency map (31~MHz). To perform this analysis we made the following assumptions:
\begin{itemize}
 \item The particle energy $U$ is equally divided between electrons and protons, setting $U_{\rm pr}/U_{\rm el} \equiv k = 1$. However, in Table~\ref{tab:equip} we also list the results for the assumption of $k = 0$, where all of the particle energy is in the relativistic electrons (and positrons), in line with the idea that the jets could be mainly composed of these constituents \citep{Reynolds96}. We show also the results for the assumption of $k = 10$, as un upper limit from what predicted by electromagnetic acceleration models involving protons and electrons \citep{Bell1978}.
 \item The equipartition magnetic field is usually computed assuming that the relativistic particle energies are confined between a minimum $\epsilon_{\rm min}$ and a maximum $\epsilon_{\rm max}$ value, corresponding to an observable frequency range, typically assumed to be 10~MHz -- 100~GHz \citep{Pacholczyk70}. However, a fixed frequency range corresponds to an energy range that depends on the magnetic field, which varies in different parts of the source. Therefore, we decided to put limits directly on the electron population energies \citep{Brunetti97, Beck05}. This approach, compared to the standard one, provides slightly higher $B$ values \citep[see Appendix A,][]{Brunetti97}. \cite{Reynolds96} and \cite{Dunn2006} put constraints on the maximum value of $\gamma_{\rm min}$ noting that the synchrotron self-Compton flux density generated in the source core, which depends on $\gamma_{\rm min}$, cannot exceed the observed X-ray flux density. They obtained $1 \lesssim \gamma_{\rm min} \lesssim 100$ for an electron-positron jet and $50 \lesssim \gamma_{\rm min} \lesssim 100$ for an electron-proton jet. \cite{Falcke95} also argue for radio loud AGN a $\gamma_{\rm min} \simeq 100$. We repeated our analysis for two values of $\gamma_{\rm min}$: 10 (with $k = 0$) and 100 (with $k = 1$ and with $k = 10$), corresponding to $\epsilon_{\rm min}$ of 5 and 50~MeV respectively. Above $\gamma_{\rm min} \simeq 1000$ we would expect a turnover in the low frequency part of our spectra that we do not detect. The $\epsilon_{\rm max}$ value does not affect the results, and we used an arbitrarily high value of 5~GeV ($\gamma_{\rm max} \simeq 10000$).
 \item For each zone we assumed a cylindrical configuration and we repeated the computation for two depth $D=20$~kpc and $D=40$~kpc. In the rest of the paper the flow zones are assumed to have a depth of 20~kpc and the halo zones are assumed to have a depth of 40~kpc. The depth of the core is assumed to be 5~kpc. 
 \item The low frequency spectrum slopes have been assumed to be equal to $-0.85$, as observed. 
\end{itemize}

The equipartition magnetic fields, the minimum pressures and the corresponding synchrotron ageing times are listed in Table~\ref{tab:equip}. In the first part, we list the equipartition analysis results assuming that all of the energy resides in the electrons and positrons only ($k = 0$) and that $\gamma_{\rm min} = 10$. In the middle and third parts of the table, we assume a $\gamma_{\rm min}$ values of 100 and we relax the electrons to protons energy ratio to $k = 1$ and $k=10$ respectively. In the rest of the paper we will refer to the values in the second part ($k = 1$ and $\gamma_{\rm min} = 100$) only, while we notice that the other two combinations of parameters provide ages so small that the plasma bubbles should move at a velocity higher than the sound speed ($\simeq 900\un km\ s^{-1}$ in the outskirts).

The source age is obtained by \citep[see e.g.][]{Murgia11}:
\begin{equation}
t_{\rm s} = 1590\,\frac{B^{0.5}}{ \left( B^2+B_{\rm IC}^2 \right) \left[\left( 1+z \right) \nu_{\rm b} \right]^{0.5}},
\label{eq:time}
\end{equation} 
where the synchrotron age $t_{\rm s}$ is in Myr, the magnetic field strength in $\mu$G and the break frequency $\nu_{\rm b}$ in GHz, while $B_{\rm IC} = 3.25(1+z)^2\un \mu G$ is the inverse Compton equivalent magnetic field strength with energy density equal to that of the CMB \citep{Slee01}. The break frequencies were obtained from the fit of the JP model (see Table~\ref{tab:spfit}). We assumed a constant and uniform magnetic field strength and neglected any influence on the spectra from e.g. expansion or local re-energization of electrons.

\begin{table*}
\centering
\caption{Equipartition analysis}
\label{tab:equip}
\begin{tabular}{lccccccccccc}
\hline\hline
 & & \multicolumn{3}{c}{\sout{\hfill}\ $\gamma_{\rm min} = 10$, $k = 0$\ \sout{\hfill}} & \multicolumn{3}{c}{\sout{\hfill}\ $\gamma_{\rm min} = 100$, $k = 1$\ \sout{\hfill}} & \multicolumn{3}{c}{\sout{\hfill}\ $\gamma_{\rm min} = 100$, $k = 10$\ \sout{\hfill}} & \bigstrut[t]\\
Reg.   & D 
       & $B_{\rm eq}$ & $p_{\rm min}$ & t \
       & $B_{\rm eq}$ & $p_{\rm min}$ & t \
       & $B_{\rm eq}$ & $p_{\rm min}$ & t
       & $p_{\rm th}$\\
       & [kpc]
       & $[\rm \mu G]$ & $\left[10^{-12}\rm \frac{dyn}{cm^2}\right]$ & [Myr] \
       & $[\rm \mu G]$ & $\left[10^{-12}\rm \frac{dyn}{cm^2}\right]$ & [Myr] \
       & $[\rm \mu G]$ & $\left[10^{-12}\rm \frac{dyn}{cm^2}\right]$ & [Myr] \
       & $\left[10^{-12}\rm \frac{dyn}{cm^2}\right]$ \bigstrut[b]\\
\hline

C & 5 & 55.1 & 83.9 & -- & 36.0 & 35.7 & -- & 56.0 & 86.5 & -- & 640  \bigstrut[t]\\
W1 & 20 & 21.9 & 13.2 & 3.9 $^{+ 0.7 } _{ -0.1 }$ & 14.3 & 5.6 & 7.1 $^{+ 1.3 } _{ -0.2 }$ & 22.2 & 13.6 & 3.8 $^{+ 0.7 } _{ -0.1 }$ & 104 \\
W1 & 40 & 18.3 & 9.2 & 5.0 $^{+ 0.9 } _{ -0.1 }$ & 11.9 & 3.9 & 9.1 $^{+ 1.6 } _{ -0.2 }$ & 18.6 & 9.5 & 4.9 $^{+ 0.9 } _{ -0.1 }$ & '' \\
W2 & 20 & 20.0 & 11.0 & 5.8 $^{+ 0.7 } _{ -0.4 }$ & 13.0 & 4.7 & 10.6 $^{+ 1.3 } _{ -0.7 }$ & 20.3 & 11.4 & 5.6 $^{+ 0.7 } _{ -0.4 }$ & 53 \\
W2 & 40 & 16.7 & 7.7 & 7.5 $^{+ 0.9 } _{ -0.5 }$ & 10.9 & 3.3 & 13.5 $^{+ 1.7 } _{ -0.9 }$ & 17.0 & 7.9 & 7.3 $^{+ 0.9 } _{ -0.5 }$ & '' \\
W3 & 20 & 20.9 & 12.1 & 5.2 $^{+ 0.7 } _{ -0.3 }$ & 13.7 & 5.2 & 9.5 $^{+ 1.2 } _{ -0.5 }$ & 21.3 & 12.5 & 5.1 $^{+ 0.6 } _{ -0.3 }$ & 49 \\
W3 & 40 & 17.5 & 8.4 & 6.7 $^{+ 0.8 } _{ -0.4 }$ & 11.4 & 3.6 & 12.2 $^{+ 1.5 } _{ -0.7 }$ & 17.8 & 8.7 & 6.6 $^{+ 0.8 } _{ -0.4 }$ & '' \\
W4 & 20 & 19.0 & 10.0 & 6.6 $^{+ 0.9 } _{ -0.5 }$ & 12.4 & 4.2 & 12.0 $^{+ 1.6 } _{ -0.8 }$ & 19.3 & 10.3 & 6.4 $^{+ 0.9 } _{ -0.4 }$ & 58 \\
W4 & 40 & 15.9 & 7.0 & 8.5 $^{+ 1.1 } _{ -0.6 }$ & 10.4 & 3.0 & 15.3 $^{+ 2.1 } _{ -1.1 }$ & 16.1 & 7.2 & 8.3 $^{+ 1.1 } _{ -0.6 }$ & '' \\
E1 & 20 & 23.9 & 15.7 & 4.9 $^{+ 0.5 } _{ -0.1 }$ & 15.6 & 6.7 & 9.0 $^{+ 0.9 } _{ -0.1 }$ & 24.2 & 16.2 & 4.8 $^{+ 0.5 } _{ -0.1 }$ & 81 \\
E1 & 40 & 19.9 & 11.0 & 6.3 $^{+ 0.6 } _{ -0.1 }$ & 13.0 & 4.7 & 11.6 $^{+ 1.2 } _{ -0.2 }$ & 20.2 & 11.3 & 6.2 $^{+ 0.6 } _{ -0.1 }$ & '' \\
E2 & 20 & 20.2 & 11.2 & 5.5 $^{+ 0.7 } _{ -0.1 }$ & 13.2 & 4.8 & 10.1 $^{+ 1.3 } _{ -0.2 }$ & 20.5 & 11.6 & 5.4 $^{+ 0.7 } _{ -0.1 }$ & 68 \\
E2 & 40 & 16.9 & 7.8 & 7.1 $^{+ 0.9 } _{ -0.1 }$ & 11.0 & 3.3 & 12.9 $^{+ 1.6 } _{ -0.2 }$ & 17.1 & 8.1 & 7.0 $^{+ 0.9 } _{ -0.1 }$ & '' \\
H1 & 20 & 16.0 & 7.1 & 8.2 $^{+ 1.0 } _{ -0.5 }$ & 10.5 & 3.0 & 14.8 $^{+ 1.8 } _{ -1.0 }$ & 16.3 & 7.3 & 8.1 $^{+ 1.0 } _{ -0.5 }$ & 59 \\
H1 & 40 & 13.4 & 5.0 & 10.6 $^{+ 1.3 } _{ -0.7 }$ & 8.7 & 2.1 & 18.7 $^{+ 2.3 } _{ -1.2 }$ & 13.6 & 5.1 & 10.4 $^{+ 1.3 } _{ -0.7 }$ & '' \\
H2 & 20 & 20.0 & 11.1 & 5.1 $^{+ 0.7 } _{ -0.2 }$ & 13.1 & 4.7 & 9.4 $^{+ 1.3 } _{ -0.3 }$ & 20.3 & 11.4 & 5.0 $^{+ 0.7 } _{ -0.2 }$ & 65 \\
H2 & 40 & 16.7 & 7.7 & 6.6 $^{+ 0.9 } _{ -0.2 }$ & 10.9 & 3.3 & 12.0 $^{+ 1.6 } _{ -0.4 }$ & 17.0 & 8.0 & 6.5 $^{+ 0.9 } _{ -0.2 }$ & '' \\
H3 & 20 & 17.0 & 7.9 & 7.2 $^{+ 0.9 } _{ -0.3 }$ & 11.1 & 3.4 & 13.1 $^{+ 1.6 } _{ -0.6 }$ & 17.2 & 8.2 & 7.1 $^{+ 0.9 } _{ -0.3 }$ & 46 \\
H3 & 40 & 14.2 & 5.5 & 9.3 $^{+ 1.2 } _{ -0.4 }$ & 9.2 & 2.4 & 16.6 $^{+ 2.1 } _{ -0.7 }$ & 14.4 & 5.7 & 9.1 $^{+ 1.1 } _{ -0.4 }$ & '' \bigstrut[b]\\
\hline
\end{tabular}
\tablefoot{D is the depth of the region assuming a cylindrical configuration. $B_{\rm eq}$ and $p_{\rm min}$ are the magnetic field and the pressure from the equipartition analysis. $t$ is the estimated zone age. Errors on $t$ are derived from errors on $\nu_{\rm b}$ in Table~\ref{tab:spfit}.}
\end{table*}

The equipartition analysis provides us with reasonable values for the lifetimes of the bubbles. For example, following the various zones sampled in the west flow (W1, W2, W3, and W4), a bubble escapes from the source cocoon after $\simeq 7\un Myr$ \citep[][estimated $\sim 10\un Myr$ with simulations]{Churazov2001} and reaches the outer edge of the lobe after $\simeq 12-15\un Myr$ (zone W4). This confirms the picture of a flow continuously replenished with fresh particles. In the east flow the centre of the lobe (zone E2) is reached after $10-13$~Myr. This lifetime is about a fourth of what is derived by dynamical models. A difference between these age estimations and the global halo age ($\simeq 40$~Myr) is expected, as the latter takes into account many regions which have a lower break frequency than those in the flow zones.

\begin{figure}
\includegraphics[width=\columnwidth]{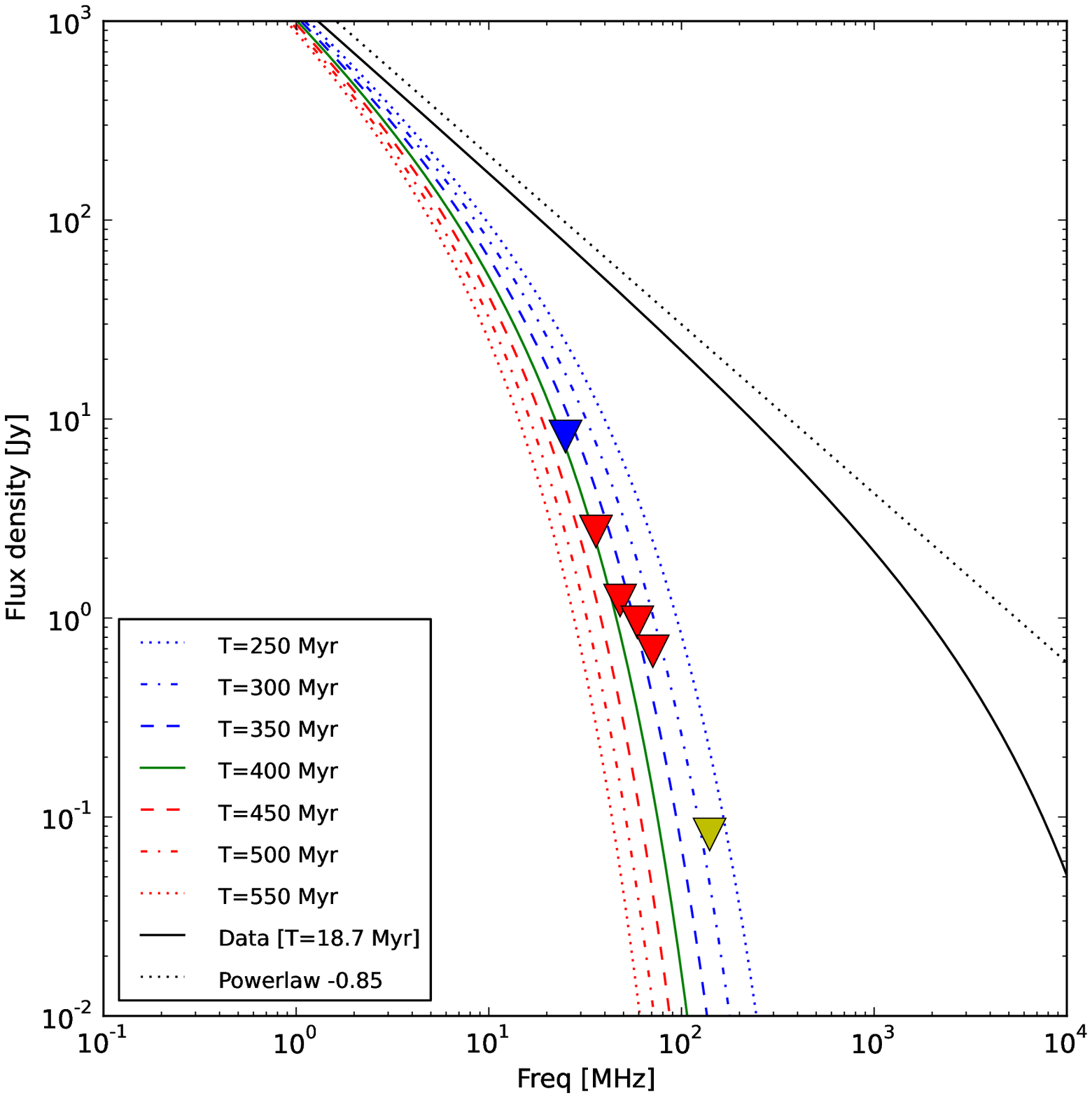}
\caption{To estimate our capability to detect ancient outbursts we determined the maximum age of an emission we would be able to detect given the sensitivity of our maps. We took the zone H1 (Fig.~\ref{fig:zones}) as representative for an old and inactive zone in the halo. Using its $\nu_{\rm b} = 8.3\un GHz$ and its estimated age of 18.7~Myr we let its spectrum evolve. In this figure we show the fit to the observed data as a black line. Coloured lines show the evolution of the expected spectral shape as time passes by (from top to bottom, after 250, 300, 350, 400, 450, 500 and 550~Myr, respectively, after the electron injection). Triangles indicates the $3 \sigma$ detection limit for our maps. Blue is for the LBA-low map (Fig.~\ref{fig:VirA-LLBA}), red for the four LBA-high maps (Fig.~\ref{fig:VirA-LBA}), while yellow for the HBA map (Fig.~\ref{fig:VirA-HBA}).}
\label{fig:oldburst}
\end{figure}

In Fig.~\ref{fig:oldburst} we plot the theoretical temporal evolution of the radio spectrum in the halo zone H1 using the standard JP model. We simulated how the spectrum of a zone with these characteristics can evolve and how old it must be in order to go undetected in our maps. We find that in this illustrative example we would be able to detect emission as old as $\sim 400$~Myr. This number is rather optimistic and it would decrease if adiabatic expansion had played an important role or if we had overestimated the current age of the zone. Finally, given the confinement of the source discussed in Sect.~\ref{sec:flux}, it is possible that particles from older AGN events were mixed with those from recent events. This can also play a role in steepening the low-frequency end of the lobes' spectrum. Such a scenario was observed in simulation by \cite{Morsony2010}, where the authors simulated AGN driven jets in a dynamic, cosmologically evolved galaxy cluster. They found that largest scale reached by AGN jets 
is only 
proportional to the AGN power ($R \propto P_{\rm j}^{1/3}$) and does not depend on the activity time. In this case the estimated halo age should be interpreted as a lower limit.

\section{Discussion}
\label{sec:discussion}

Radio morphological evidence (Sect.~\ref{sec:images}) and spectral analysis (Sect.~\ref{sec:spectral}), show that the Virgo~A halo is an active part of the source and not a relic of past activities \citep[as already pointed out by][]{Owen2000}. Its radio emission is confined inside sharp boundaries.

The low-frequency spectral index map is fairly uniform, apart from a flattening in the central cocoon and in the northern lobe, and a steepening in the regions where the flow activity is fading. Thanks to the LOFAR data, together with high frequency observations up to 10~GHz, we were able to extract wide-band radio spectra of the source halo to an unprecedented detail. A continuous injection model applied to the whole halo shows cut-off frequencies at $\sim 1.3$~GHz, which provides an estimation of the halo age of $\simeq 40\un Myr$.

We performed a detailed spectral analysis of nine different zones in the halo. Leaving all the parameters free to vary, we obtain a good fit only assuming $\alpha_{\rm inj} \simeq -0.85$ (see Fig.~\ref{fig:spectra}). In order to reduce the number of free parameters we tried to set the slope of the energy distribution of the injected electron population to $\delta_{\rm inj} = -2.2$ ($\alpha_{\rm inj} = -0.6$), as we observe in the source central region, and which is also in perfect agreement with the broad-band spectrum of the jet of M87 \citep{Perlman05}. However, in this case the JP and KP models fail to reproduce the observed spectra, while fixing $\alpha_{\rm inj} \simeq -0.85$ produces reasonable results with both models (see Fig.~\ref{fig:spectra-alphafix}). Observationally, this reflects the presence of a steep low-frequency end in the spectra of all regions in the halo, even those just outside the central cocoon. We speculate that this steep spectrum is the consequence of the strong adiabatic expansion of a mix of plasmas at different ages that takes place as soon as the plasma bubbles leave the dense central area, shifting their already steepened high-frequency part of their spectrum down to the MHz region. Once we assume $\alpha_{\rm inj} \simeq -0.85$, we are indeed able to follow the ageing of the plasma bubbles along their path. For example, in the west flow we observe a shift of the break frequency ($\rm 15~GHz \rightarrow 9~GHz \rightarrow 8~GHz$), which reflects an ageing of the electrons ($\rm 7~Myr \rightarrow 10~Myr \rightarrow 12~Myr$), as we move farther from the source centre along the flow. It is worth noting that between region W2 and W3 the results with error-bars are consistent with no-ageing, we can speculate that those two regions are, in projection, at the a similar distance from W1 but on different fronts of the flow, which is then creating a mushroom-like structure close to the edge of the halo (as in the east flow) instead of flowing along it. This picture can also explain the spectral index map in the southern lobe, where the flatter values, which seem to be connected with the active flow and initially follow it, end directly at the edge of the lobe that is where the flow should separate and create the top of the mushroom-shaped lobe. This picture is very much in line with the source three dimensional reconstruction proposed in \cite{Churazov2001}, but requires more detailed analyses to be confirmed.

The minimum-pressure analysis in nine different zones in the source provides equipartition magnetic field values and relative pressures. These values are compatible with those found by \cite{Owen2000}, although some of the initial assumptions were different: a different low-energy cut-off, different spectral slope and a different model for the synchrotron emissivity. Nevertheless, the net balance of these modifications reduces only slightly the values of the magnetic field strength we estimated for the Virgo~A halo. It is worth pointing out that all the obtained $B_{\rm eq}$ values are a few times above $3.25\un \mu G$, below which the Inverse Compton scattering of CMB photons is the dominant effect compared to synchrotron losses.

We compared our minimum pressure values with those derived from XMM observations \citep{Matsushita02}, listed in the last column of Table~\ref{tab:equip}. The thermal gas pressure is always more than an order of magnitude higher than the minimum pressure obtained from equipartition analysis. This could imply that the regions where plasma electrons and magnetic fields are co-located are much smaller than the assumed zone volumes (a ``filling factor'' $< 1$), or that much of the energy is contained in the thermal gas, or that the source is not at the equipartition. Yet another possibility is that much of the energy is stored in the relativistic protons \citep[see e.g.][]{Dunn2004,Birzan2008}. In this case, to keep the equipartition principle valid, we would have to assume a $k\gtrsim1000$, although this would generate much higher magnetic field strengths and therefore unrealistic short plasma ages. It is well known that inside the inner cocoon radio jets create cavities in the thermal gas, whereas on the larger halo scales there is evidence for the uplift of cold thermal gas \citep[see e.g.][]{Forman07,Million2010} along the radio-flows. This has been explained by \cite{Churazov2001} as gas uplifted by buoyant bubbles from the cold cluster core. In such a scenario the amount of pressure provided by thermal gas cannot be neglected and accounts for a not-negligible fraction of the total pressure. Furthermore, using X-ray and optical data, \cite{Churazov2008} also found that the combined contribution of cosmic rays, magnetic fields and micro-turbulence to the total pressure in the core of M87 is $\sim10\%$ of the gas thermal pressure. This is in line with our findings.

\subsection{Plasma age and dynamical time}
\label{sec:ages}

Hydro-dynamic simulations of buoyant bubbles suggest that the halo outskirts can be reached in $42-67\un Myr$ \citep[][the proper number depending on the orientation and assumed distance]{Churazov2001}. These estimations are a factor of $\sim 4$ greater than what was found in our synchrotron spectral analysis. \cite{Gull73} first noticed that the lifetime of synchrotron-emitting particles is short compared with dynamical time-scales. Some reasons that can explain this discrepancy are presented in \cite{Churazov2001}, \cite{Blundell2000}, and \cite{Owen2000}: (i) the bubbles may be filled with a mix of weak and strong magnetic fields, the relativistic electrons may survive for a long time in the weak magnetic fields and then radiate most of their energies as soon as they diffuse in the strong magnetic field regions. This picture would also account for the filamentary structure visible in the radio images. (ii) In situ acceleration of particles may play an important role \citep{Parma99,Prieto02}. In both these cases the age of the radiating particles can hardly be determined 
by radio observations. (iii) New plasma can also flow along pre-existing channels and replenish aged plasma. 

Finally, the buoyant rise time may not necessarily be the correct time, the radio plasma that forms bubbles comes from a jet which likely leaves it with a momentum. \cite{Bruggen2002} simulated this situation in an environment compatible with that of Virgo and found in this case that the plasma can reach the distance of 20 kpc in $\simeq 15\un Myr$, which is very much similar to what we measure.

\subsection{Energetics}
\label{sec:energetics}

In \cite{Owen2000} the authors derived an estimate of the halo age using energetic conservation arguments, following \cite{Eilek89}, from
\begin{equation}
 \frac{{\rm d}U_{\rm int}}{{\rm d}t} = P_j - p \frac{{\rm d}V}{{\rm d}t} - L_{\rm rad}.
\end{equation}
where $p$ the pressure of the halo plasma, $V$ the volume, $U_{\rm int}$ the internal energy, $P_j$ the jet power and $L_{\rm rad}$ is the radiative losses from the halo, which is small compared to $P_j$ and can be neglected. We can solve the equation assuming that the halo is spherically symmetric with a radius $R$ and that it expands due to its own internal energy up to $R = 35$~kpc. We also assumed that $P_j$ is time-independent. If the expansion is slow, the pressure of the bubble can be approximated with the pressure of the surrounding medium, which we obtained from XMM temperature and density profiles \citep{Matsushita02}. For a non-relativistic plasma (most of the bubble content is thermal, with $\Gamma = 5/3$), we obtain a halo age of $\sim 250 \left(P_{j}/10^{44}\right)^{-1}\un Myr$. For a bubble dominated by relativistic particles and magnetic fields ($\Gamma = 4/3$), we obtain $\sim 400 \left(P_{j}/10^{44}\right)^{-1}\un Myr$. It is important to notice that a shorter burst of energetic particles would require a much higher $P_j$ (although for a shorter duration) than a long continuous injection of particles. In this simple model, our measurements of the halo age based on equipartition ($t \simeq 40\un Myr$), provide an estimate of the jet power of $P_j \simeq 6 \times 10^{44}\un erg\ s^{-1}$ for $\Gamma = 5/3$ and of $P_j \simeq 10 \times 10^{44}\un erg\ s^{-1}$ for $\Gamma = 4/3$. This result is consistent with the conclusions of \cite{Owen2000} where the jet power $P_j$ is estimated to be $\sim \rm few \times 10^{44}\un erg\ s^{-1}$ and with \cite{DiMatteo2003} which predict a $P_j \simeq 5 \times 10^{44}\un erg\ s^{-1}$ for accretion at the Bondi rate. The result is instead above what has been found by other authors, e.g.~\cite{Reynolds96} find $P_j\simeq 10^{43}\un erg\ s^{-1}$. This energy supply is $\sim100$ times higher than the X-ray luminosity of the cooling flow region around M87, that is $\sim 3 \times 10^{43}\un erg\ s^{-1}$~\citep{Mushotzky1980}. Therefore, even a modest efficiency of energy dissipation into heat, is able to exceed the radiative cooling of the gas. Finally, we notice that an energetically proton-dominated plasma ($k \simeq 1000$) would lower by a factor of $\sim 10$ the age of the halo, boosting the necessary jet power up by the same amount which would drive the source at quasar luminosity \citep{Falcke95}, which is not observed.

\subsection{Ultra high energy cosmic rays}
\label{sec:uhecr}

A debated argument is the possibility for a radio galaxy like M87 to accelerate ultra high energy cosmic rays (UHECR) of $\sim 10^{20}\un eV$ directly in the radio lobes. The \cite{Auger07} found a correlation between the arrival directions of cosmic rays with energy
above $6 \times 10^{19}\un eV$ and the positions of AGN within $\sim 75\un Mpc$, with a small excess of detections in the direction of Centaurus~A, but no event detected in the direction of Virgo~A. \cite{Ghisellini08} correlated the position of UHECR events with the directions of H\textsc{i}-selected galaxies, and proposed that the UHECR coming from the direction of Centaurus~A instead originate from the more distant Centaurus cluster, whose galaxies are in fact richer in H\textsc{i} than the galaxies of the Virgo cluster, explaining why there is no UHECR event from the Virgo direction. This opens the possibility that UHECR are produced by gamma ray bursts or newly born magnetars, but does not rule out an AGN origin.

In Centaurus~A, the lobe radius ($\sim 100\un kpc$ wide) and estimated magnetic field strength \citep[$B \simeq 1\un \mu G$,][]{Hardcastle09} satisfy the Hillas argument \citep{Hillas84}, which requires that the Larmor radius $r_L$ of the accelerated particles fits within the source. The maximum particle energy $E_p = E_{20}\ 10^{20}\un eV$ that can be accelerated inside the Virgo~A lobes is 
\[
E_{20} = Z\ e\ r_L\ B \simeq Z\ r_{100}\ B_{-6}\un eV\,,
\]
where $r_{100} = r_L\ 100\un kpc$ and $B_{-6}=B\ 10^{-6}\un G$. Using an average magnetic field of $10\un \mu G$ and a lobe radius of 20~kpc, we estimate the maximum particle energies to be $E_p \simeq Z\ 2 \times 10^{20}\un eV$, where $Z$ is the atomic number. However, efficient stochastic acceleration processes due to resonant interaction between particles and magnetic field turbulent perturbations require lobe contents to be predominantly relativistic plasma, and this is probably not the case for Virgo~A.

\section{Conclusions}
\label{sec:conclusions}

In this paper we presented the first LOFAR observation of Virgo~A in the $15-162$~MHz frequency range and an analysis of its radio spectra. The major results are:

\begin{itemize}
 \item Virgo~A halo is an active part of the source and not a relic of past activities \citep[as already pointed out by][]{Owen2000}.
 \item Going down to 25~MHz no previously unseen steep-spectrum features were detected. The source appears instead to be well confined within boundaries that are identical at all frequencies.
 \item A low-frequency spectral index map of Virgo~A shows no obvious relation between spectral index and brightness. Generally, a steepening of the spectral index is present where the morphological evidences of flow activities are reduced in the southern lobe and at the position of the north-east X-ray cavity. A flattening of the spectral index, instead, is visible at the northern lobe towards west and in the position of enhanced radio flux density.
 \item A spectral analysis of the extended halo was performed. With the assumption of $\alpha_{\rm inj} = -0.6$ as in the core region, characterized by the flatter spectral slope, none of models tested is able to fit the data. Instead, $\alpha_{\rm inj} = -0.85$ is required, outside the central cocoon, to explain the spectra. This steepening in the low-frequency end of the spectra can be connected to a strong adiabatic expansion of the plasma bubbles that happens as soon as they leave the dense central region.
 \item An equipartition analysis was conducted and an average magnetic field strength of $\simeq 13\un \mu G$ was found in the flow regions, while a magnetic field of $\simeq 10\un \mu G$ is present in the halo regions. In the inner cocoon the average magnetic field reaches $\simeq 30\un \mu G$.
 \item A synchrotron ageing analysis provided a global halo age of $\simeq 40\un Myr$. The particle age tends to increase with distance from the centre in the flow regions and reaches a maximum of $\simeq 15\un Myr$ where the flows end. This age is about a factor of $\sim 4$ less than the dynamic time of a buoyantly raising bubble.
 \item Minimum pressure analysis reveals that, given our assumptions, the pressure generated by the plasma and the magnetic fields is much less than what is required to sustain the halo against external pressure. Probably, thermal gas is also playing a role in sustaining the halo. On the other hand, some of the parameters we assumed, or the equipartition hypothesis itself, may not be correct.
 \item We estimate the jet power $P_j$ to be $6-10 \times 10^{44}\un erg\ s^{-1}$. This energy supply is 10 to 100 times higher than the X-ray luminosity of the cooling flow region around M87 ($\sim 3 \times 10^{43}\un erg\ s^{-1}$) and is therefore able to exceed the radiative cooling of the gas.
\end{itemize}

The extended radio-halo of Virgo~A is a composite of many plasma bubbles of different ages. They are inflated in the central cocoon, where clear cavities in the thermal gas are visible, by the source's powerful jets. As soon as they leave the dense central area, a strong adiabatic expansion shifts their radio emission towards lower frequencies and lower fluxes. Then they buoyantly rise towards the halo outskirts where they disperse. During their motion, the travelling bubbles lift up cold X-ray emitting gas from the centre of the cluster.

To better constrain different synchrotron models and ultimately draw conclusions about the underlying mechanisms, further observations are necessary. Steps forward in the study of Virgo~A require: (i) high angular resolution maps of the Virgo~A halo at frequencies above 4 GHz, (ii) higher angular resolution and (iii) higher dynamic range on low frequency ($<200\un MHz$) maps. Once these requirements are fulfilled, wide-frequency spectral studies can be performed on regions of the halo that are small enough to disentangle the diffuse emission from that originating in the filaments.

Point (i) will be soon feasible with the VLA and next-generation single-dish radio telescopes, like the Sardinia Radio Telescope\footnote{http://www.srt.inaf.it.}. In the C-band ($4-8$~GHz) the VLA will require mosaicing and single dish observations to fill in the shortest baselines, the results of these observations will be presented in forthcoming papers. At higher frequencies, single dish radio-telescopes have the required sensitivity to detect the extended halo and they can be used at the price of lower resolution, currently $\sim 40\arcsec$ at $20\un GHz$.

Requirements (ii) and (iii) will be soon available with new LOFAR data and presented in forthcoming papers. The present antenna configuration (excluding international stations) can provide baselines of 80 km, which allow an angular resolution of $25\arcsec$ at 30~MHz. Furthermore, the important technical improvements achieved during commissioning will provide better knowledge of the telescope behaviour while the 8 new stations now available will give a much better \textit{uv}-coverage. With these improvements we will be able to greatly reduce  deconvolution errors and improve the dynamic range of future maps.

\begin{acknowledgements}
The authors wish to thank Eugene Churazov and Hans B\"ohringer for many interesting discussions, Frazer Owen for providing the 325~MHz map and Helge Rottmann for providing the 10.55~GHz map.
LOFAR, the Low-Frequency Array designed and constructed by ASTRON, has facilities in several countries, that are owned by various parties (each with their own funding sources), and that are collectively operated by the International LOFAR Telescope (ILT) foundation under a joint scientific policy.
C.~Ferrari and G.~Macario acknowledge financial support by the ``Agence Nationale de la Recherche'' through grant ANR-09-JCJC-0001. 
\end{acknowledgements}

\bibliographystyle{aa}
\bibliography{lofar-virgo}

\label{lastpage}

\end{document}